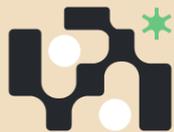

# Frequency-Histogram Coarse Graining in Elementary Cellular Automata and 2D CA

Sanyam Jain [1], Stefano Nichele [2],[3]

1. *Dept. of Computer Science and Communication, Østfold University College, Halden, Norway*
2. *Dept. of Computer Science, Oslo Metropolitan University, Oslo, Norway*
3. *Dept. of Computer Science and Communication, Østfold University College, Halden, Norway*

**Abstract**
Cellular automata and other discrete dynamical systems have long been studied as models of emergent complexity. Recently, neural cellular automata have been proposed as models to investigate the emerge of a more general artificial intelligence, thanks to their propensity to support properties such as self-organization, emergence, and open-endedness. However, understanding emergent complexity in large scale systems is an open challenge. How can the important computations leading to emergent complex structures and behaviors be identified? In this work, we systematically investigate a form of dimensionality reduction for 1-dimensional and 2-dimensional cellular automata based on coarse-graining of macrostates into smaller blocks. We discuss selected examples and provide the entire exploration of coarse graining with different filtering levels in the appendix (available also digitally at this link: https://s4nyam.github.io/eca88/). We argue that being able to capture emergent complexity in AI systems may pave the way to open-ended evolution, a plausible path to reach artificial general intelligence.

**Keywords:** Artificial Intelligence; Cellular Automata; Complexity; Artificial Life; Coarse Graining

## Introduction

Recent advances in deep learning, such as large language models, have shown promise in some tasks thought to be possible only in biological intelligence. However, such systems remain rather rigid if compared to the adaptivity and the open-endedness of living organisms [1]. In fact, natural intelligence may be considered a form of spontaneously self-organizing and ever evolving complex system producing increasing levels of emergent complexity [2]. Clune [3] describes the AI generating algorithm approach as a possible alternative to creating machines that can "automatically learn how to produce general AI", as opposed to the "manual AI approach" which is predominant in the AI community nowadays. *Natural evolution is the only generating algorithm that has been able to produce general intelligence.* Artificial Life frameworks such as Cellular Automata (CA) have long been used as models to understand computation, life, and evolution.

Recent advances of such models, including continuous CA [4, 5] and neural CA [6, 7], have been proposed as substrates to study the emergence of a general intelligence [8]. However, an open question is how to quantify the emergent complexity of such artificial systems at different scales, in order to guide evolution towards higher emergent complexity. In this work, we consider Elementary Cellular Automata (ECA), 2D CA based on Game of Life rules, and an evolved multiple neighborhoods CA (MNCA) rule. We perform a dimensionality reduction process using a coarse-graining technique [9], in order to identify complex pattern formations and highlight them while filtering out uninteresting computations happening in the background. We utilize a frequency-based technique that scans the space-time diagrams and collects frequencies of different blocks, which are then used for coarse-graining by effectively replacing blocks with frequency below a certain threshold with a meaningful state, in fact reducing the dimensionality of the CA. Additionally, such coarse-graining technique may allow identifying emergent complexity at different scales. We highlight some example results in our discussion (while leaving the full results for the Appendix), with the aim of discussing the implications of this approach for the long-term goal of understanding how AI systems may become more open-ended, adaptive, and general, through the emergence of complex behaviors out of simple rules.

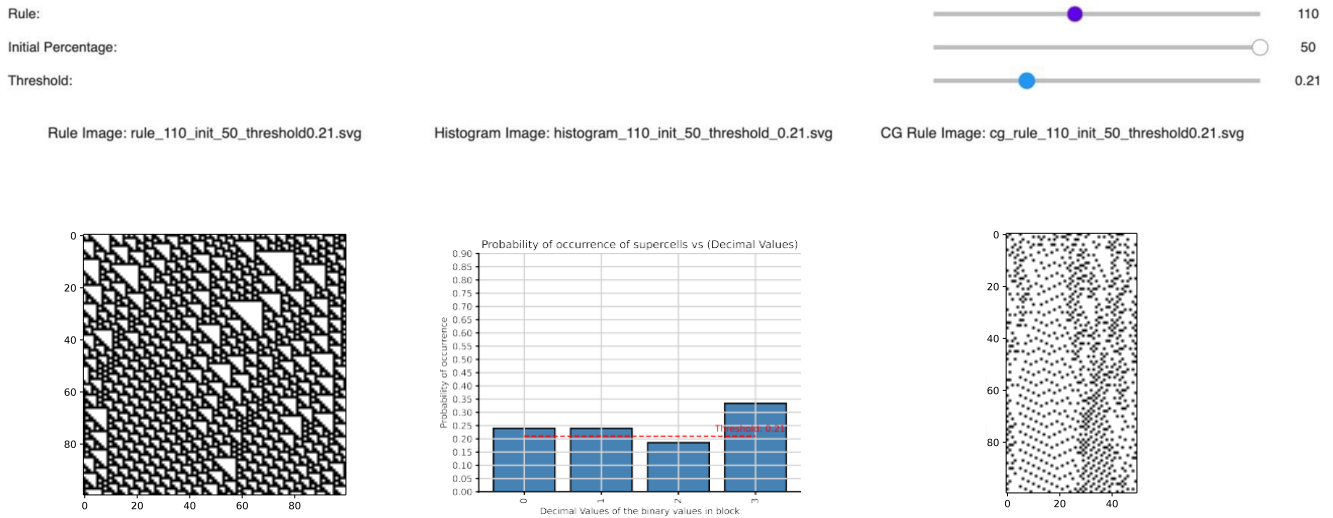

Figure 1: Snapshot from the platform to study Binary FHCG for ECA. In this example, Rule 110 is depicted on the left. In the centre, the frequency histogram for blocks of size 2x1 is shown. On the right, the resulting coarse graining with the selected threshold. The rule, the initial state, and the threshold can be selected from the sliders at the top.

**Background and Related Work**

*Cellular Automata*

Cellular Automata are computational models characterized by a regular grid of cells, each capable of existing in a discrete number of states. CA evole through discrete time steps guided by simple rules dependent on neighboring cell states. Among them, Elementary Cellular Automata (ECA) stand out with their one-dimensional grid, featuring binary cell states (on or off). Despite their simplicity, ECAs exhibit remarkable complexity [10] as the future state of each cell relies solely on its current state and the states of its two neighboring cells, leading to the emergence of intricate patterns. Stephen Wolfram [11] classified their behavior in four qualitative classes: 1) evolving to homogeneous states; 2) evolving to simple and periodic patterns; 3) evolving to disordered patterns; 4) evolving to complex localized structures. It has been demonstrated that, due to different types of equivalences, there exist a set of 88 unique ECA rules (out of the 256 possible rules) [12].

The Game of Life (GoL) [13] is one of the most renowned CA instances, operating on a two-dimensional grid with cells in either alive or dead states. GoL evolves through discrete generations based on specific rules involving the eight adjacent neighbors' states, leading to intriguing phenomena such as self-replicating structures and gliders. Advancing further, Multi-Neighbor Cellular Automata (MNCA) [14] enable interactions with a broader range of neighboring cells, allowing simulations of complex phenomena like urban growth patterns, traffic flow, and ecological dynamics, where local interactions with multiple neighbors significantly influence macroscopic behavior and emergent properties. These diverse branches of cellular automata find applications across disciplines, offering valuable insights into complexity and emergent behavior, showcasing how elegance and simplicity can yield a rich tapestry of patterns and behaviors and contributing to our understanding of complex systems in both natural and artificial domains [10].

*Complexity and Open-Endedness*

Gaining insight into the emergent computation of a complex system is not trivial, in particular by looking at the activity of the simple components of the system. In [15], Melanie Mitchell explains how a "computational mechanics" framework may be employed to explain the emergent complexity of a CA. In CA, particles and their interactions may describe the computation embedded in the space-time behavior, providing insights into how the decentralized processing happens. Mitchell argues that "the language of particles and their interactions form an explanatory vocabulary for decentralised computation in the context of one-dimensional cellular automata". One popular way to approximate the behavior of a CA and assess its complexity is through compression. For example, in [16] they partition the 1D CA space through compression. In [17], compression is used also for 2D and 3D CA. However, compression as proxy for complexity has some issues, such as the fact that random or chaotic states are the most incompressible, while complex localized patterns have somewhat "intermediate" levels of compressibility. For a review of different proposed complexity metrics for CA and other complex systems, please refer to [18], where the notion of complexity as degree of hierarchy is of particular interest when complexity emerging at different scales is of relevance. One notable example of equal complexity at different hierarchical scales is the implementation of the Game of Life inside the Game of Life [19]. Recent works [9, 20, 21] show more advanced methods to capture and evaluate complex emerging behaviour. Such works propose multiple approaches for both discrete CA and continuous CA. In particular, the authors investigate coarse-graining and artificial neural networks (Auto-Encoders) to quantify complexity, conducting experiments to demonstrate the metric's ability to automatically identify CA with complex emerging patterns. The coase-graining method proposed in [9], based on frequency-histograms, has been shown to qualitatively approximate selected ECA rules to the particle domain boundaries presented in [22, 18].

In this work, we systematically explore frequency-histogram coarse graining for the entire ECA space, 2D Game of Life, and an example of MNCA.

---

**Algorithm 1** Auto Thresholds & processing FHCG of 1D Cellular Automata

**Data:** List of 1D CA rules: R, State (NumPy array): S, List of initial percentages: I, $p_i$ is the probability of observing supercell i on a grid G, $s_i$ is a given supercell in the set S(in) containing all the possible blocks, and $count_G(s_i)$ is the number of blocks matching $s_i$ in G. $\delta$ is the fraction amount added to uplift probability for Thresholds.

**Result:** Modified Automatons, Thresholds

**foreach** `step_state` *in* S **do**
  chunks = blockshaped (`step_state, block_size`)
  **foreach** block *in* chunks **do**
    `supercells` ← block
  **end**
**end**
▷ At this stage you should have collection of all chunks, in ascending order, of 1D CA, for "t" steps.
**foreach** `super_cell` *in* `supercells` **do**
  $p_i = \frac{count_G(s_i)}{\sum_{j \in S(in)} count_G(s_j)}$
  `new_board` ← $p_i$
**end**
  ▷ Store unique Probabilities of each block type, block types are shown below

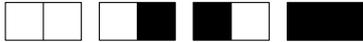

**foreach** $p_i$ *in* `unique_probabilities` **do**
  Thresholds ← sorted ($p_i + \delta$)
**end**
**foreach** $\alpha$ *in* Thresholds **do**
  $G(i) = \begin{cases} 0, & \text{if } p_i \geq \alpha \\ 1, & \text{if } p_i < \alpha \end{cases}$
  ModifiedAutomatons.append(Reshape(Automaton, coarse_shape))
**end**
**return** ModifiedAutomatons

---

*Frequency-Histogram Coarse Graining*

In order to capture and possibly quantify emergent complexity in dynamical systems, it is beneficial to reduce large systems into smaller sizes, while being able to preserve important pattern formations and filter out background computations.

The work in [9] introduces different coarse graining methods, namely Frequency-Histogram Coarse Graining (FHCG), Clustering, and Auto-Encoder based Coarse Graining. Of particular interest due to its computational simplicity is FHCG, where blocks of cells are mapped to supercell states based on the probability of their occurrence in the CA space-time evolution. The frequency of different configurations is computed and used to create a frequency histogram. Supercells are then assigned new states based on their frequencies, with a threshold determining the mapping to either state 0 or 1 (in case of binary coarse graining). It also allows for multiple-state coarse graining.

**Proposed Methods**

Our proposed coarse graining method is available here https://s4nyam.github.io/mncaportal. The experimental approach uses a Python framework to implement ECA with FHCG. The algorithm automates the simulation and visualization of 1D and 2D cellular automata rules. It defines functions to generate and apply a cellular automaton rule, simulate the automaton for a given rule and initial configuration, and process the resulting automaton into small chunks called supercells. It then calculates probabilities of occurrence for each supercell configuration and visualizes their histograms. The algorithm further defines a function to visualize the original and processed automaton for different probability thresholds, providing insights into the dynamic patterns formed by different cellular automaton rules. Finally, the algorithm applies this visualization process to a set of predefined cellular automata rules from various initial conditions to explore their behavior and pattern formations. Algorithm 1 explains the simple working of FHCG in ECA.

The `process_FHCG` function takes a 2D matrix `state` and two parameters `rule_number` and `initial_percentage`. It processes the `state` matrix as follows:

1. The matrix is divided into 2x1 blocks (The `blockshaped` function, used to divide the matrix into blocks.), and the flattened blocks are stored in `supercells_list`.

2. The occurrence probabilities of unique elements in `supercells_list` are calculated and stored in `list_probs`.

3. Histograms are plotted for each threshold value (slightly increased from probabilities) showing the probability of supercell occurrence for decimal values of binary block elements.

4. For each threshold, a modified grid is created where values less than or equal to the threshold are set to 1 and others to 0. These grids are stored in dataset.

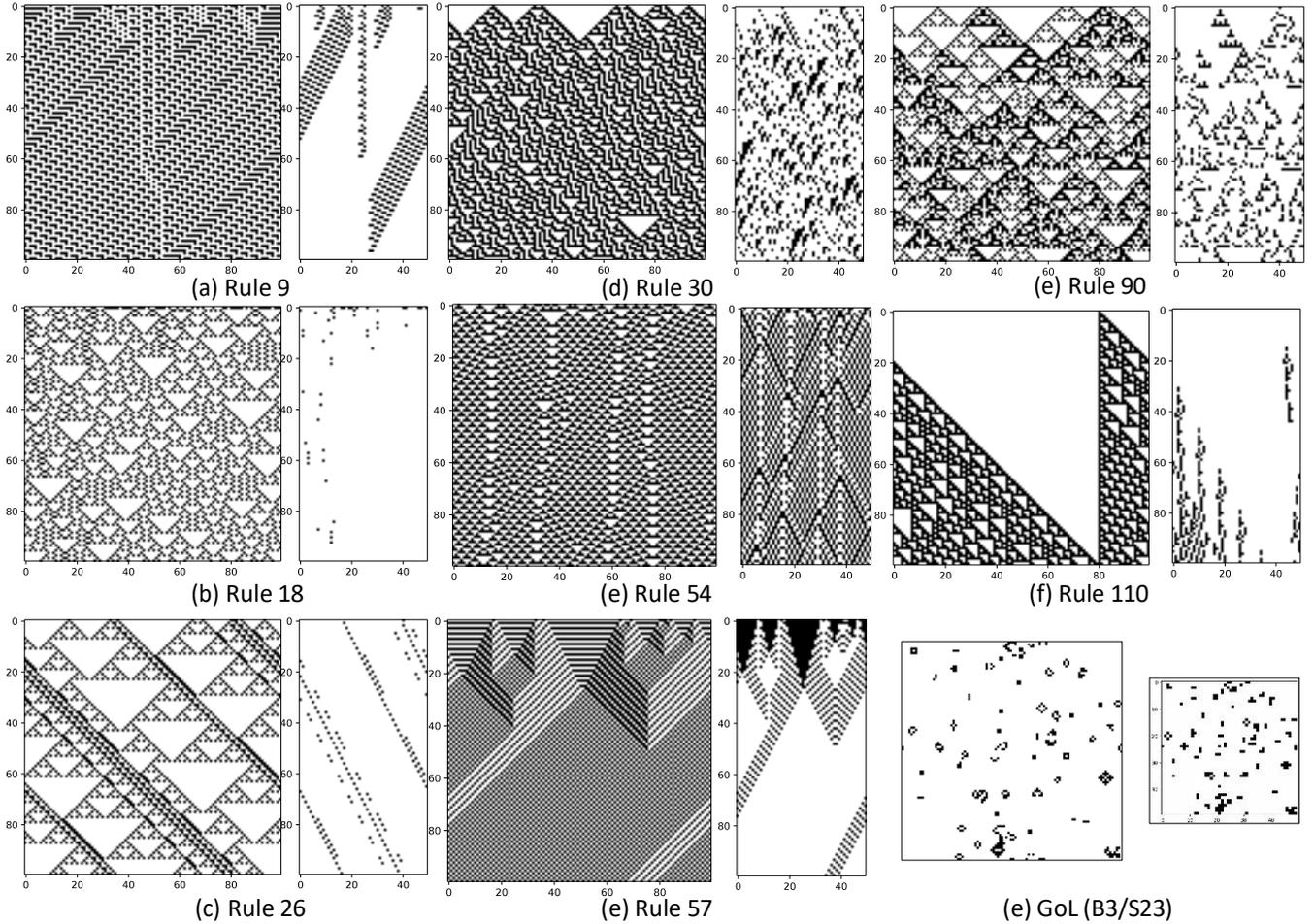

Figure 2: Example results from selected rules. Left: space-time diagram; Right: coarse-grained space-time diagram.

`process_FHCG` returns: `modified_automatons` (a list of modified 2D matrices) and `thresholds` (the list of used thresholds). We use this proposed FHCG technique to study complex behaviour in ECA for the unique 88 rules (by considering their Minimum Equivalence [23]). We propose a novel dataset consisting of all 88 rules visualised using automatic thresholding for FHCG and corresponding CA output. We let the FHCG function decide how to set the thresholds for each simulation dynamically, by keeping the probability of each block preserved. For each of the four possible blocks, we add a fraction δ (here 0.005) to the highest probability value and assign it as one of the three thresholds. A snapshot for such a platform to study ECA is shown in Figure 1.

**Discussion of Results and Conclusion**

An extract of our full results (presented in the Appendix) is depicted in Figure 2. Our method of FHCG effectively compresses the cellular automata representations while highlighting rare events, making it easier to study interesting emergent behaviors and their complexity. Such understanding is crucial for studying emergent phenomena and understanding the evolution in cellular automata and other dynamical systems.

In particular, it can be seen that hidden structures can be discovered in the underlying computation for several rules, such as rule 18, 30, and 90 (class III) or rule 110 (class IV). In rule 54 (class IV), it can be easily seen that diagonal particles are discovered. For rules with simple periodic patterns as rule 9, 26, or 57 (class II), the background computations are filtered our and the emergent patterns are extracted. In case of two-dimensions, the gliders and other emergent computational structures in the Game of Life, are simplified in a lower dimension. While some granularity is lost, it can be easily identified where the computations are ongoing (for example a glider may be replaced by a more compact structure).

These techniques can be applied recursively and in combination to study such systems at various scales, providing valuable insights into their emergent properties and overall complexity. Being able to identify and quantify emergent complexity for large-scale evolving systems may be a crucial step in achieving a more general artificial intelligence. Novel ways of guiding evolution and learning towards higher complexity levels may be discovered, ultimately allowing our artificial systems to become more open-ended and adaptive.

**References**

1. Stanley KO. Why open-endedness matters. Artificial life 2019; 25:232–5
2. Booker L. Perspectives on adaptation in natural and artificial systems. Vol. 8. Santa Fe Institute Studies on, 2005
3. Clune J. AI-GAs: AI-generating algorithms, an alternate paradigm for producing general artificial intelligence. arXiv preprint arXiv:1905.10985 2019
4. Chan BWC. Lenia-biology of artificial life. arXiv preprint arXiv:1812.05433 2018
5. Chan BWC. Lenia and Expanded Universe. *The 2020 Conference on Artificial Life*. MIT Press, 2020
6. Mordvintsev A, Randazzo E, Niklasson E, and Levin M. Growing neural cellular automata. Distill 2020; 5:e23
7. Variengien A, Nichele S, Glover T, and Pontes-Filho S. Towards self-organized control: Using neural cellular automata to robustly control a cart-pole agent. arXiv preprint arXiv:2106.15240 2021
8. Gregor K and Besse F. Self-Organizing Intelligent Matter: A blueprint for an AI generating algorithm. 2021. arXiv: 2101.07627 [cs.NE]
9. Cisneros H, Sivic J, and Mikolov T. Visualizing computation in large-scale cellular automata. *The 2020 Conference on Artificial Life*. MIT Press, 2020
10. Wolfram S et al. A new kind of science. Vol. 5. Wolfram media Champaign, IL, 2002
11. Wolfram S. Universality and complexity in cellular automata. Physica D: Nonlinear Phenomena 1984; 10:1–35
12. Wolfram S. Theory and applications of cellular automata. World Scientific 1986
13. Berlekamp ER, Conway JH, and Guy RK. Winning ways for your mathematical plays, volume 4. AK Peters/CRC Press, 2004
14. Kraakman B. Understanding Multiple Neighborhood Cellular Automata. 2021
15. Mitchell M, Crutchfield JP, Das R, et al. Evolving cellular automata with genetic algorithms: A review of recent work. *Proceedings EvCA'96*. Vol. 8. Moscow. 1996
16. Zenil H. Compression-based investigation of the dynamical properties of cellular automata and other systems. Complex Systems 2010; 19:1–28
17. Nichele S and Tufte G. Measuring Phenotypic Structural Complexity of Artificial Cellular Organisms: Approximation of Kolmogorov Complexity with Lempel-Ziv Compression. *Innovations in Bio-inspired Computing and Applications: Proceedings of the 4th International Conference on Innovations in Bio-Inspired Computing and Applications, IBICA 2013, August 22-24, 2013-Ostrava, Czech Republic*. Springer. 2014 :23–35
18. Mitchell M. Complexity: A guided tour. Oxford university press, 2009
19. Bradury P. Life in Life. Youtube. 2012. Available from: https://youtu.be/xP5-iIeKXE8
20. Cisneros H, Sivic J, and Mikolov T. Evolving structures in complex systems. *2019 IEEE Symposium Series on Computational Intelligence (SSCI)*. IEEE. 2019 :230–7
21. Jain S, Shrestha A, and Nichele S. Capturing emerging complexity in lenia. *Italian Workshop on Artificial Life and Evolutionary Computation*. Springer. 2023 :41–53
22. Hanson JE and Crutchfield JP. Computational mechanics of cellular automata: An example. Physica D: Nonlinear Phenomena 1997; 103:169–89
23. Castillo-Ramirez A and Magaña-Chavez MG. A study on the composition of elementary cellular automata. arXiv preprint arXiv:2305.02947 2023
24. Medernach D, Kowaliw T, Ryan C, and Doursat R. Long-term evolutionary dynamics in heterogeneous cellular automata. *Proceedings of the 15th annual conference on Genetic and evolutionary computation*. 2013 :231–8

**Appendix**

In the following pages of the Appendix, we are presenting a systematic investigation of coarse-graining for:

- The 88 Minimum Equivalent ECA rules
- The Game of Life
- An evolved MNCA rule.

Results for ECA are also available digitally at the following link:

- https://s4nyam.github.io/eca88.

A portal which provides the pointers to all our code implementations is available at the following link:

- https://s4nyam.github.io/mncaportal.

Starting from the next page, we present one rule per page. Each page is structured as follows:

- The first row and first image displays the original CA simulation with 1% initialisation. The CA size is 100, and the rule is applied for 100 steps.

- Further, the second image shows the histogram plot for each of the blocks, containing three different thresholds. Note that the used block size is 2x1 for 1D CA and 2x2 for 2D CA.

- Further, the next three images show the coarse-grained versions of the original CA simulation with the corresponding thresholds applied in ascending order.

- Similarly, for the second and third rows, the density of initialisation is 20% and 50% respectively.

Note that for some rules, not all blocks are present in the space-time diagram and therefore there are less bars in the histogram and less coarse-graining results visualized. The thresholds are obtained by adding a $\delta$ of $0.005$ to the calculated probabilities. Please note that for some rules, the resulting threshold includes two histogram bars with almost identical values. This results in less coarse-graining to visualize. An example is Rule 43 (Third row with $50\%$ initialisation in Table 38), where several bars have almost identical frequency.

The following Appendix in PDF is also available at the following link:

- https://s4nyam.github.io/mncaportal/appendix.pdf

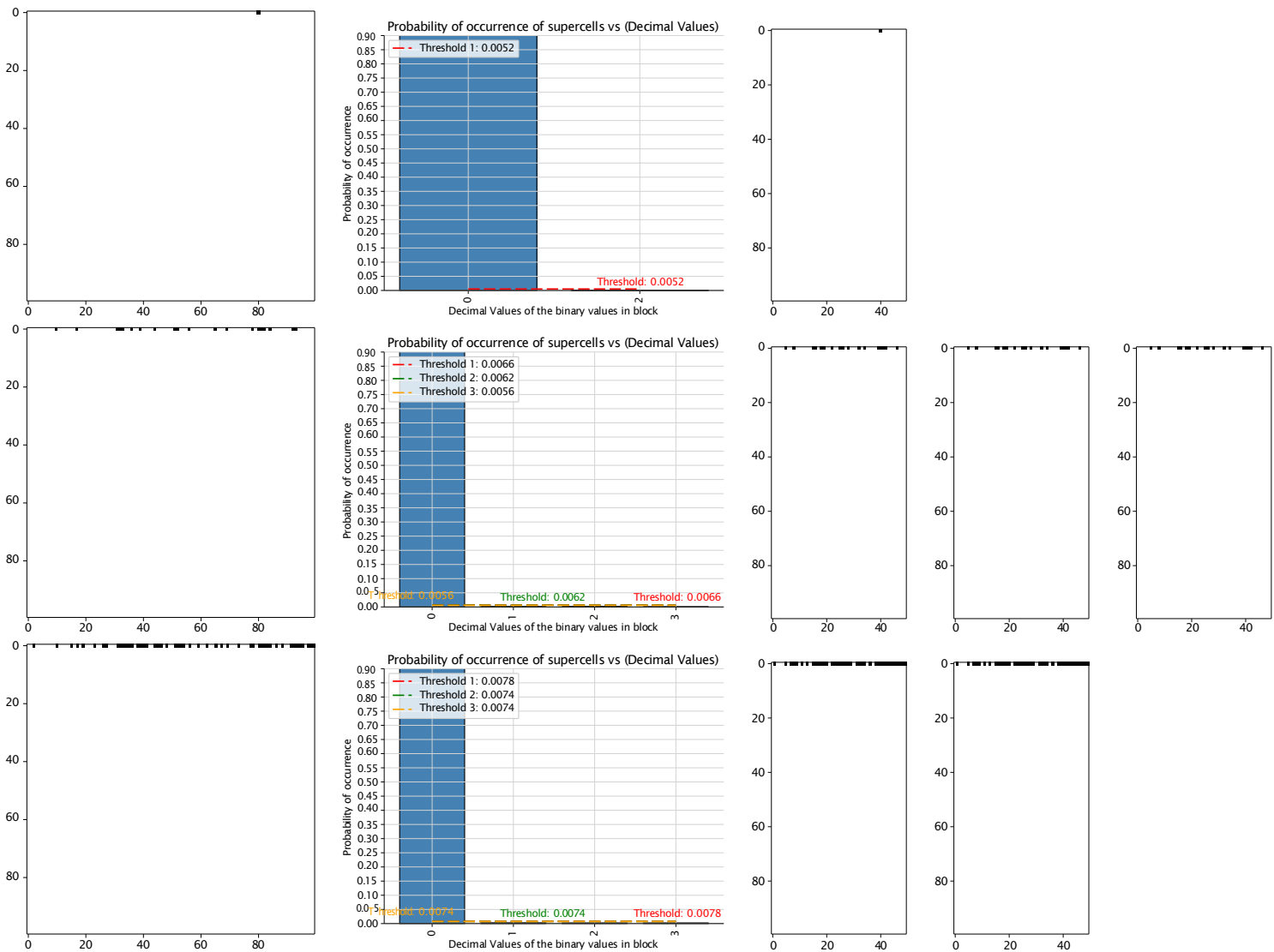

Table 1: FHCG plots for ECA Rule 0.

Table 2: FHCG plots for ECA Rule 1.

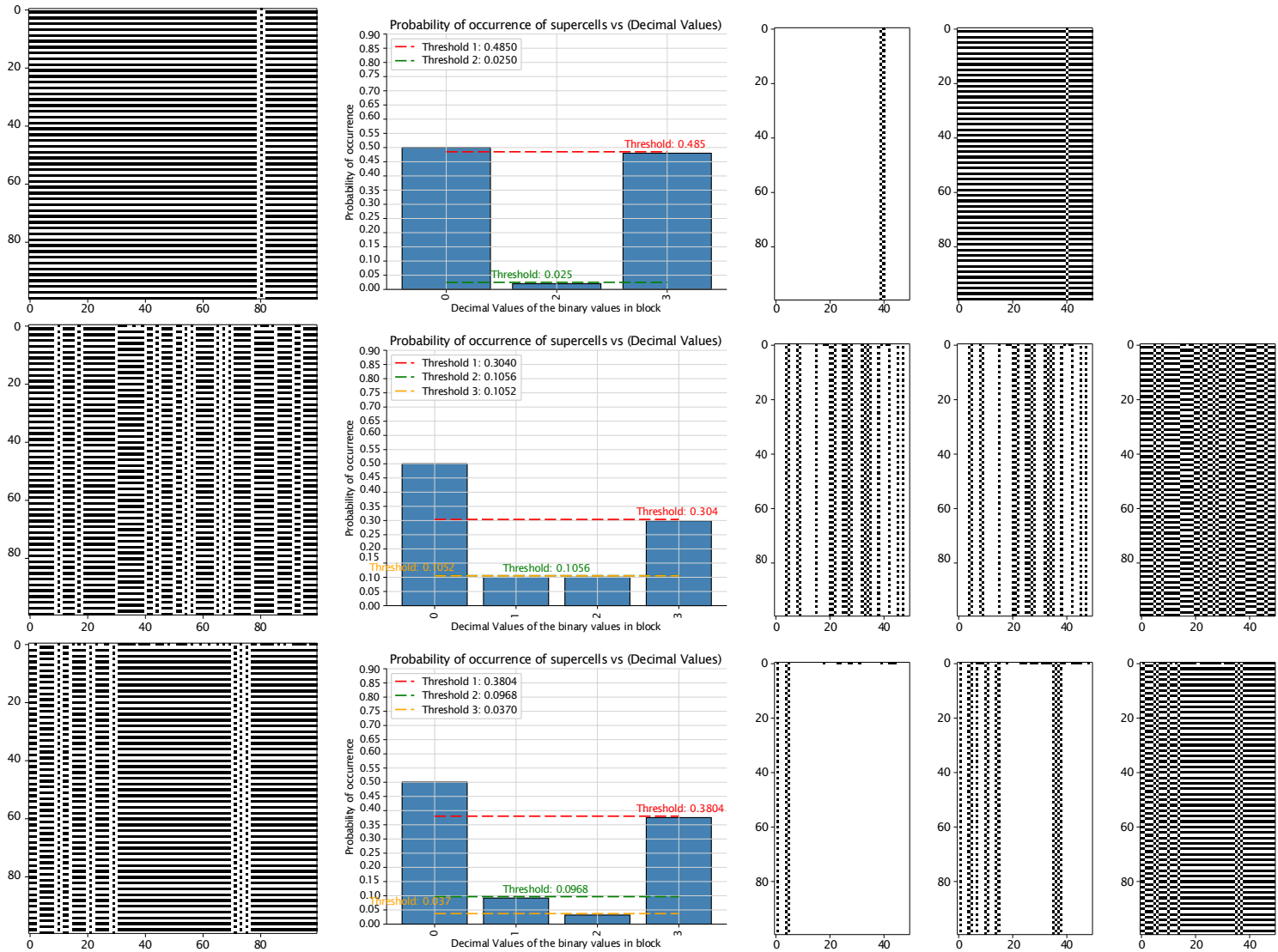

Table 3: FHCG plots for ECA Rule 3.

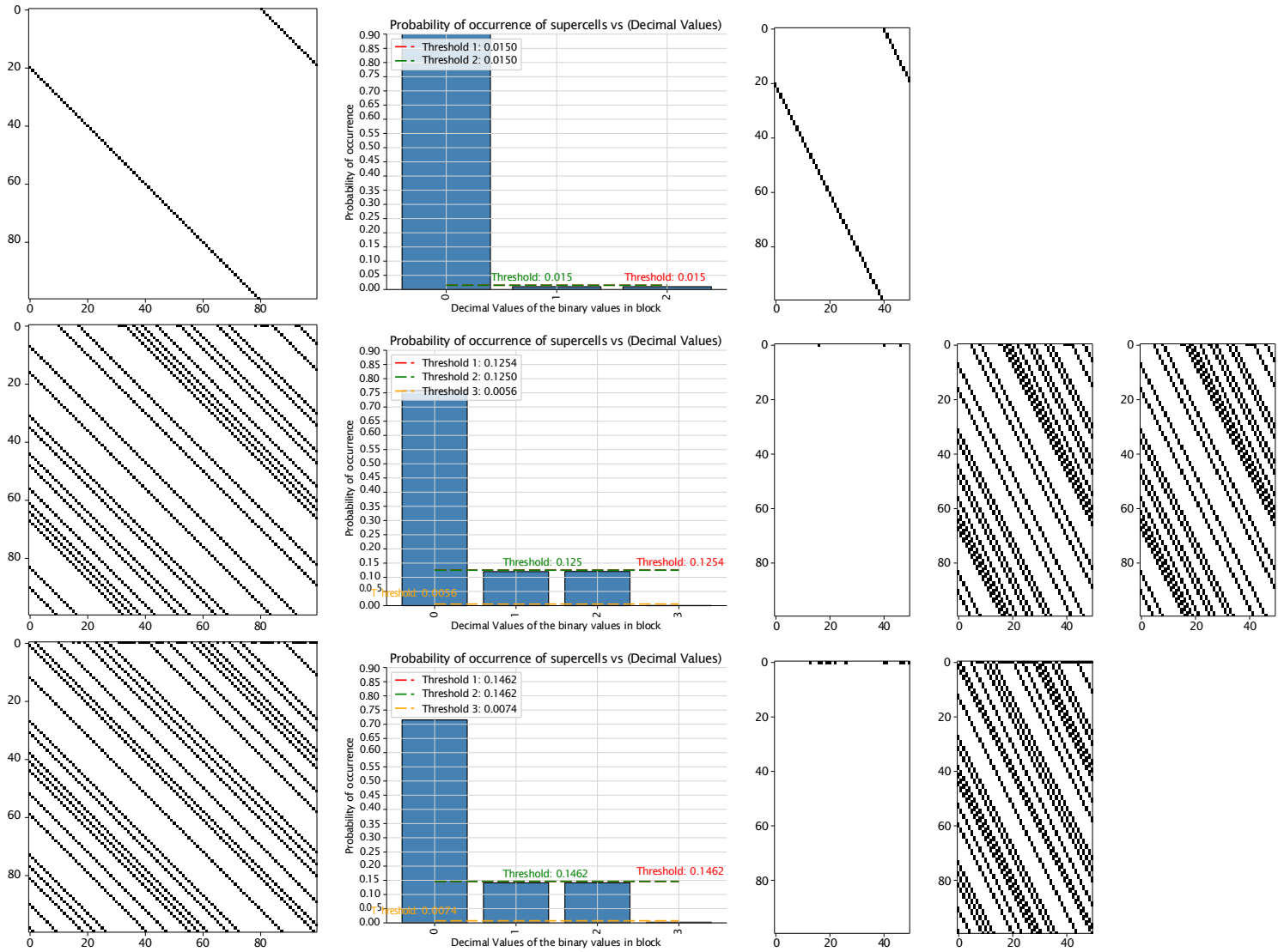

Table 4: FHCG plots for ECA Rule 4.

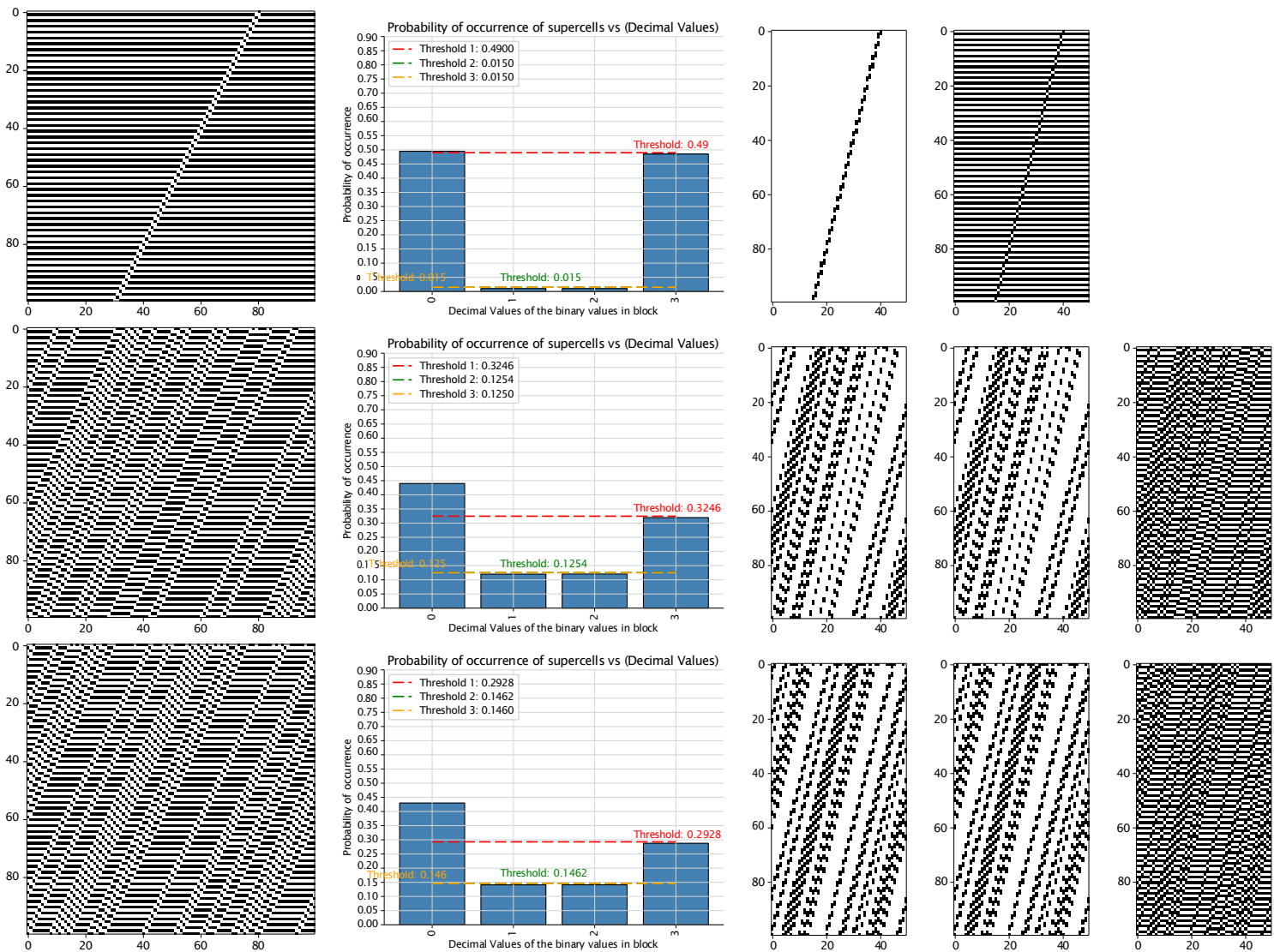

Table 5: FHCG plots for ECA Rule 5.

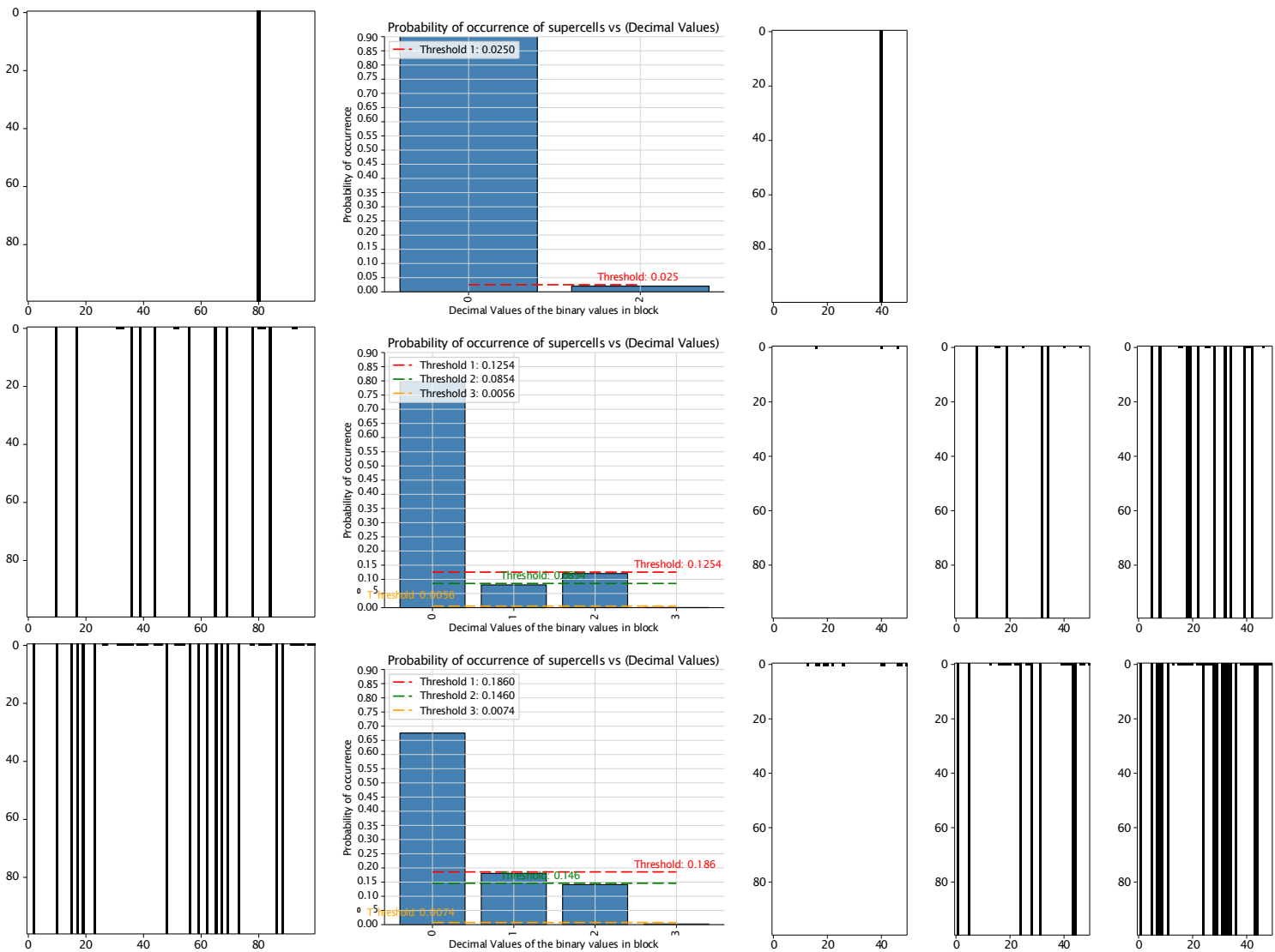

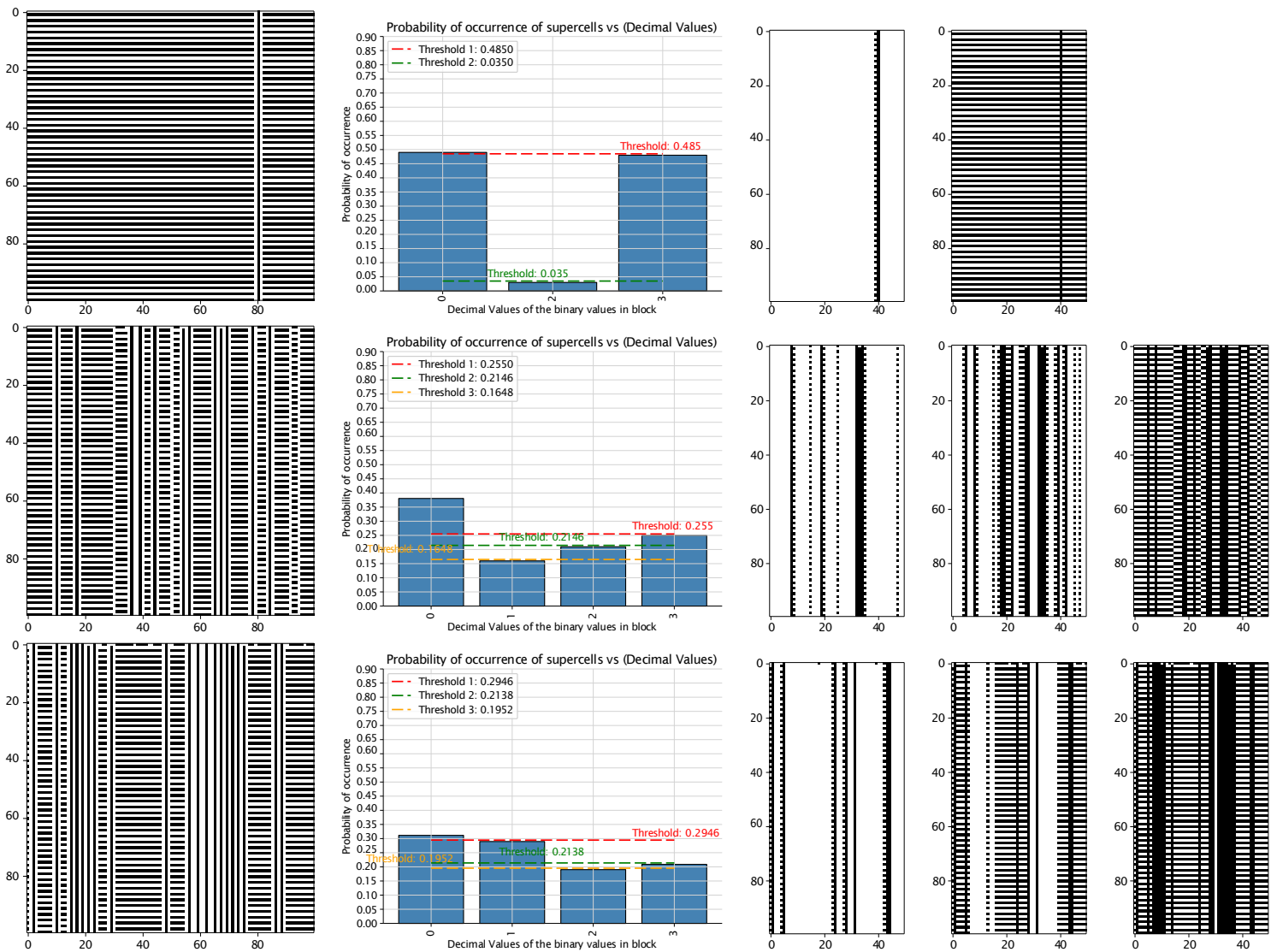

Table 7: FHCG plots for ECA Rule 7.

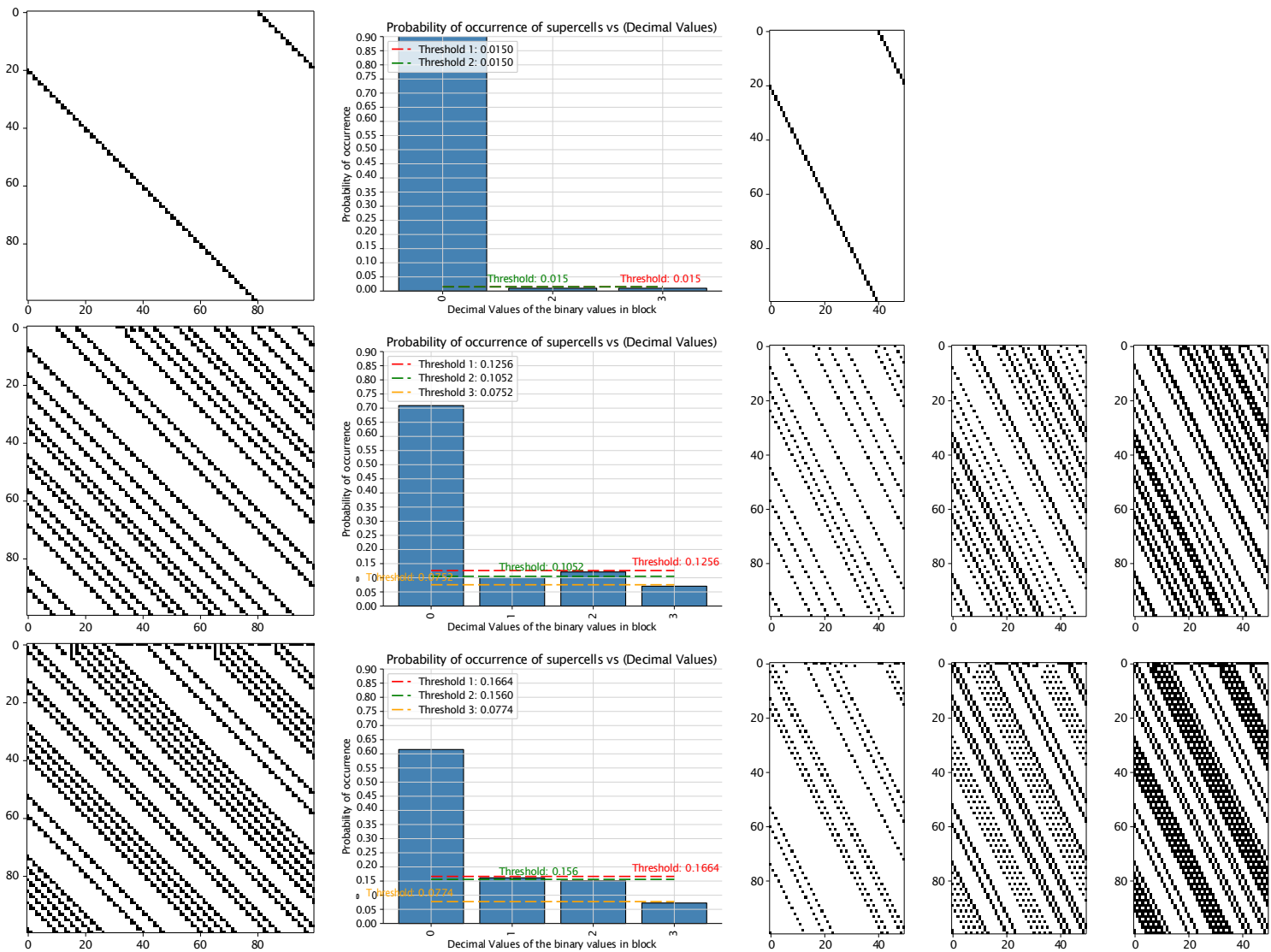

Table 8: FHCG plots for ECA Rule 8.

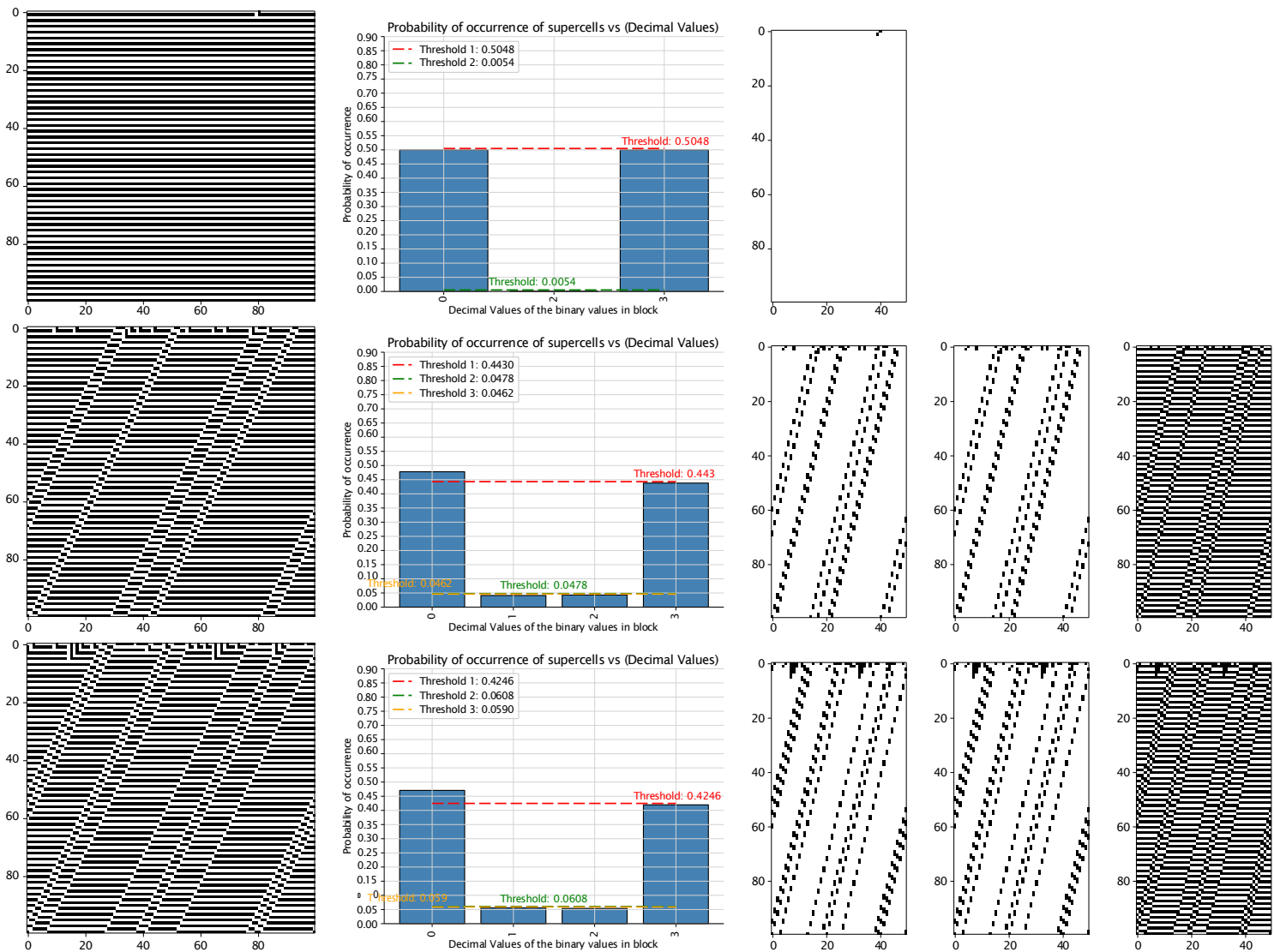

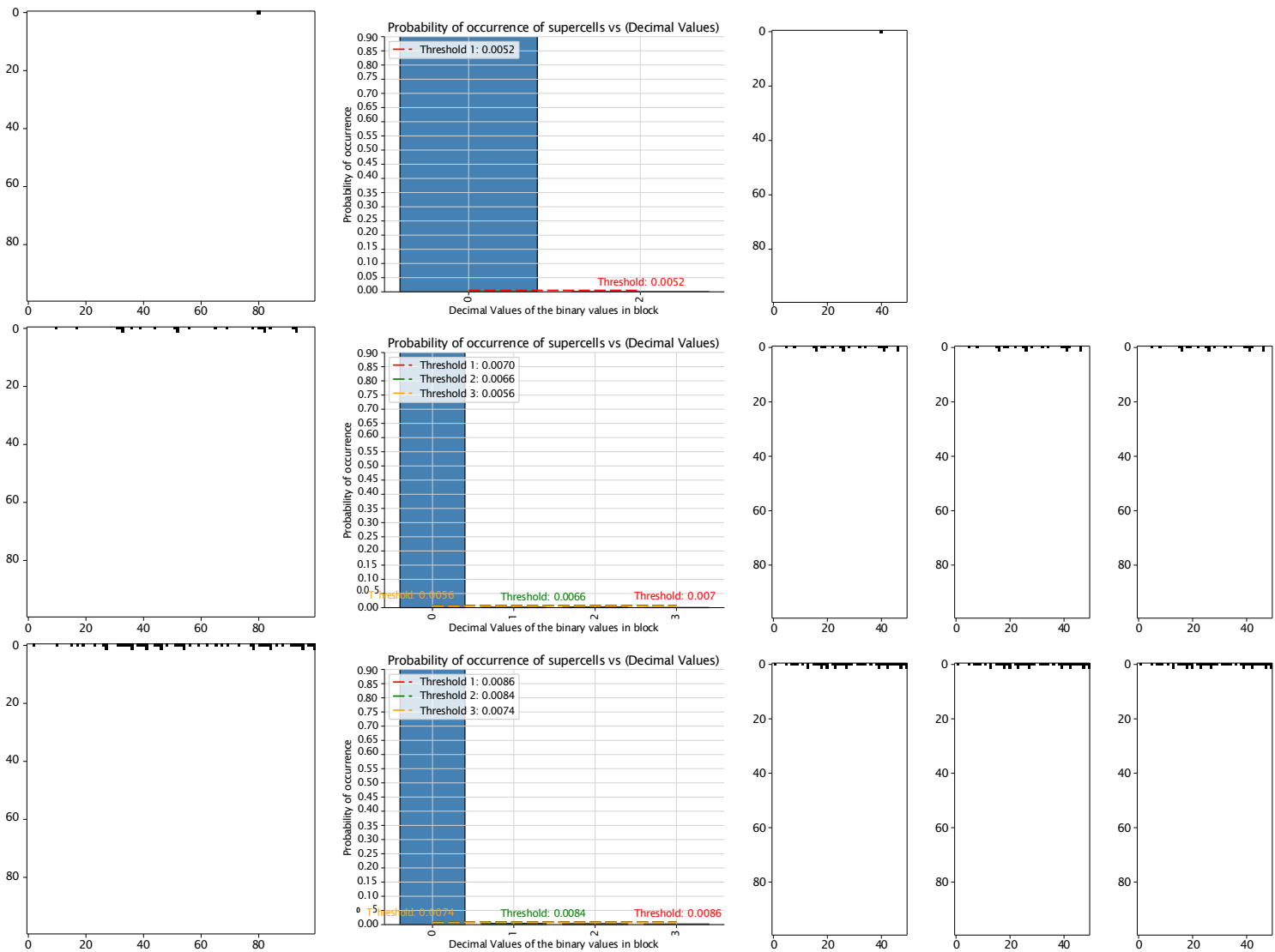

Table 9: FHCG plots for ECA Rule 9.

Table 10: FHCG plots for ECA Rule 10.

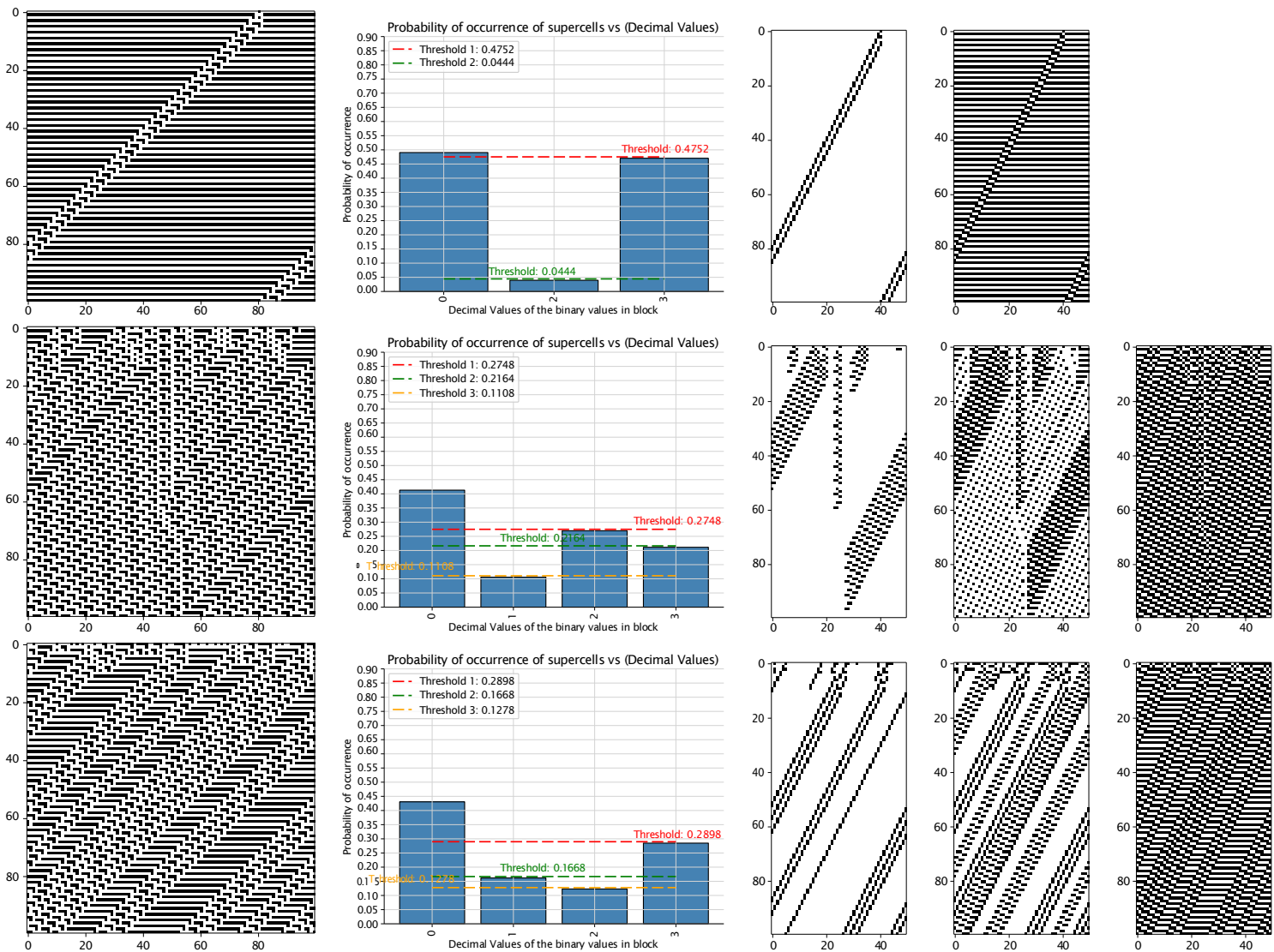

Table 11: FHCG plots for ECA Rule 11.

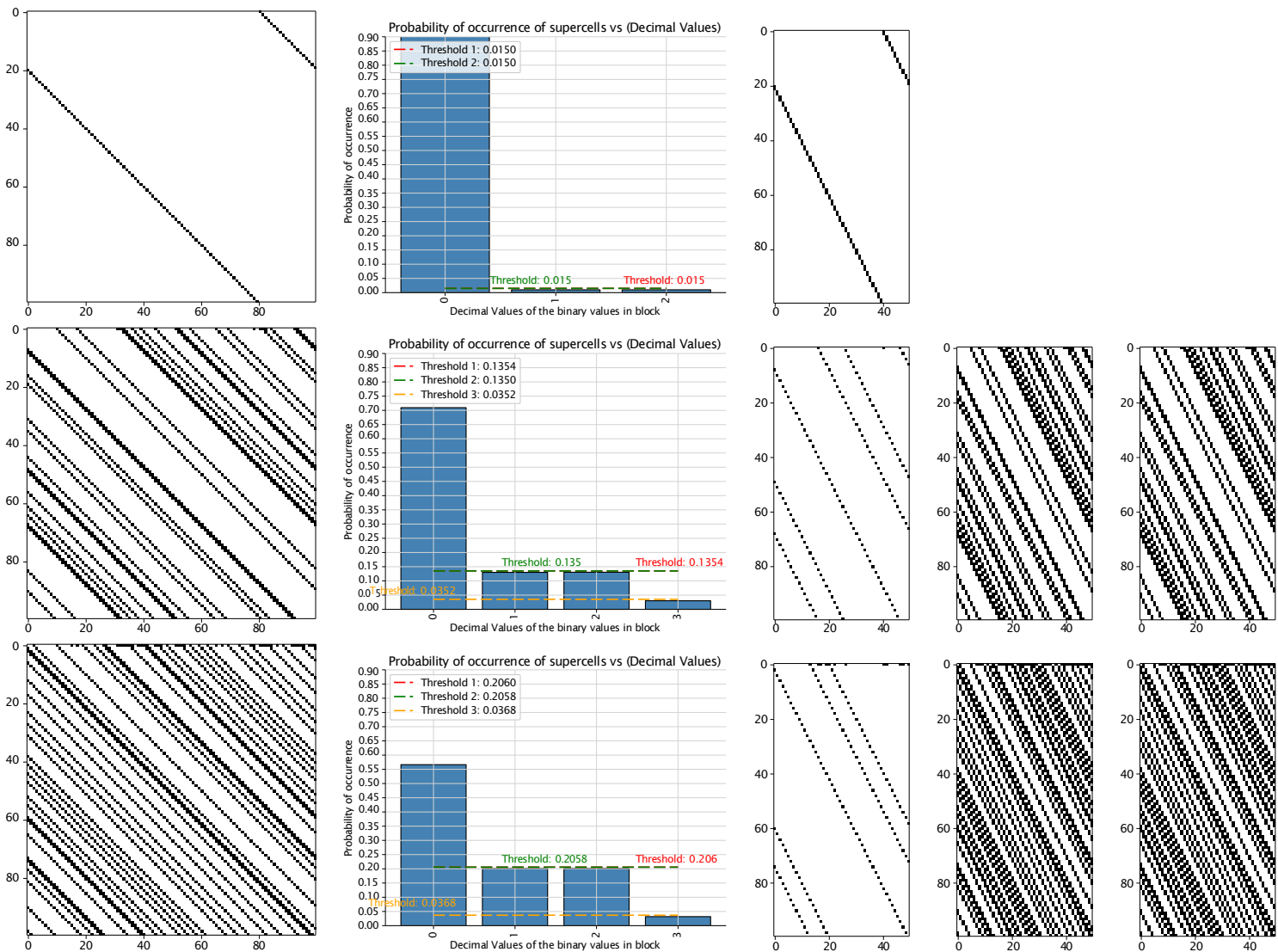

Table 12: FHCG plots for ECA Rule 12.

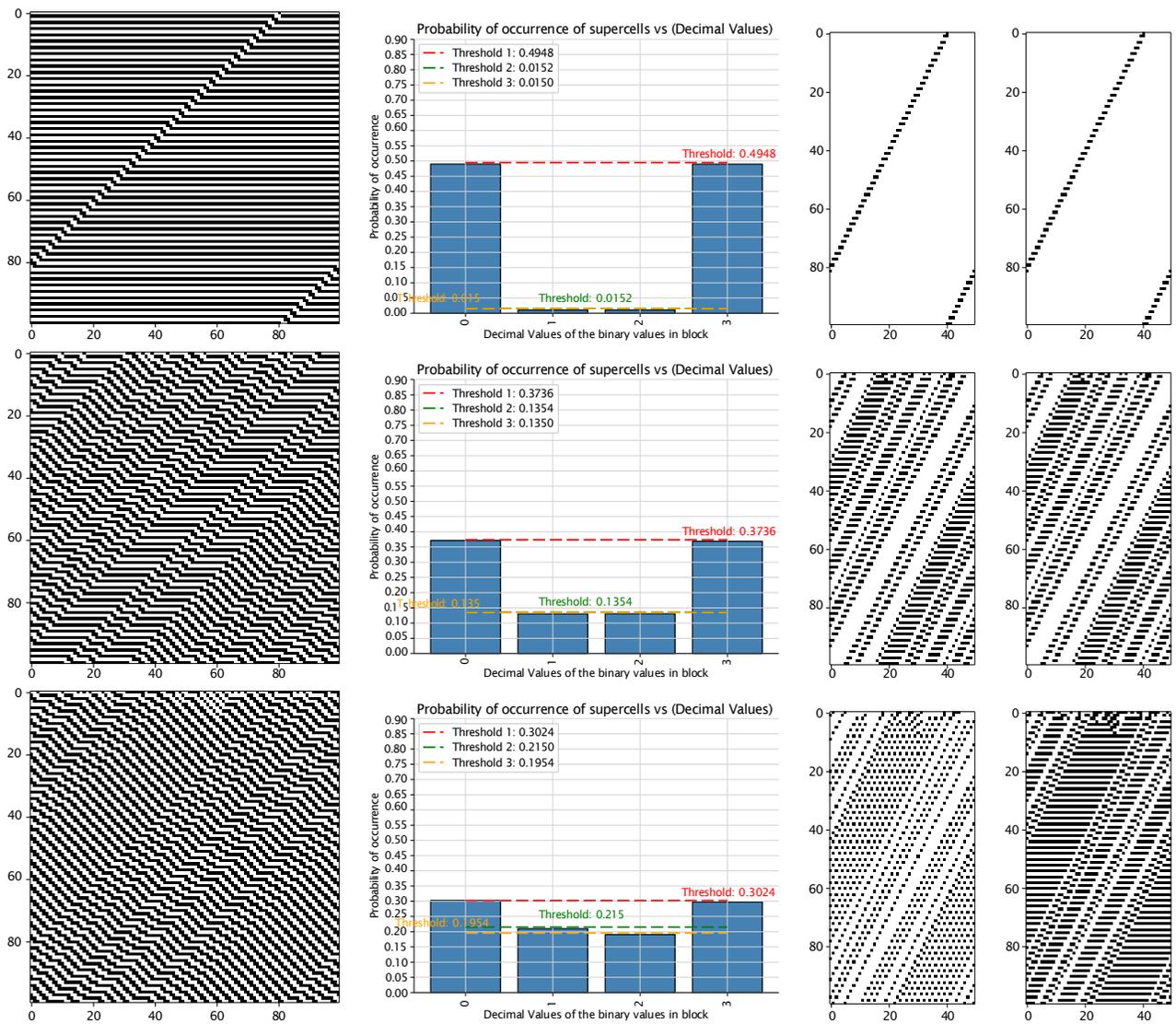

Table 13: FHCG plots for ECA Rule 13.

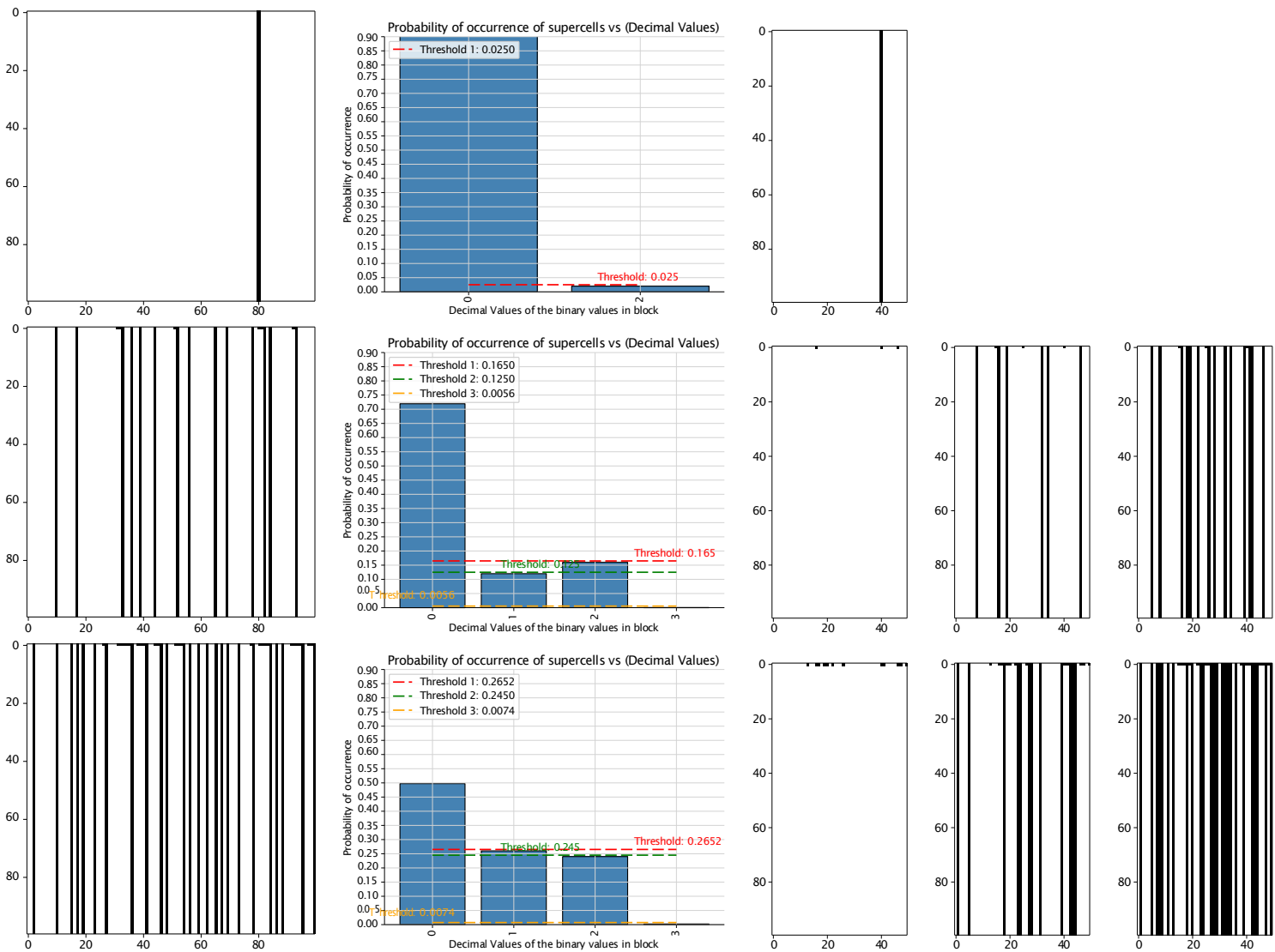

Table 14: FHCG plots for ECA Rule 14.

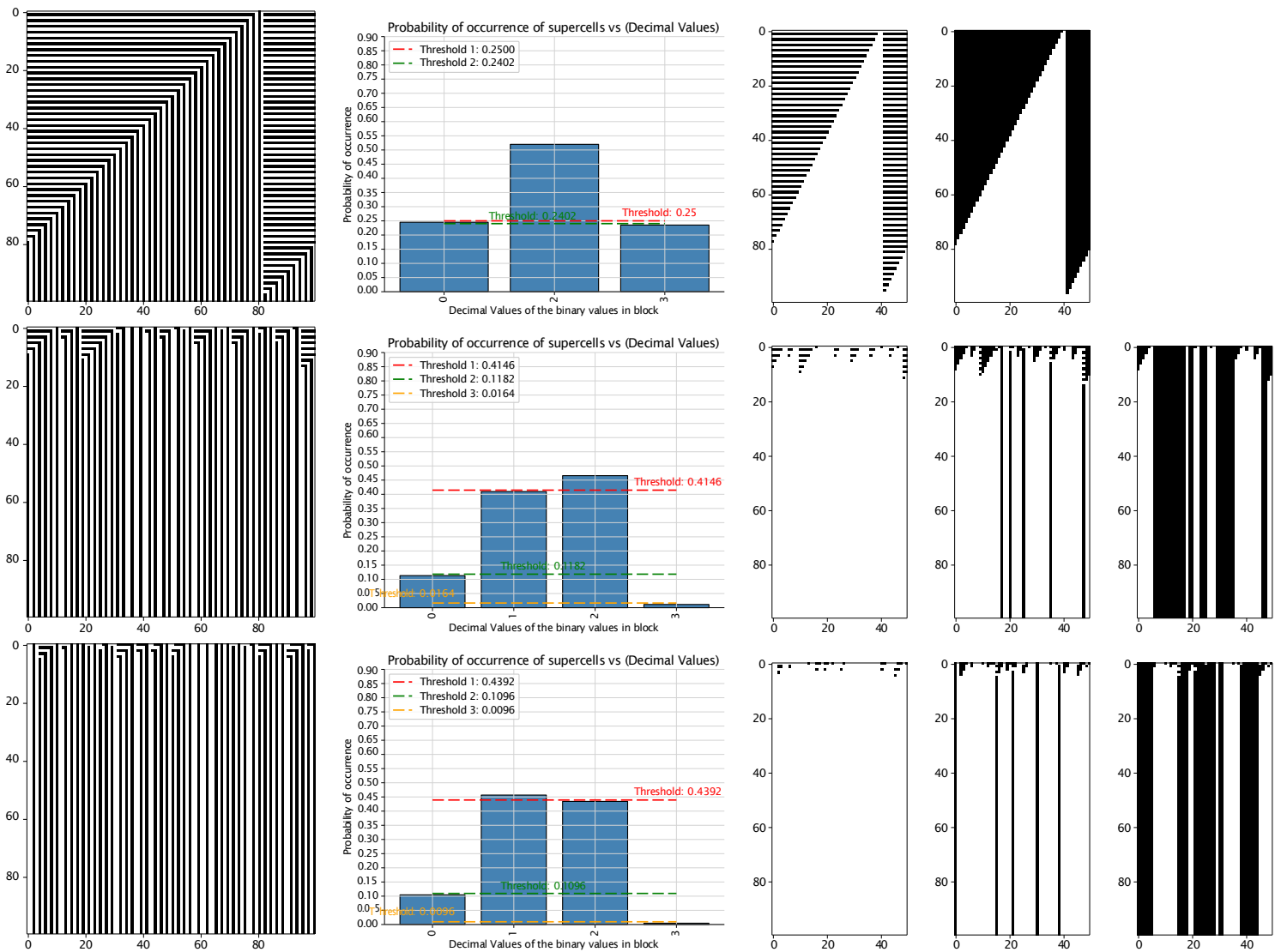

Table 15: FHCG plots for ECA Rule 15.

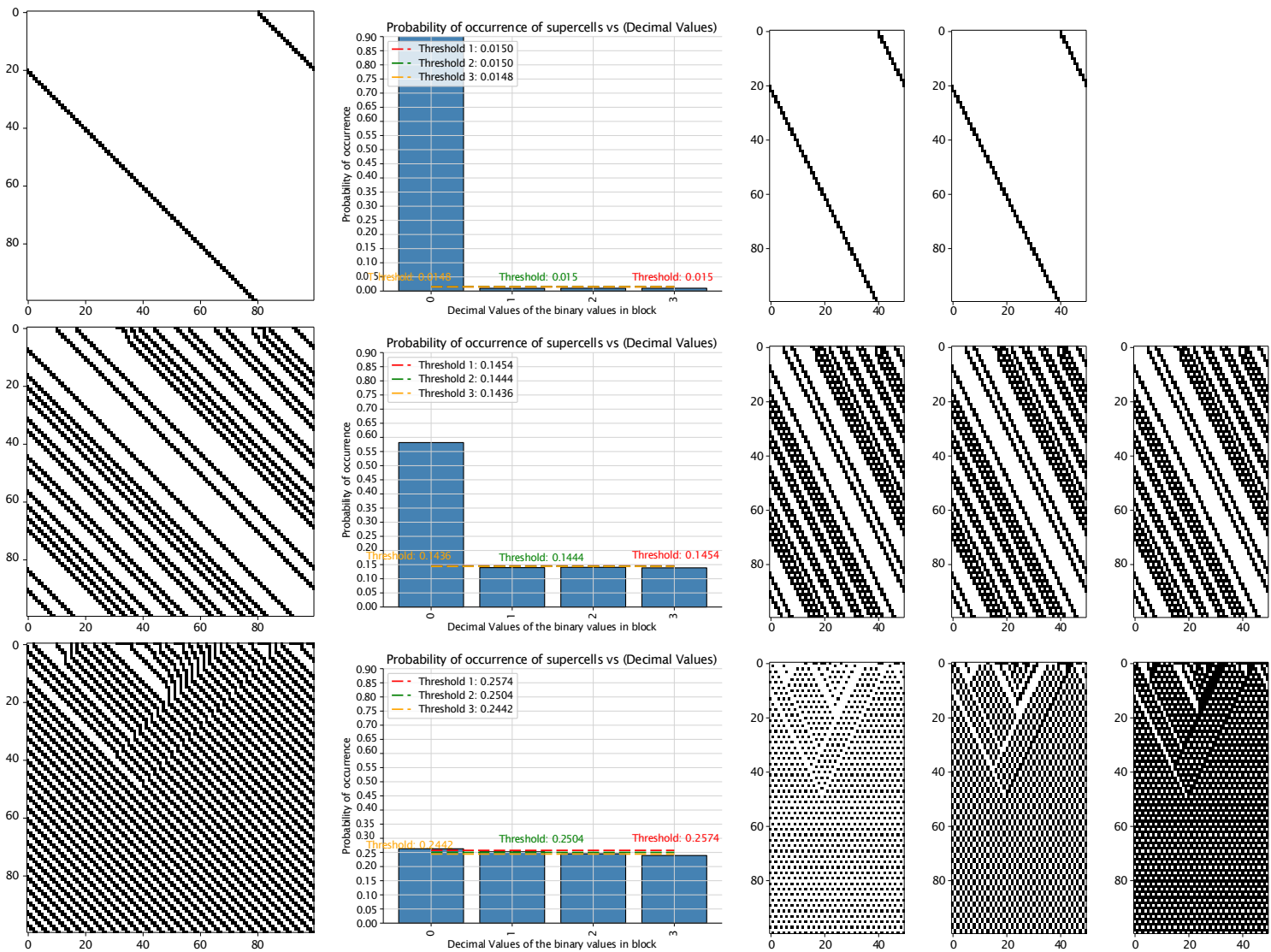

Table 16: FHCG plots for ECA Rule 16.

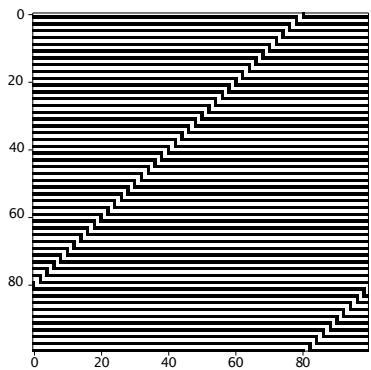 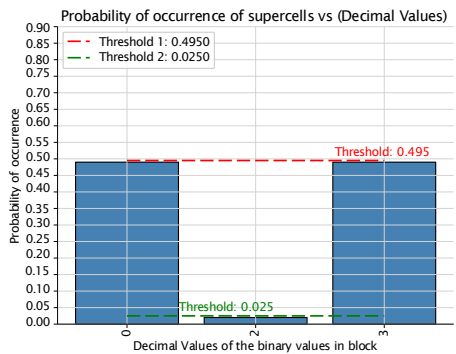 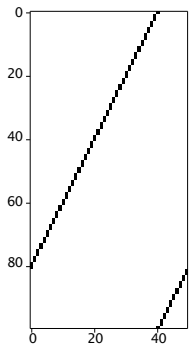

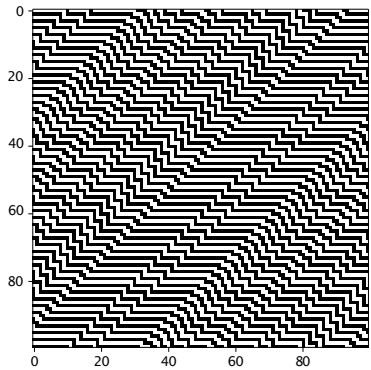 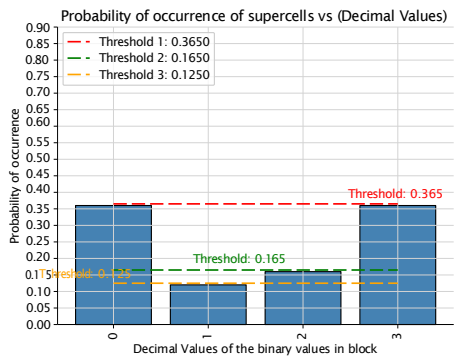 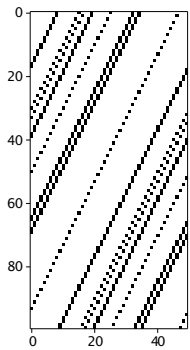 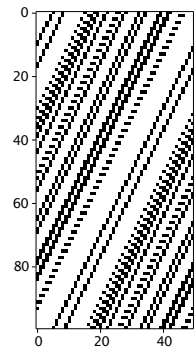

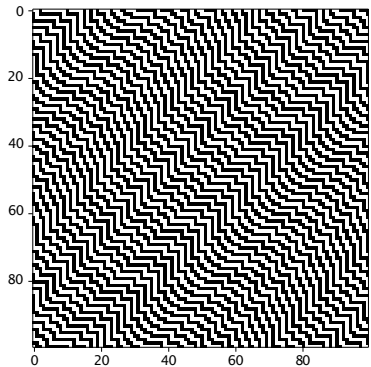 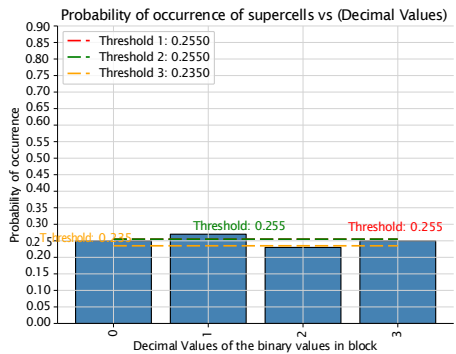 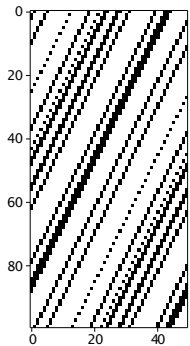 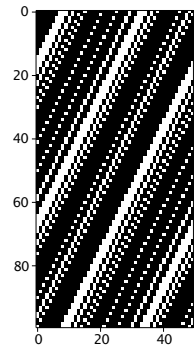

Table 17: FHCG plots for ECA Rule 17.

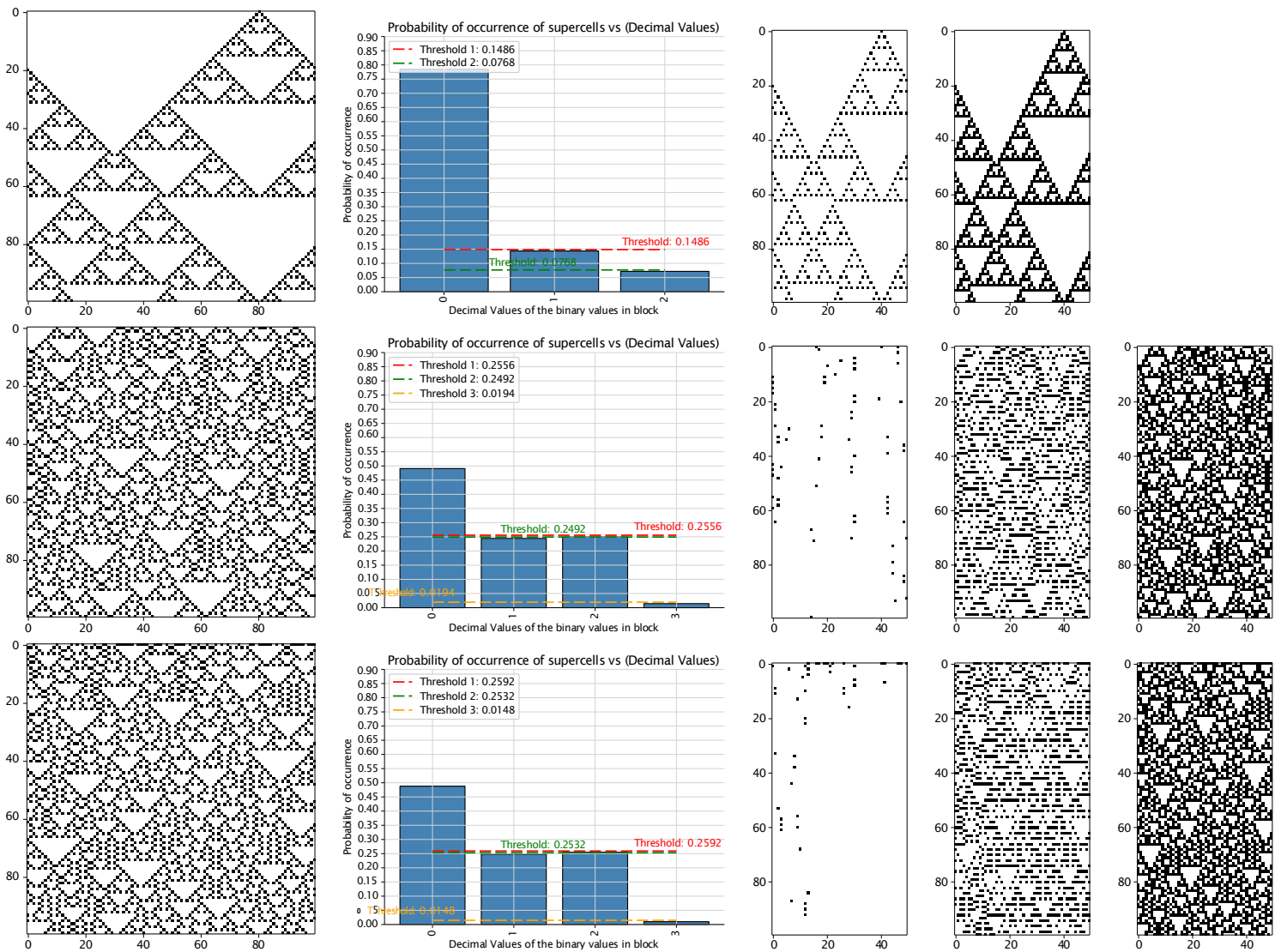

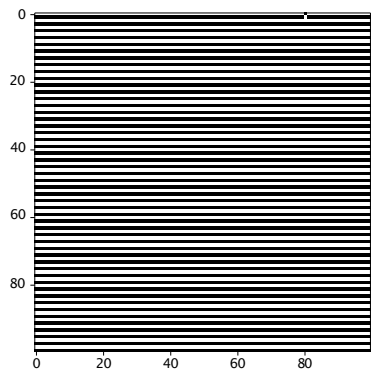
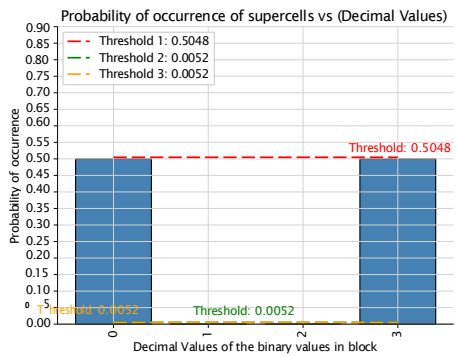
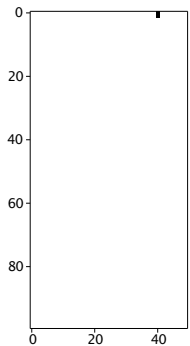
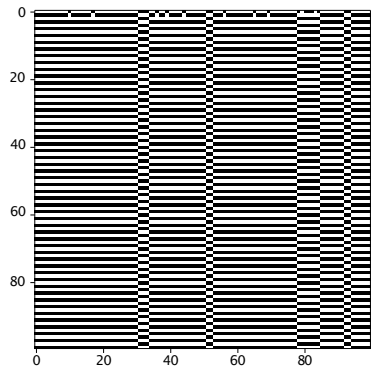
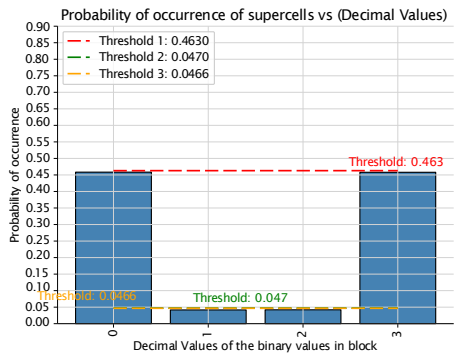
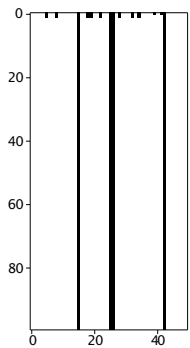
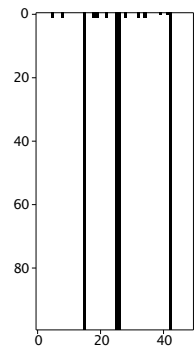
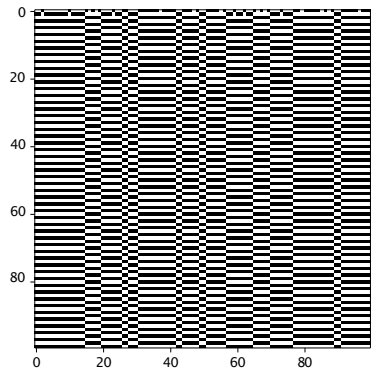
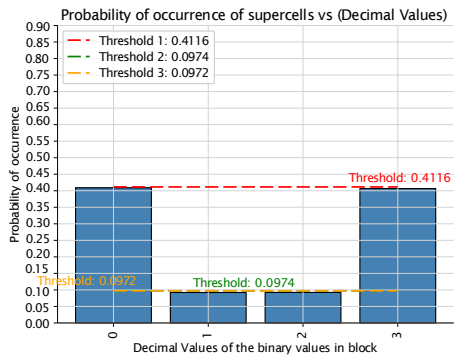
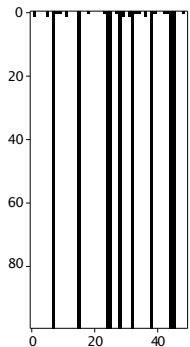
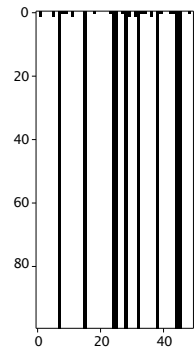

Table 19: FHCG plots for ECA Rule 19.

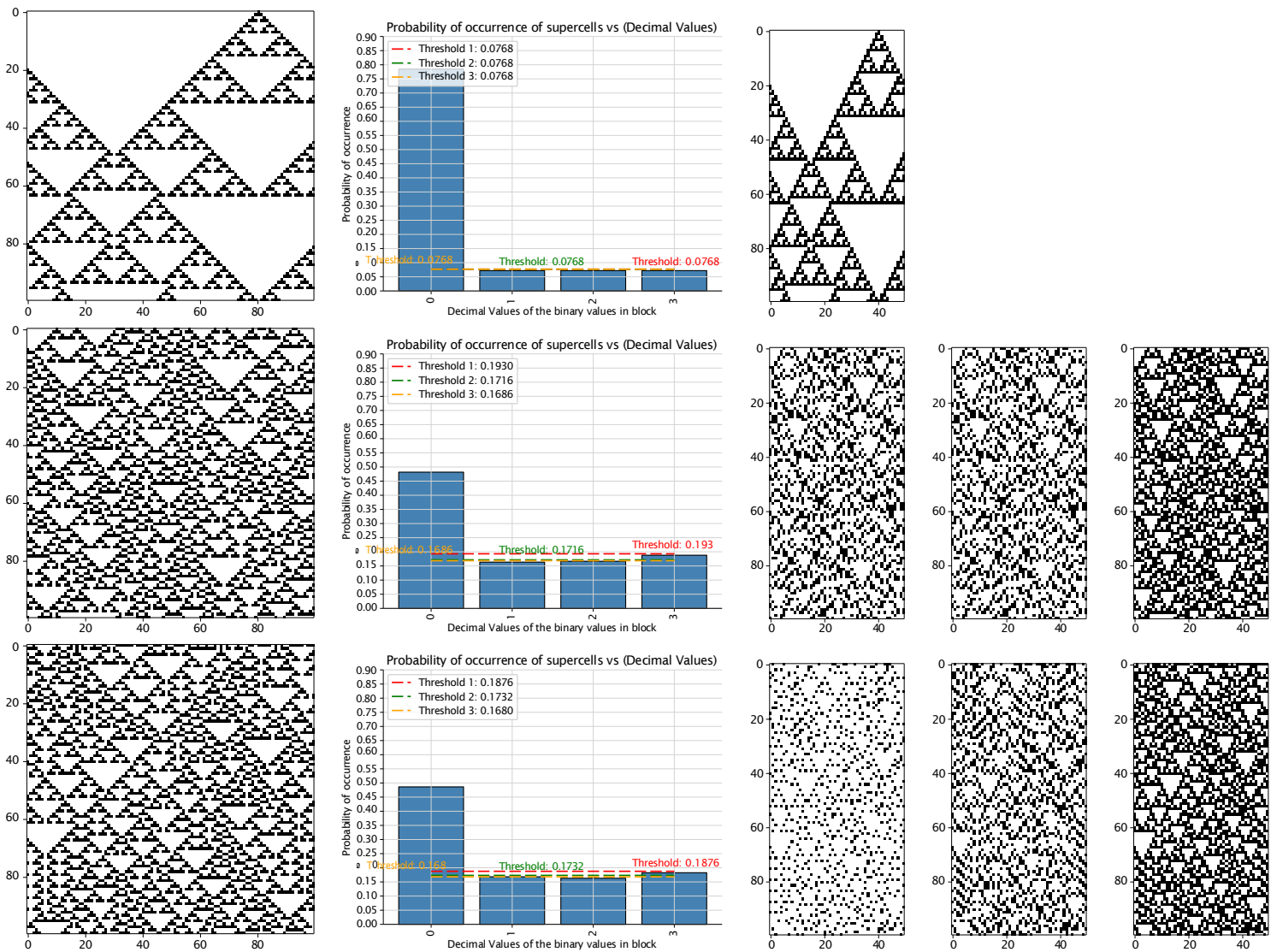

Table 20: FHCG plots for ECA Rule 20.

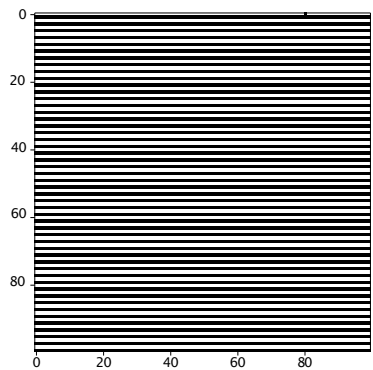 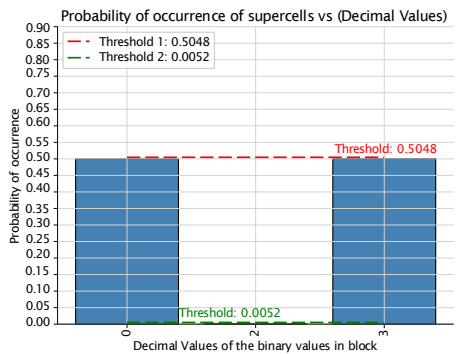 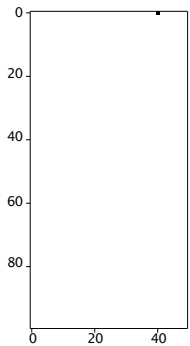

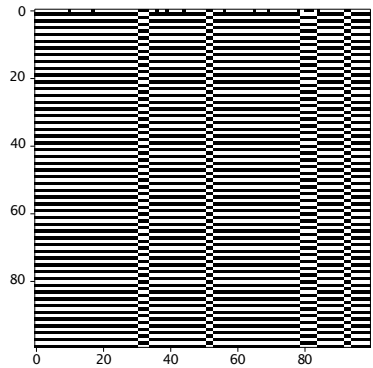 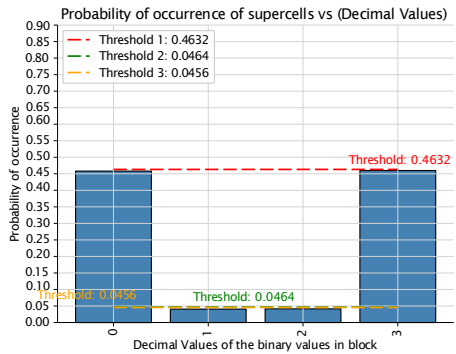 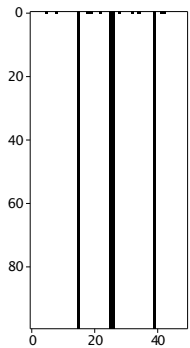 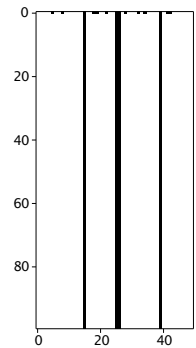

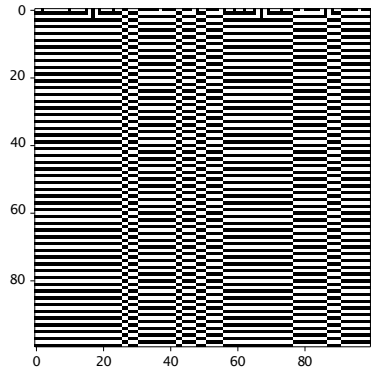 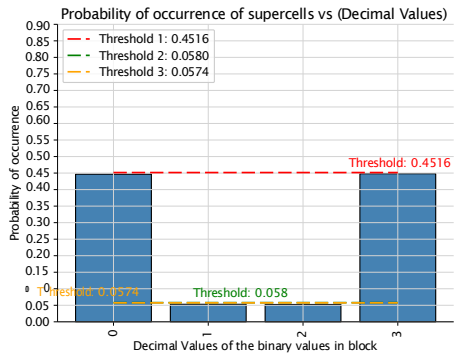 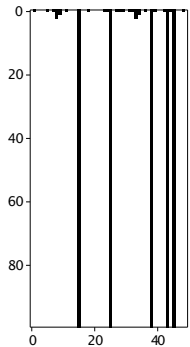 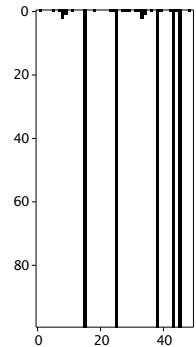

Table 21: FHCG plots for ECA Rule 21.

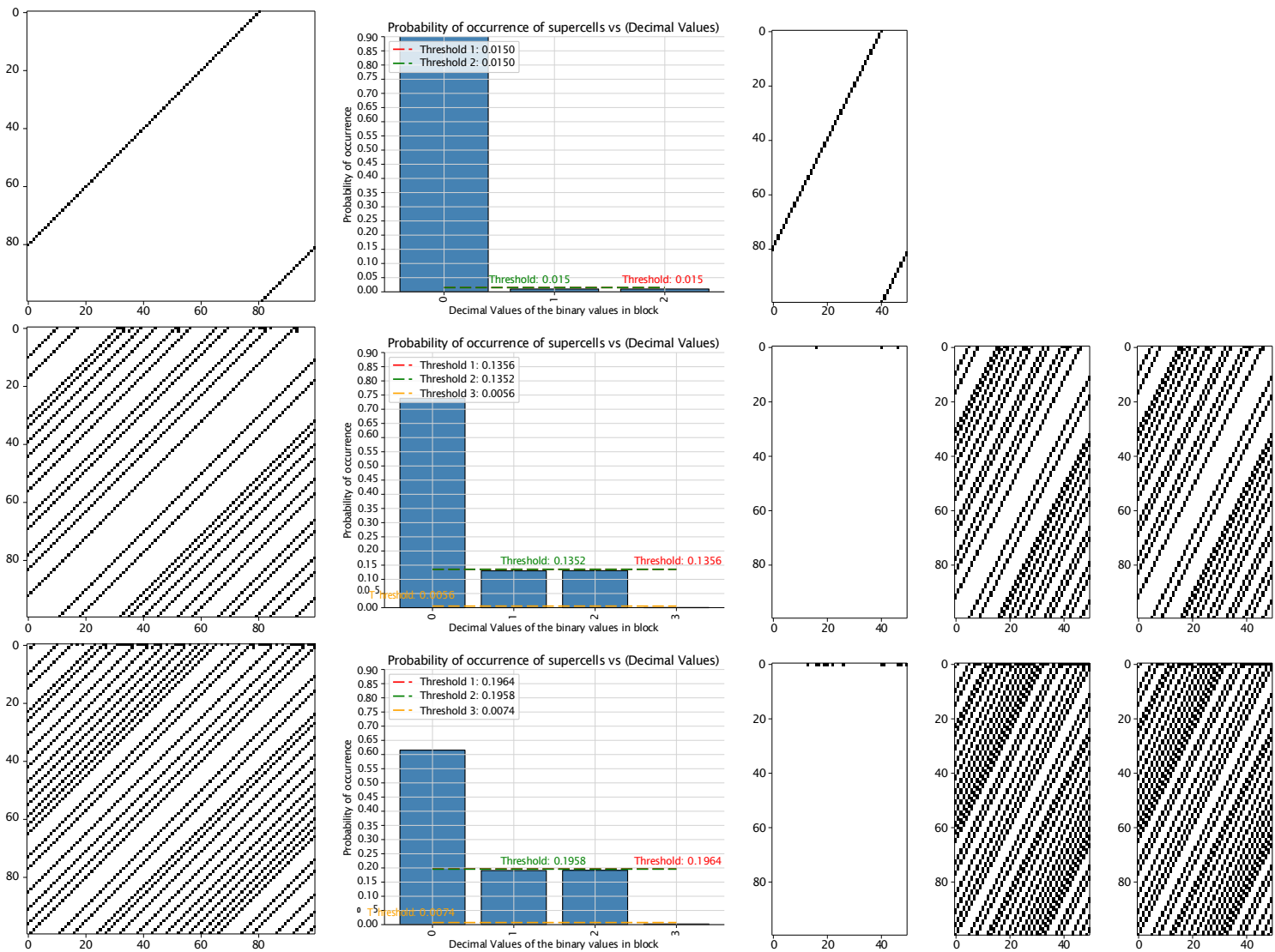

Table 22: FHCG plots for ECA Rule 22.

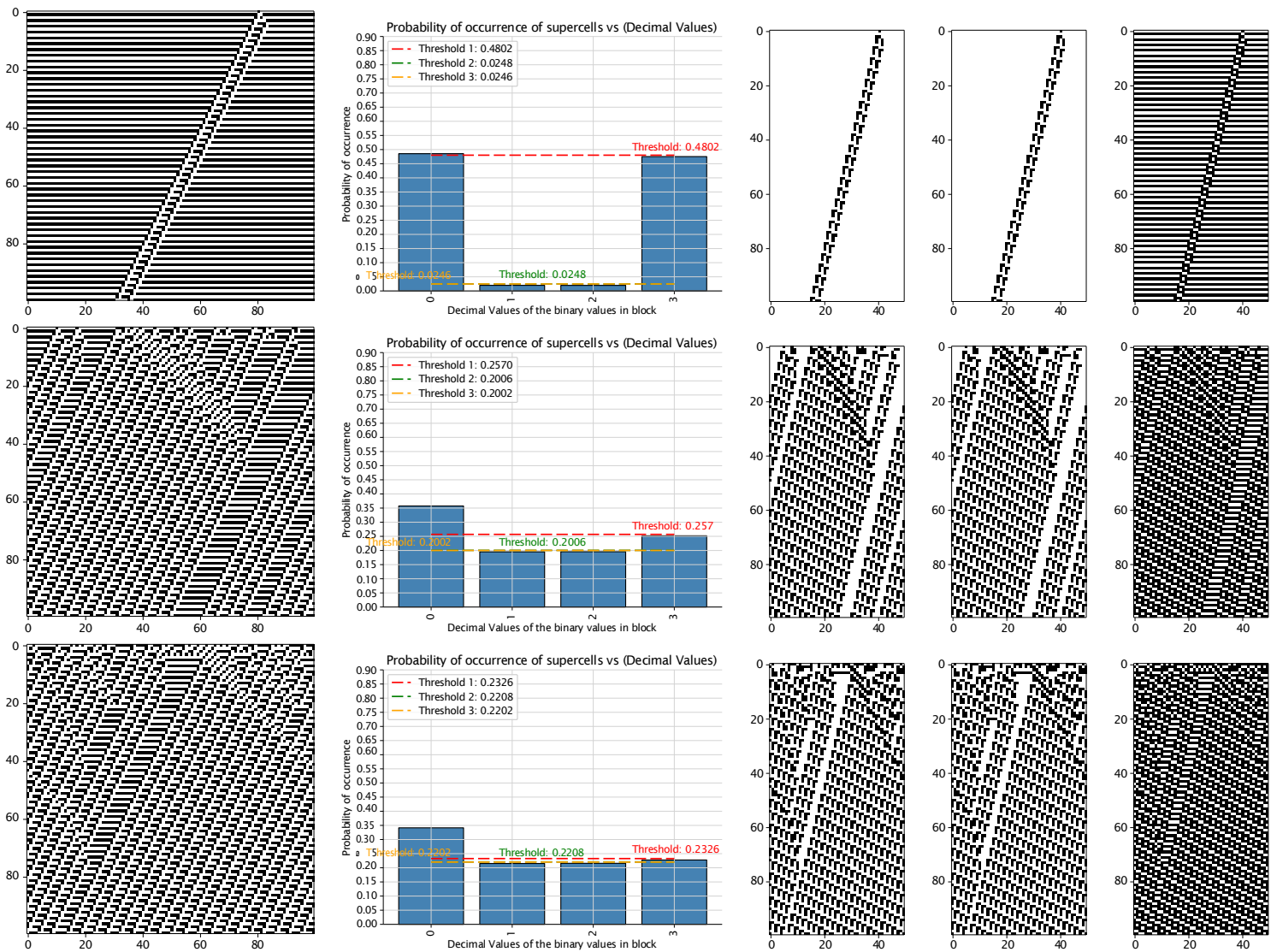

Table 23: FHCG plots for ECA Rule 23.

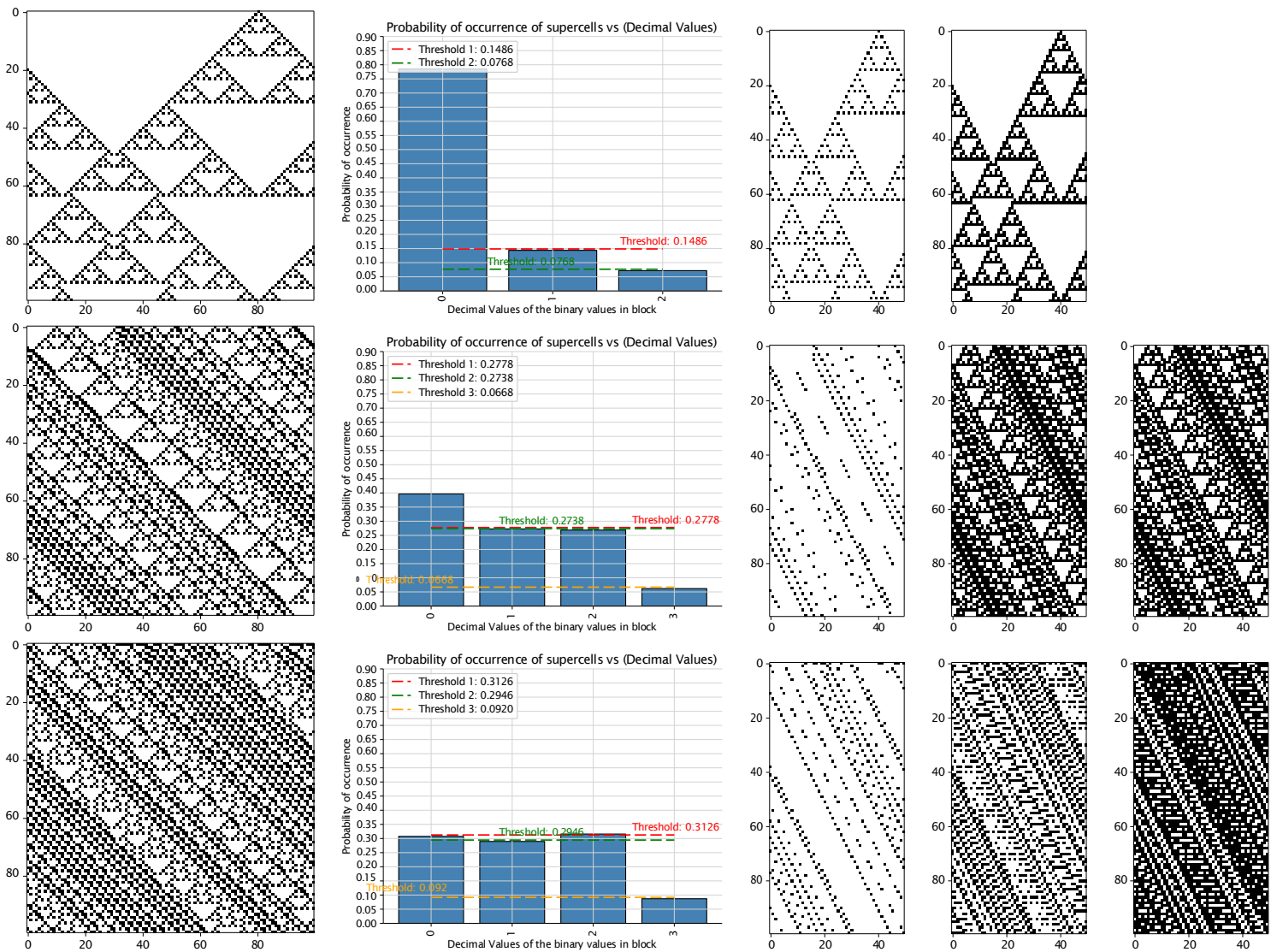

Table 24: FHCG plots for ECA Rule 24.

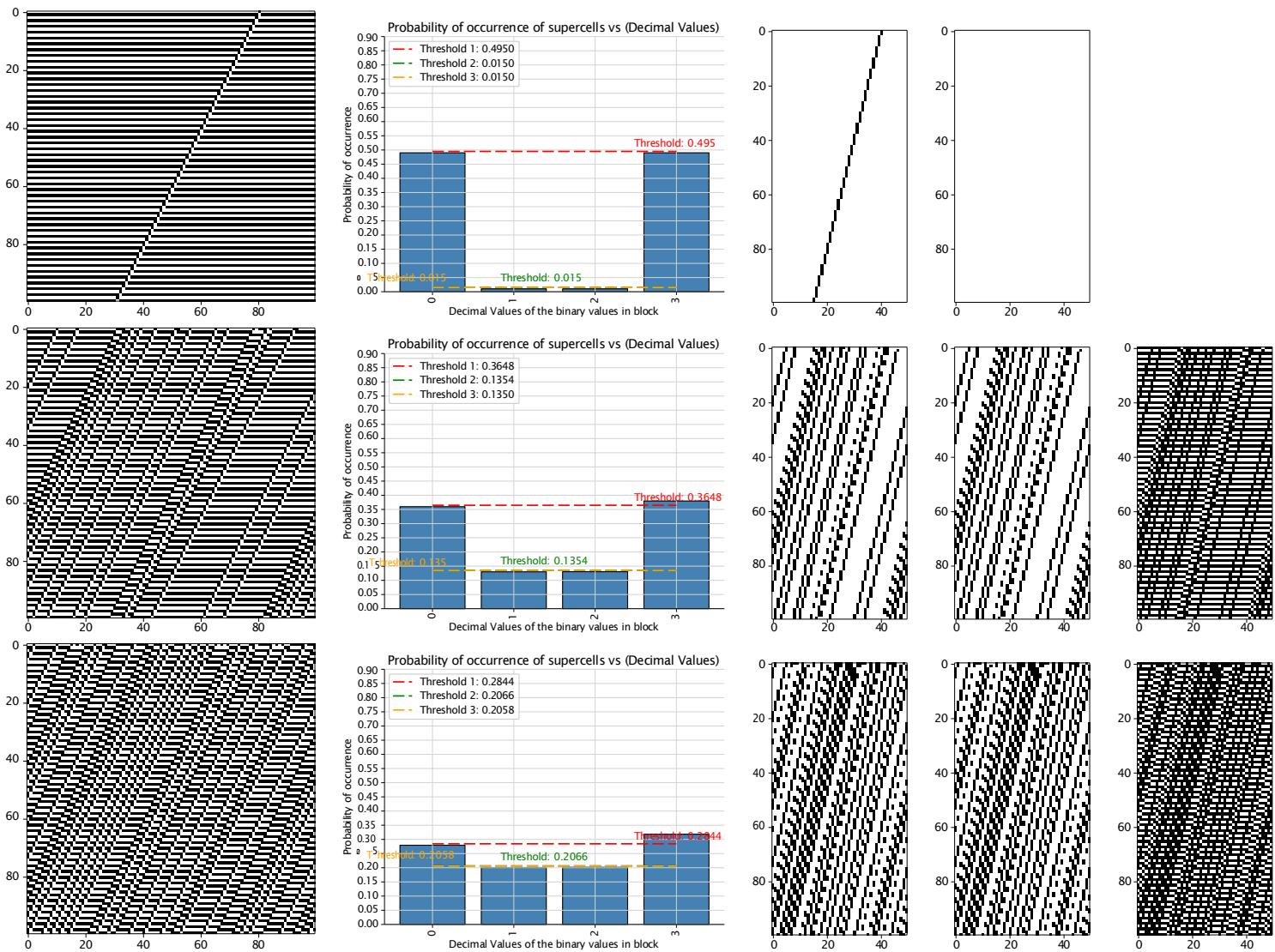

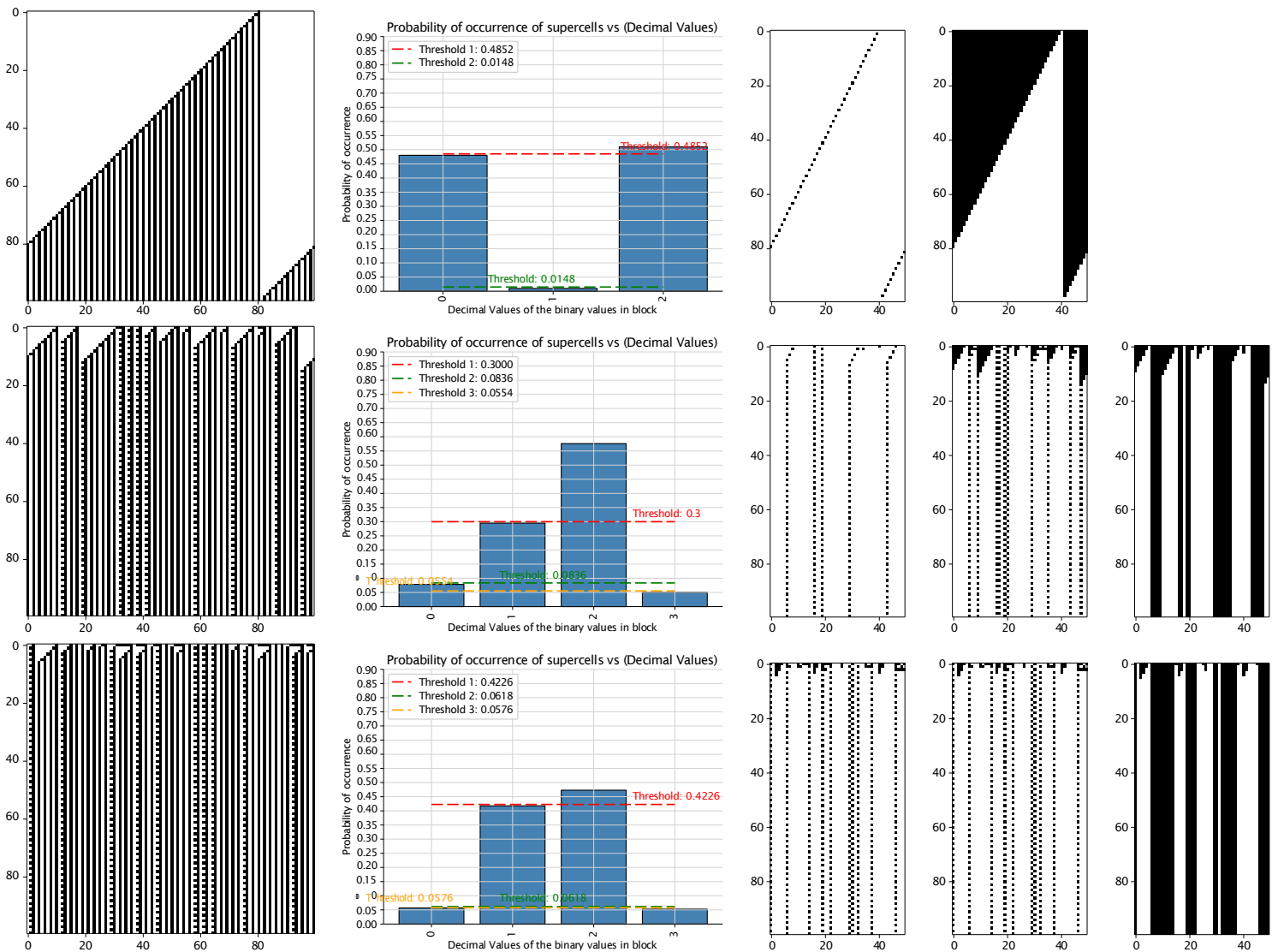

Table 25: FHCG plots for ECA Rule 25.

Table 26: FHCG plots for ECA Rule 26.

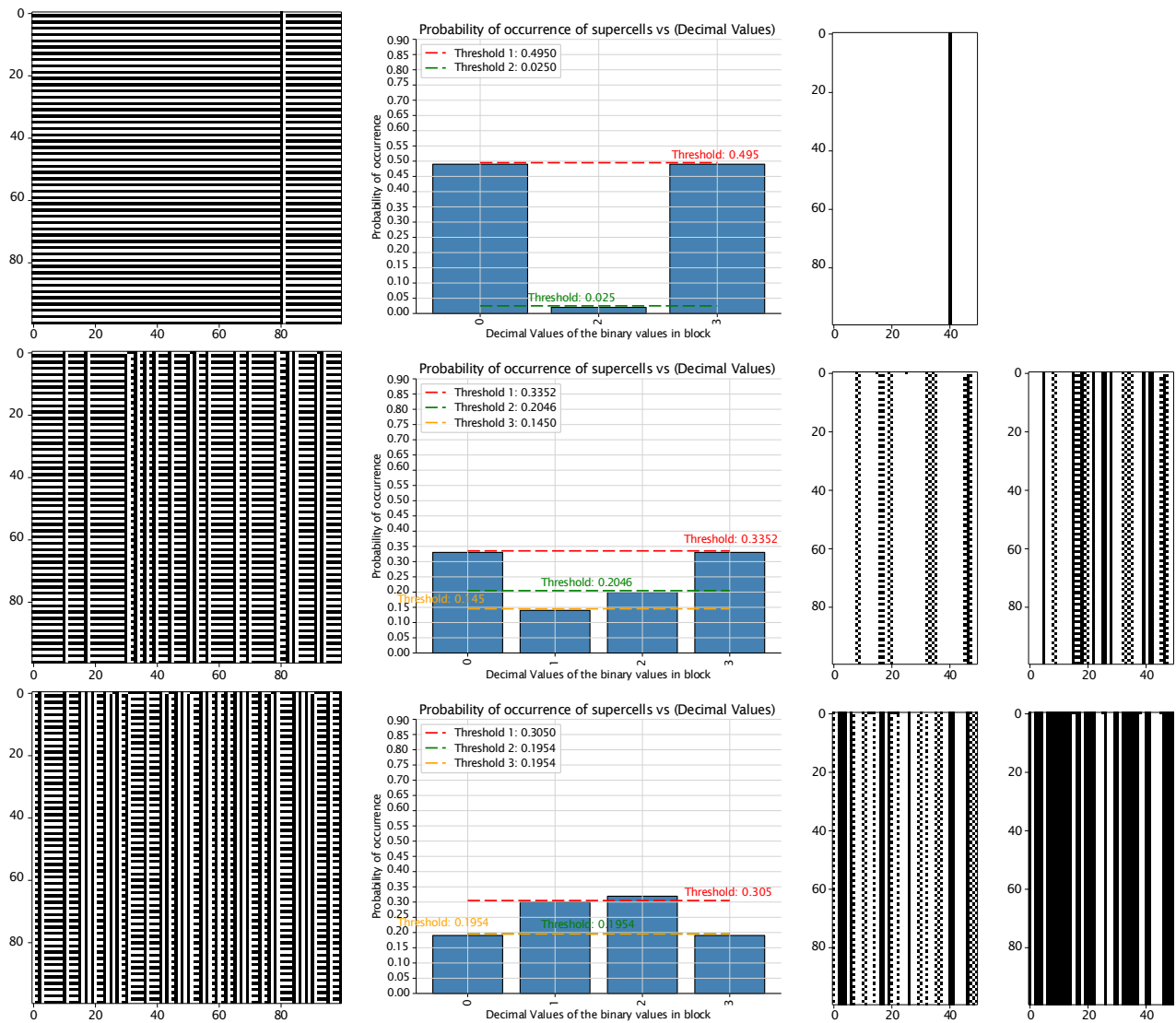

Table 27: FHCG plots for ECA Rule 27.

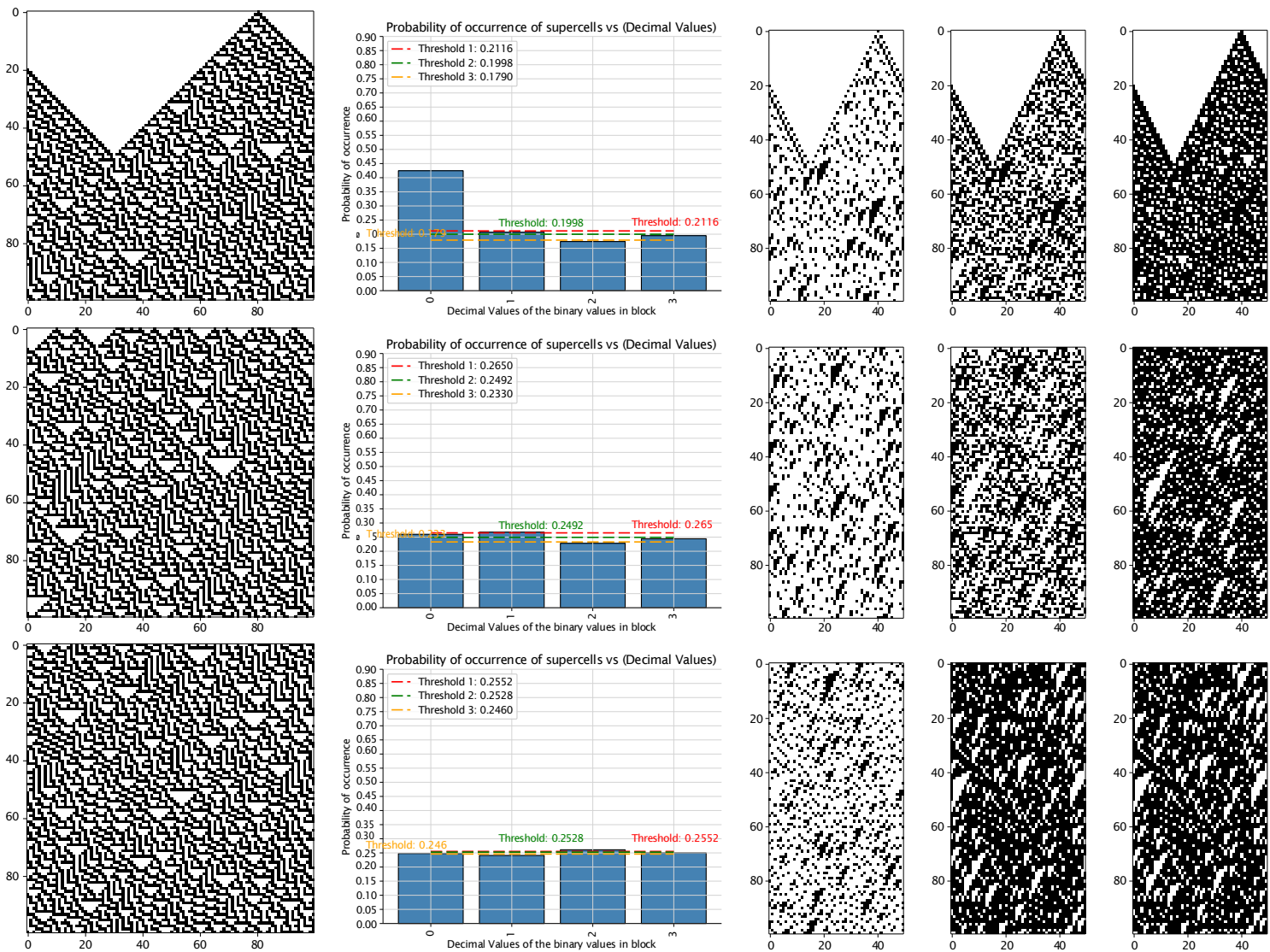

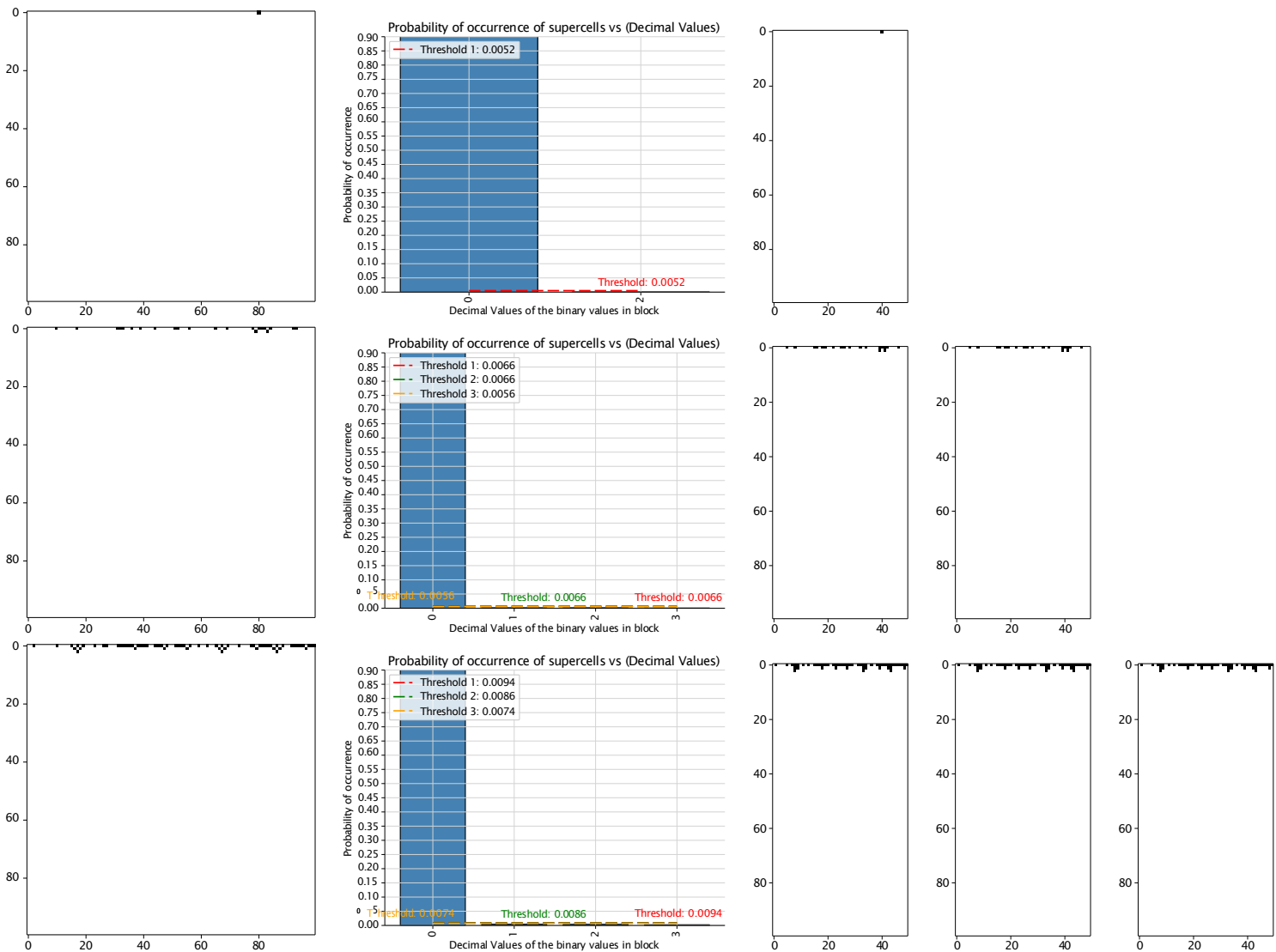

Table 28: FHCG plots for ECA Rule 28.

Table 29: FHCG plots for ECA Rule 29.

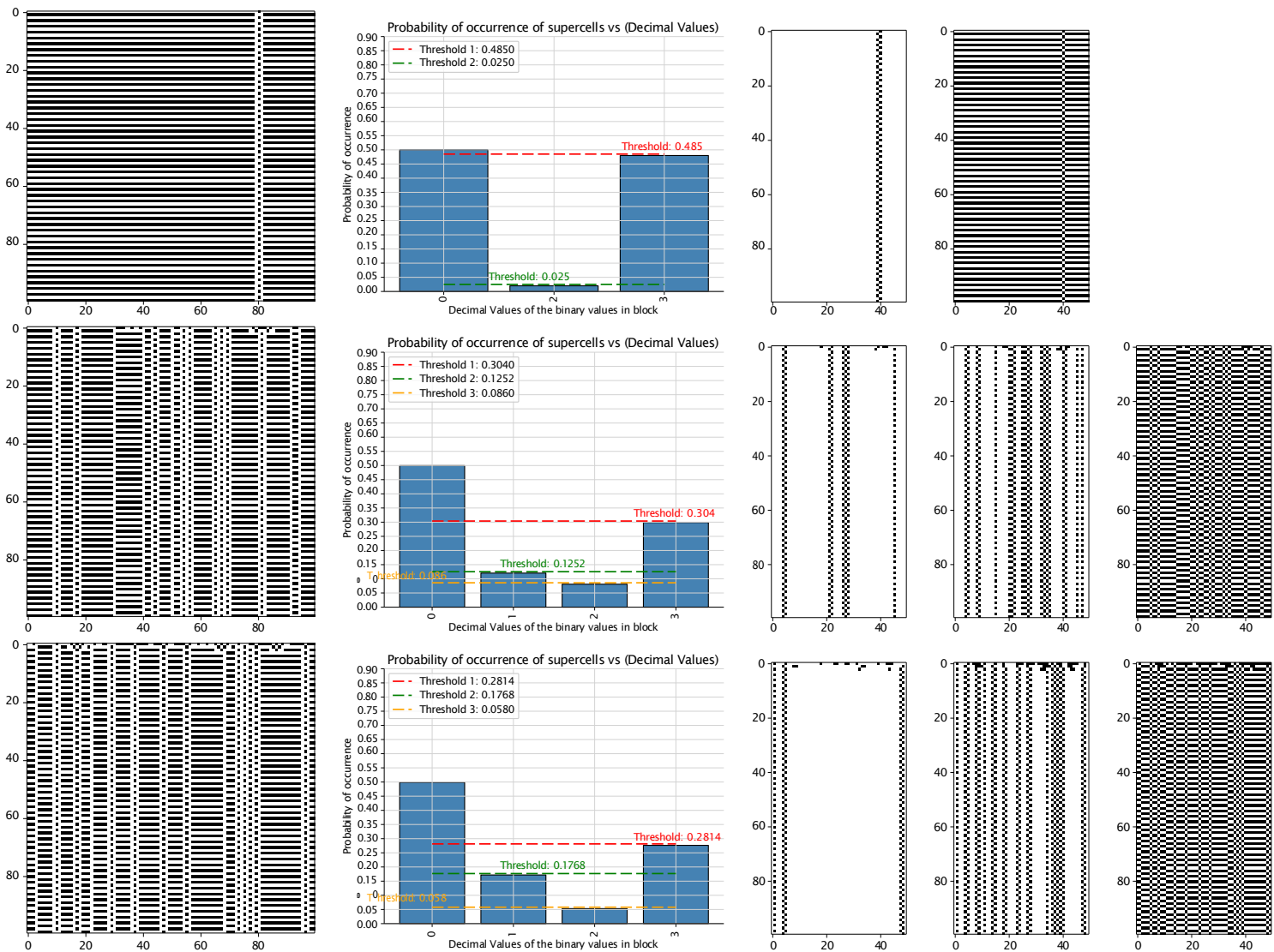

Table 30: FHCG plots for ECA Rule 30.

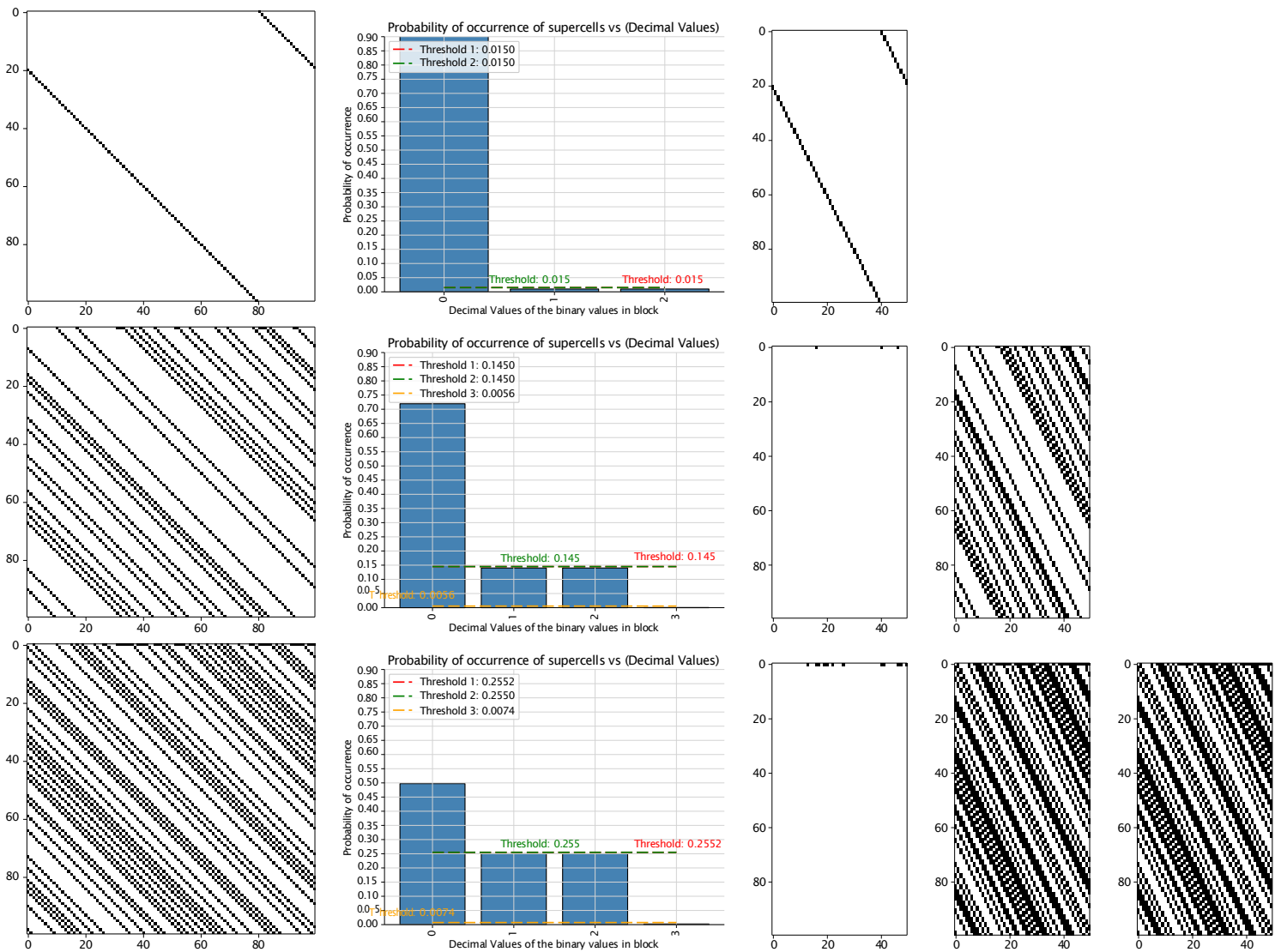

Table 31: FHCG plots for ECA Rule 31.

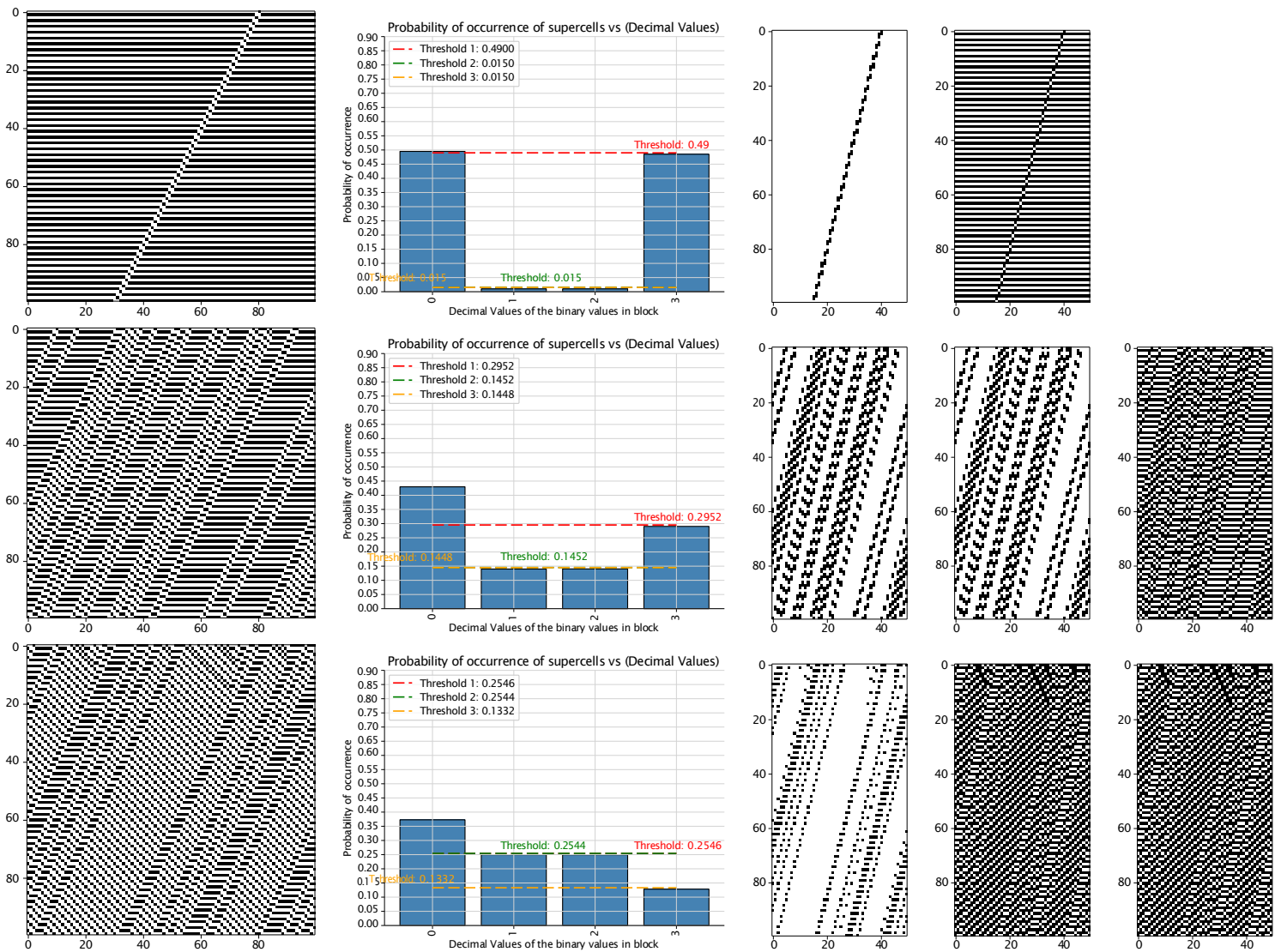

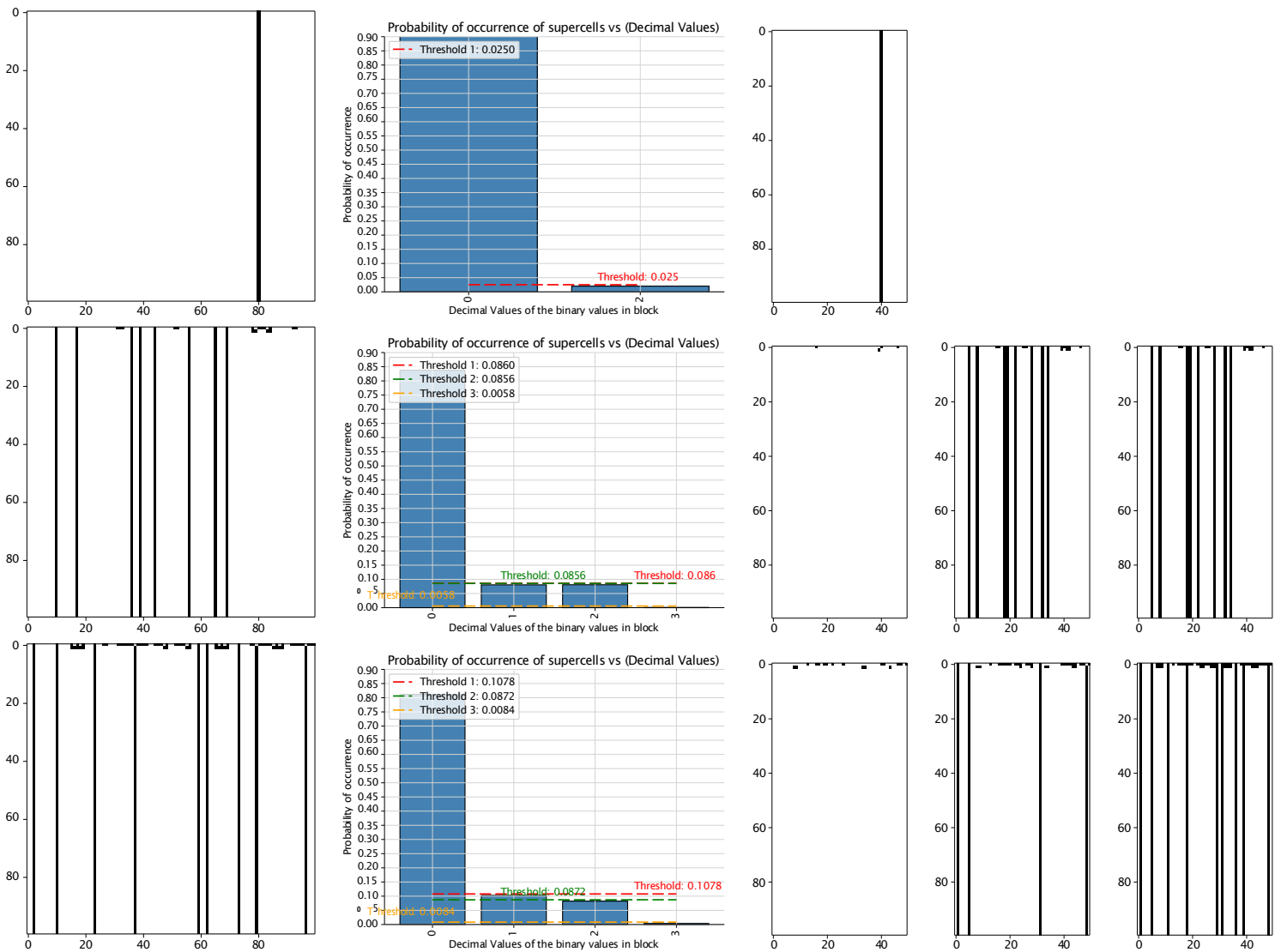

Table 32: FHCG plots for ECA Rule 32.

Table 33: FHCG plots for ECA Rule 33.

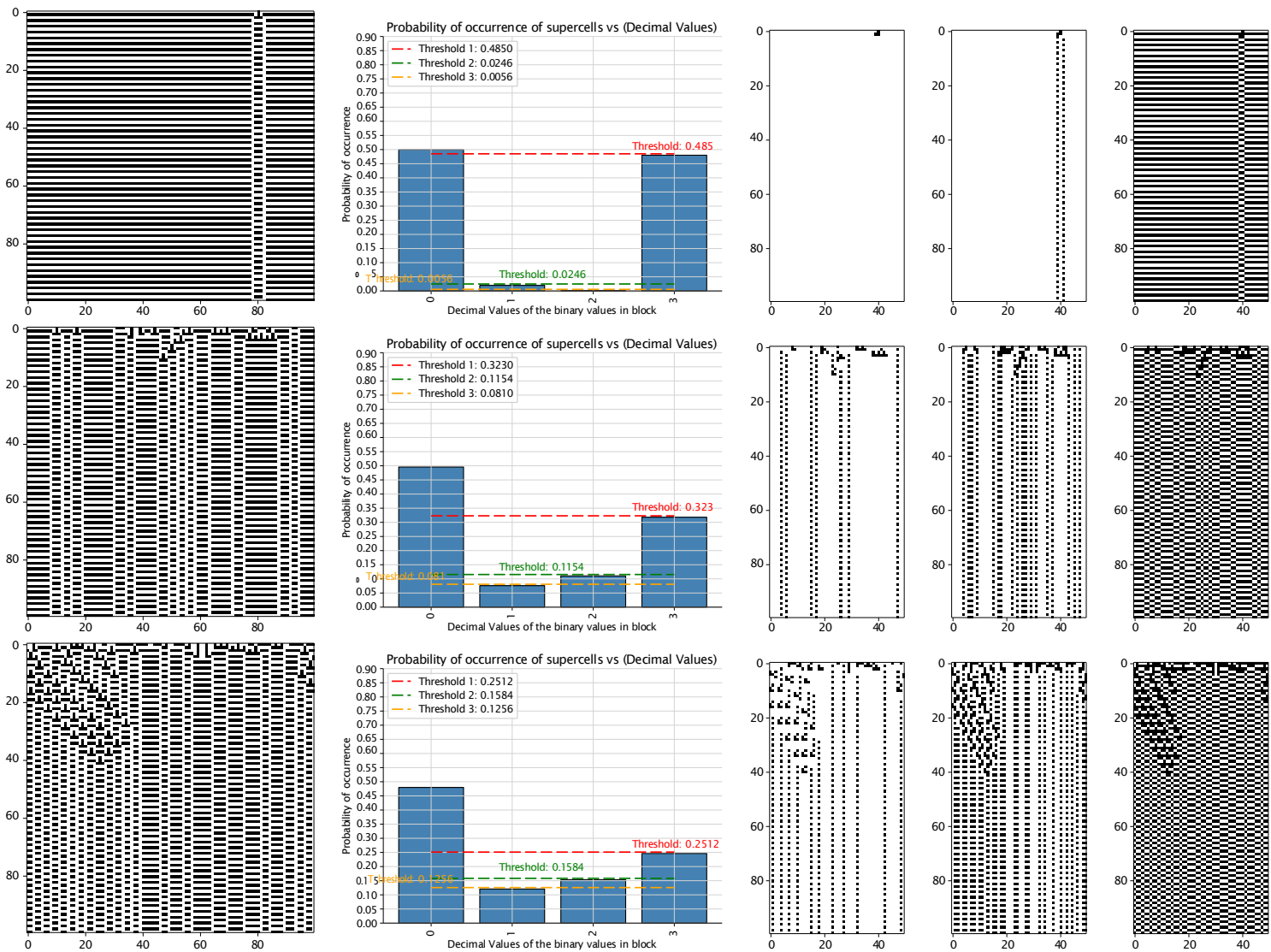

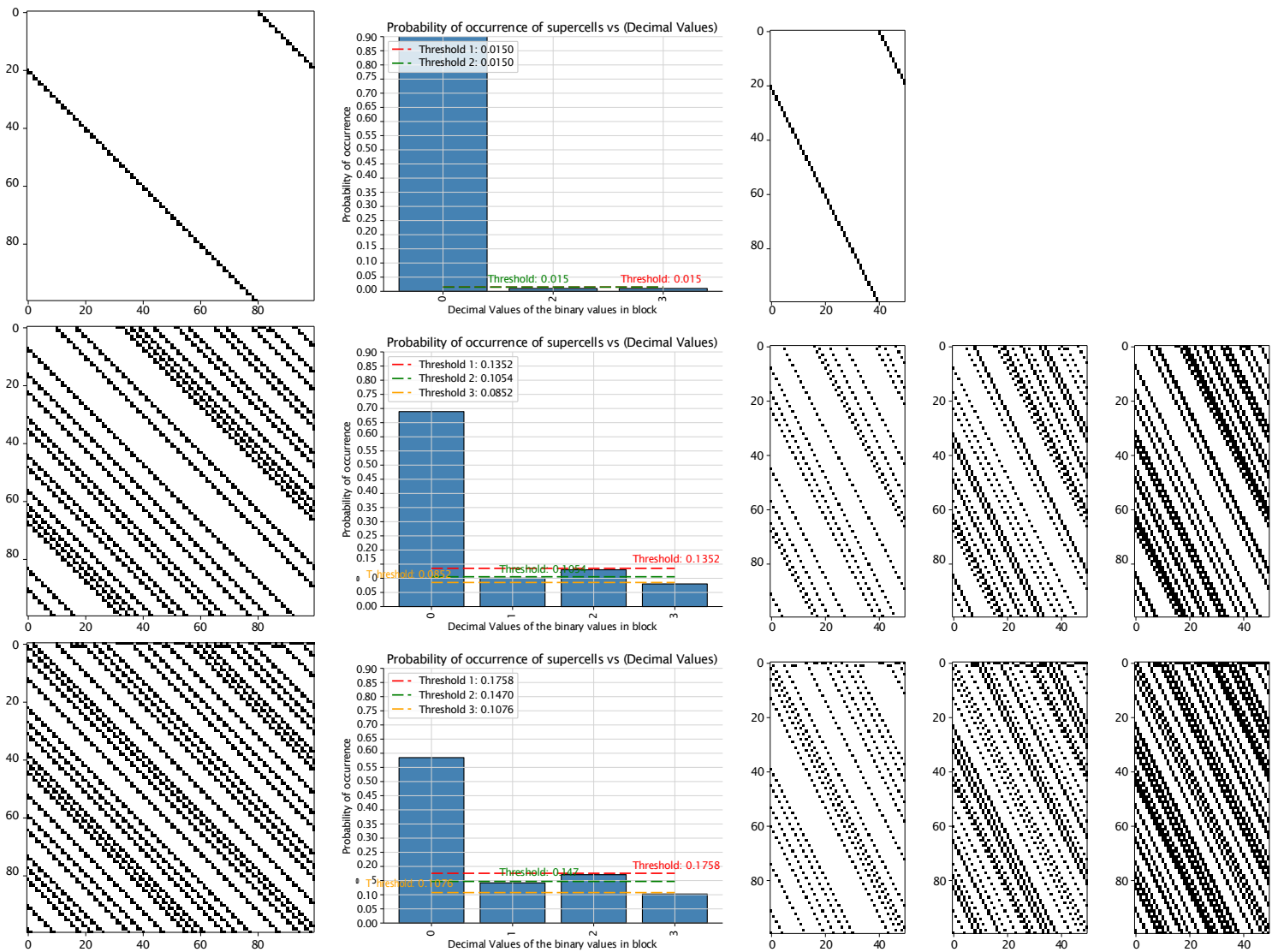

Table 34: FHCG plots for ECA Rule 34.

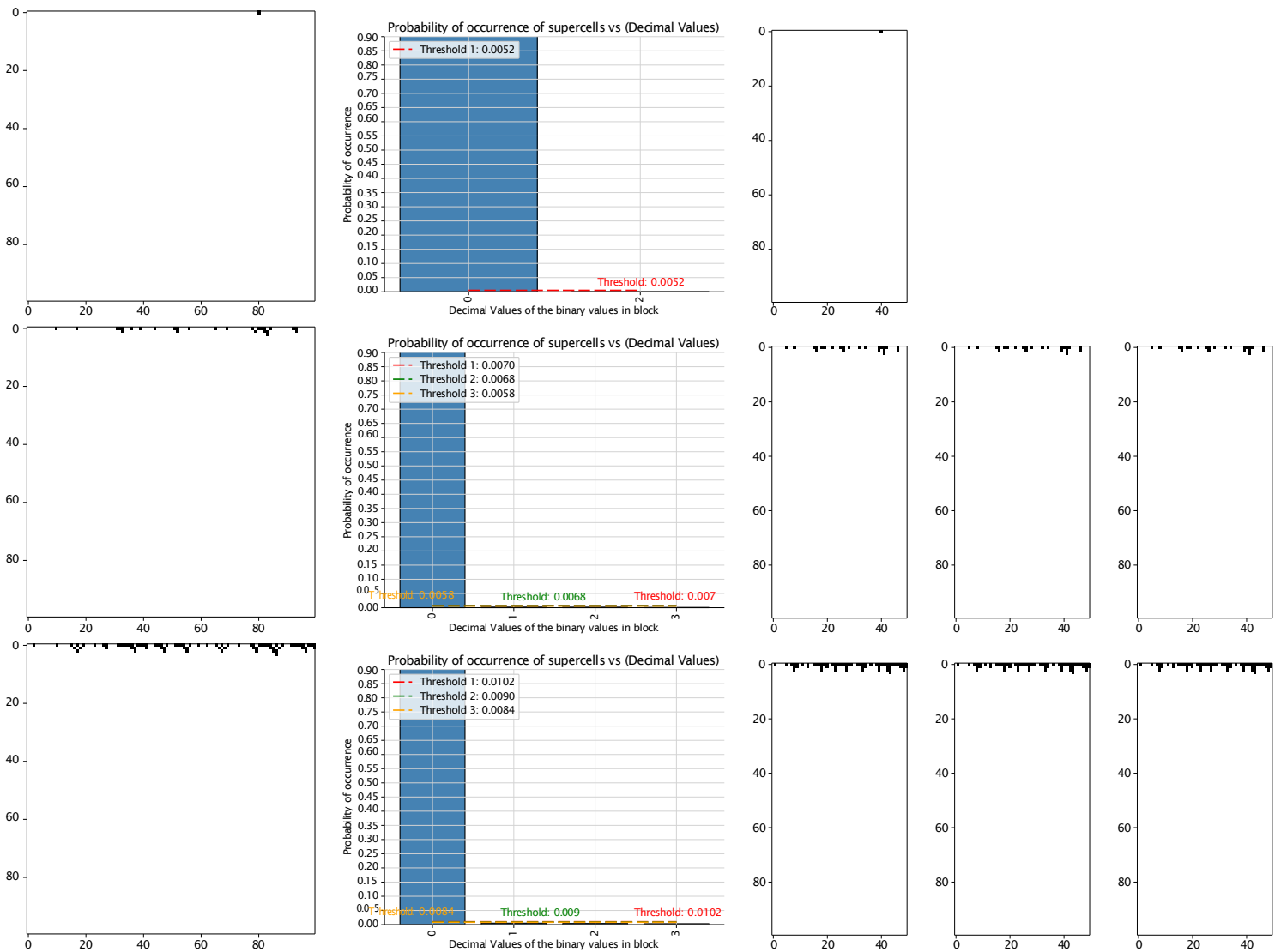

Table 35: FHCG plots for ECA Rule 35.

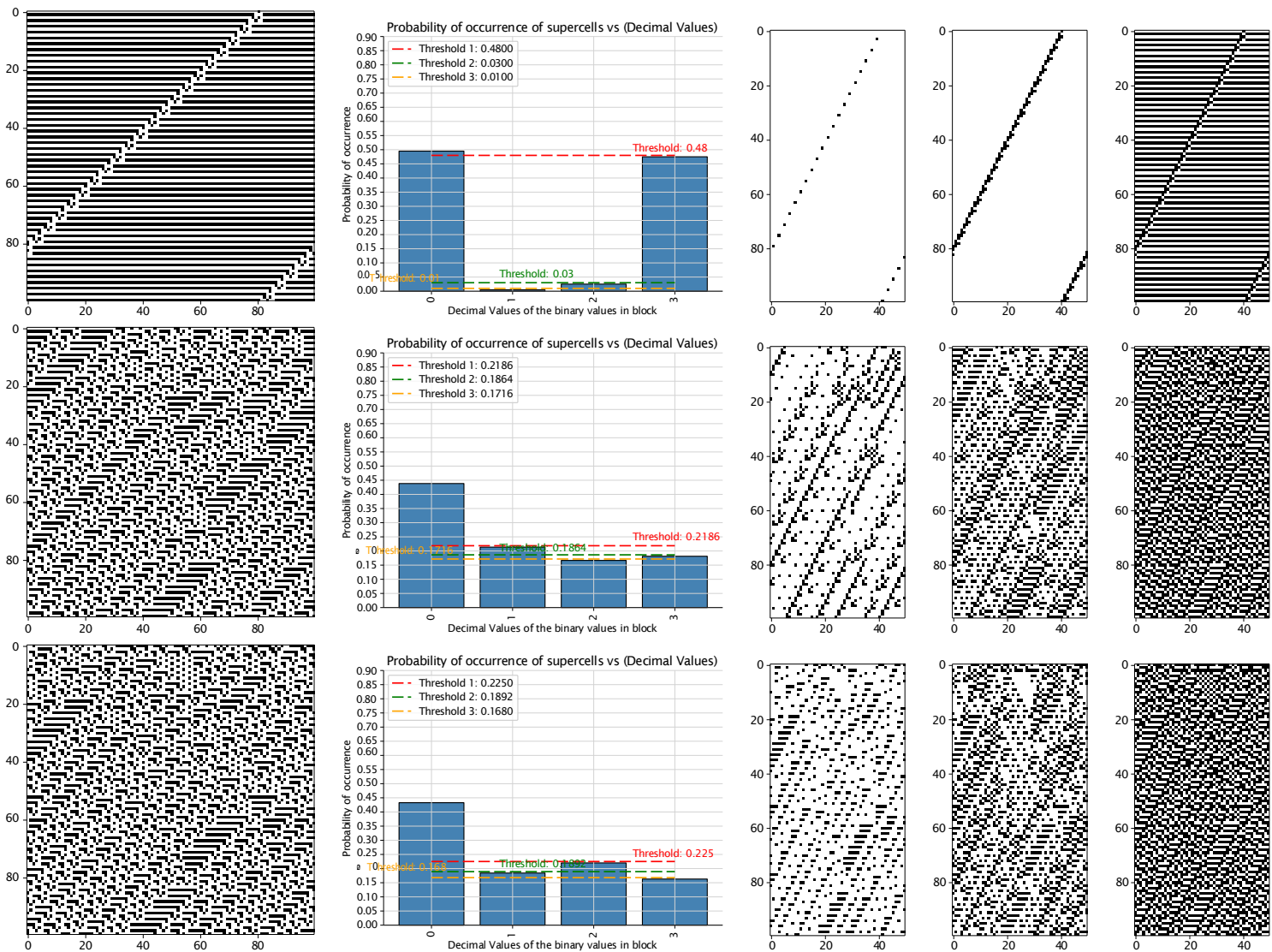

Table 36: FHCG plots for ECA Rule 36.

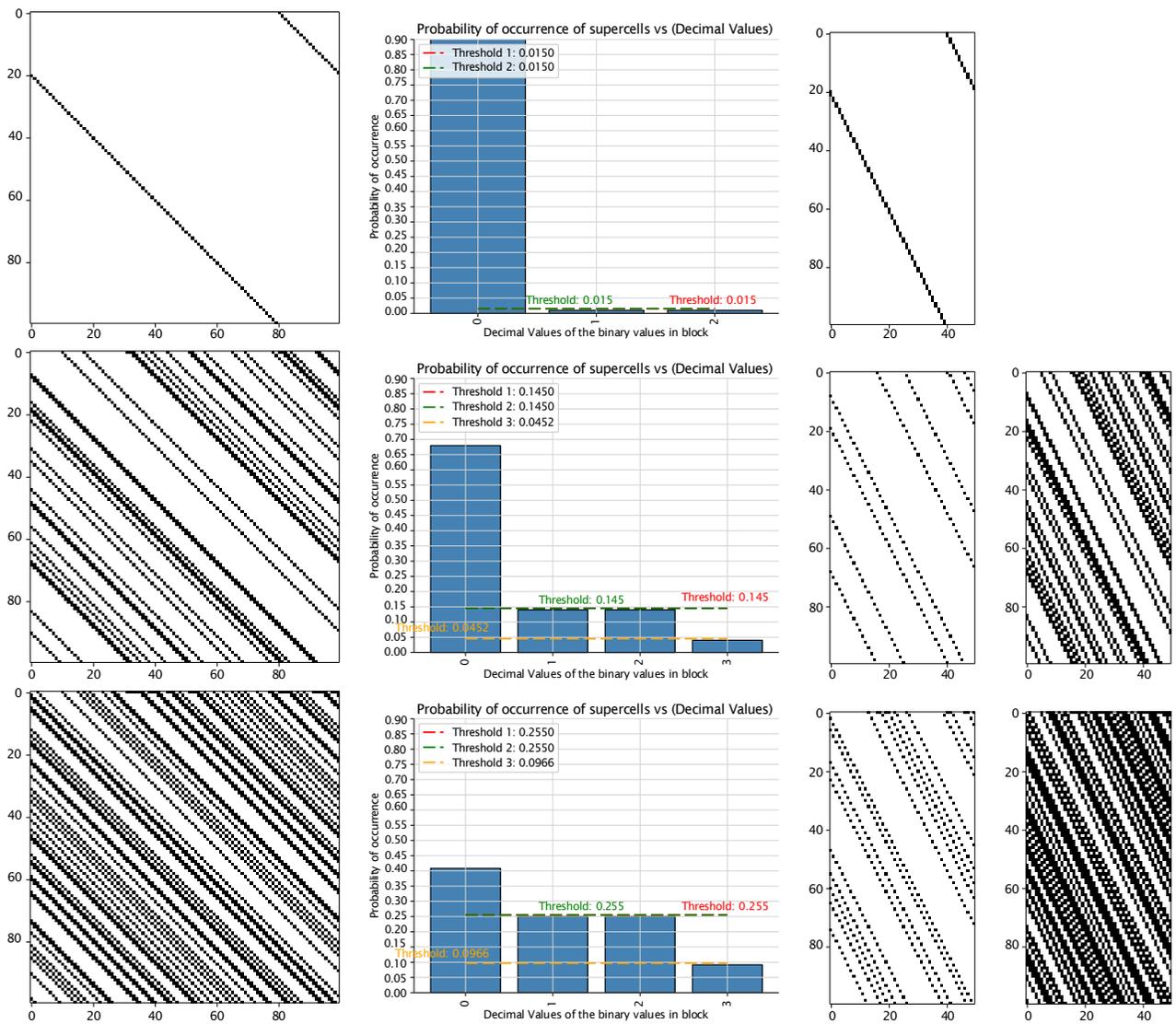

Table 37: FHCG plots for ECA Rule 37.

Table 38: FHCG plots for ECA Rule 43.

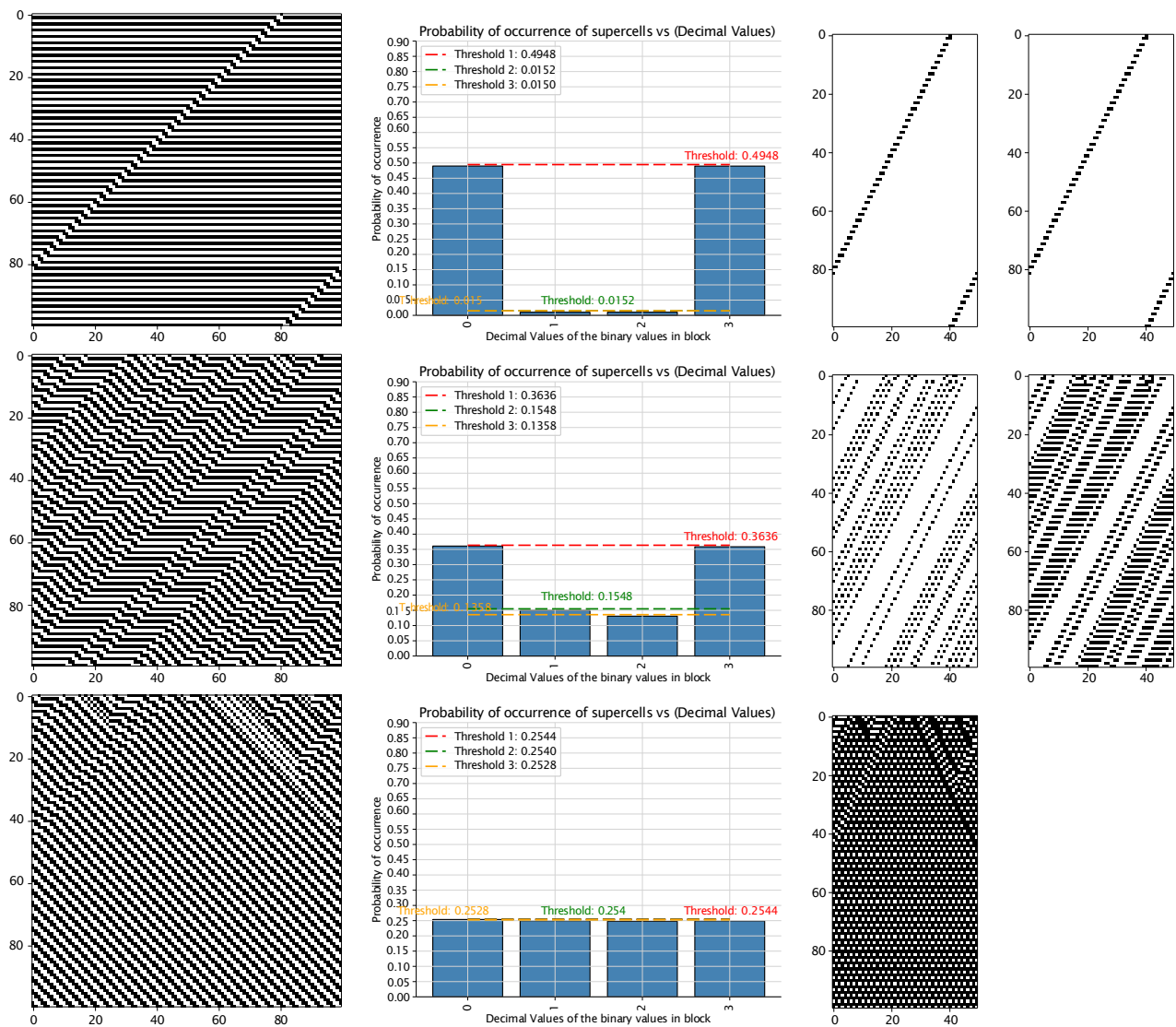

Table 39: FHCG plots for ECA Rule 44.

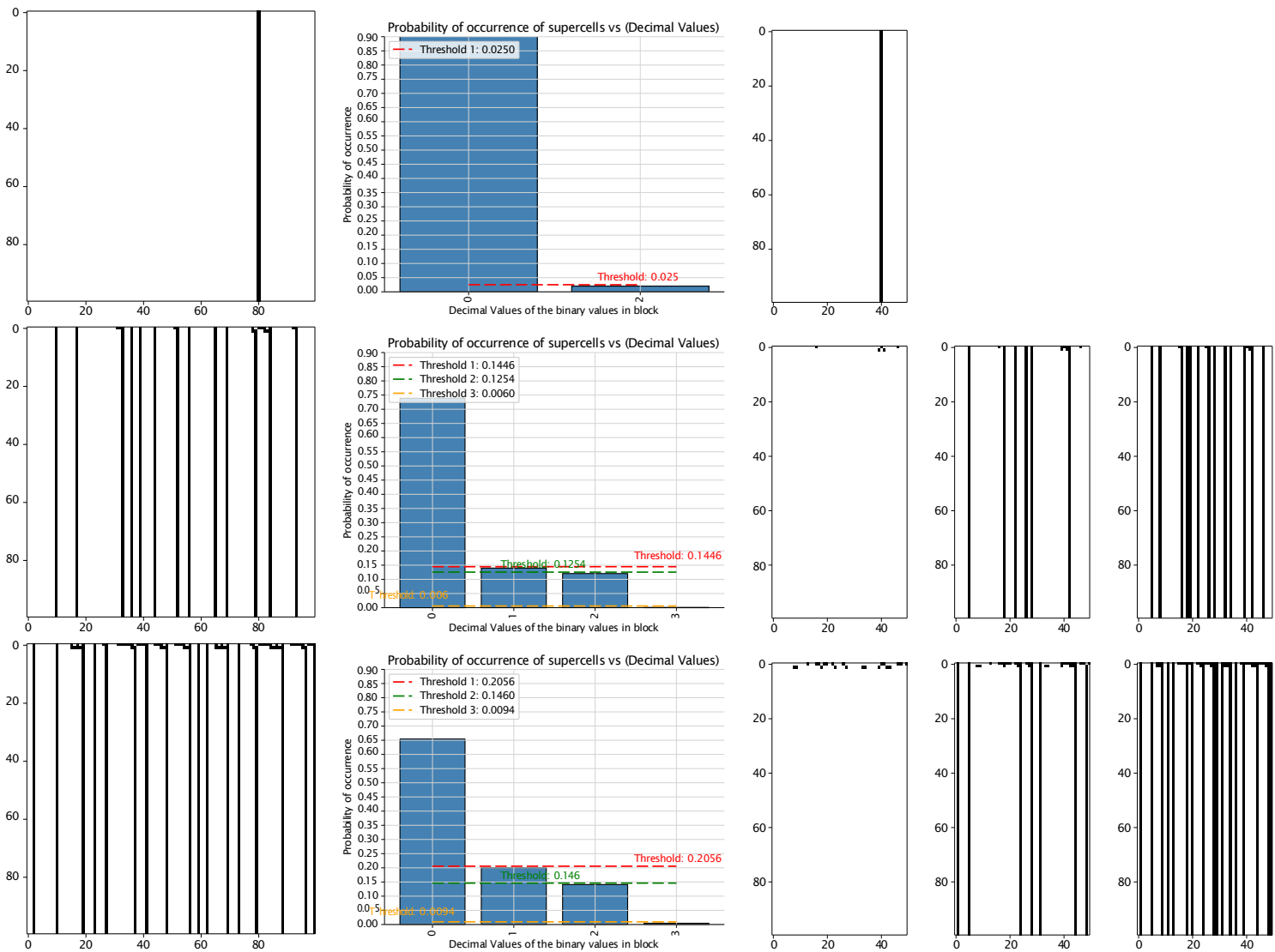

Table 40: FHCG plots for ECA Rule 45.

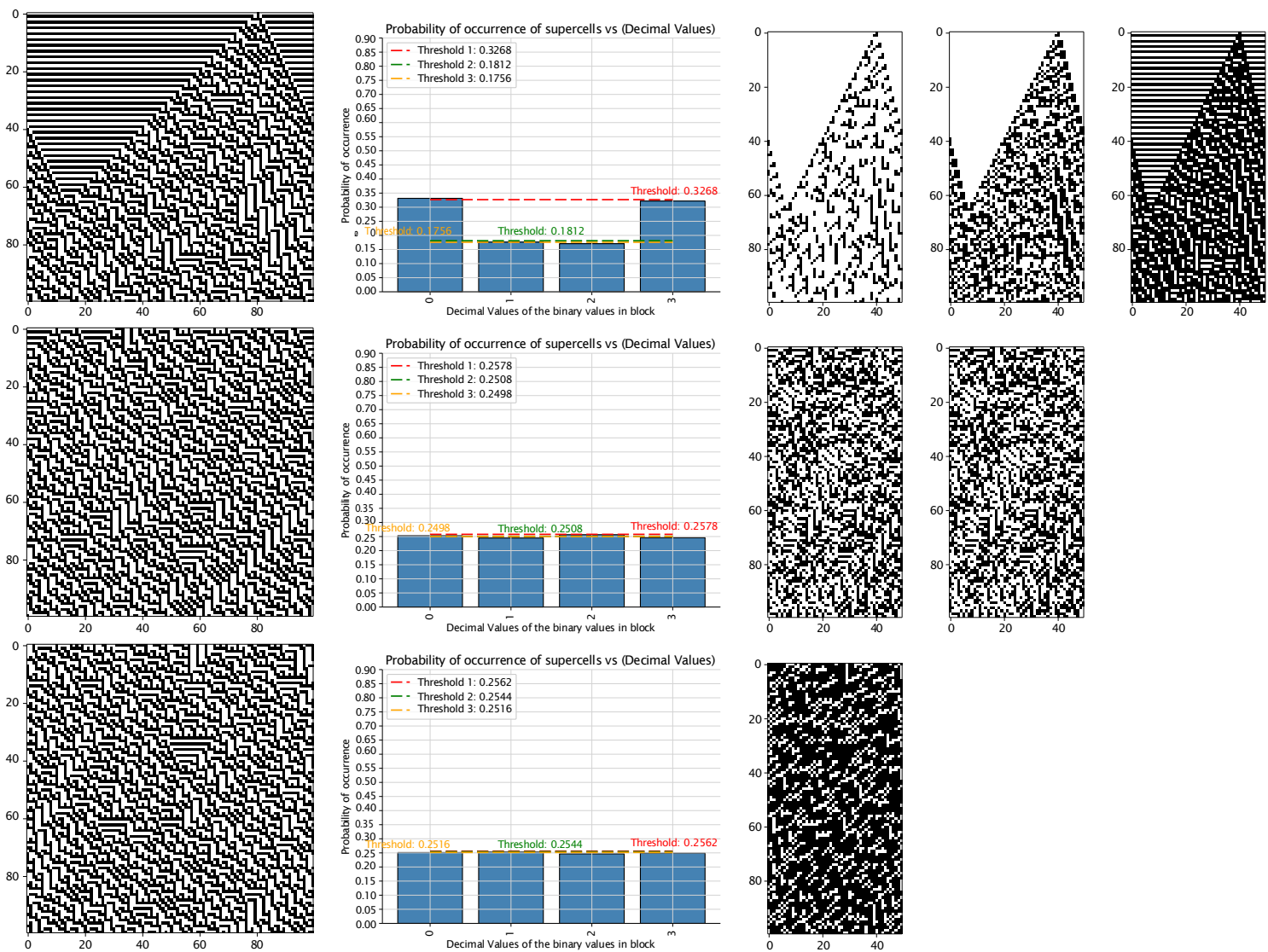

Table 41: FHCG plots for ECA Rule 46.

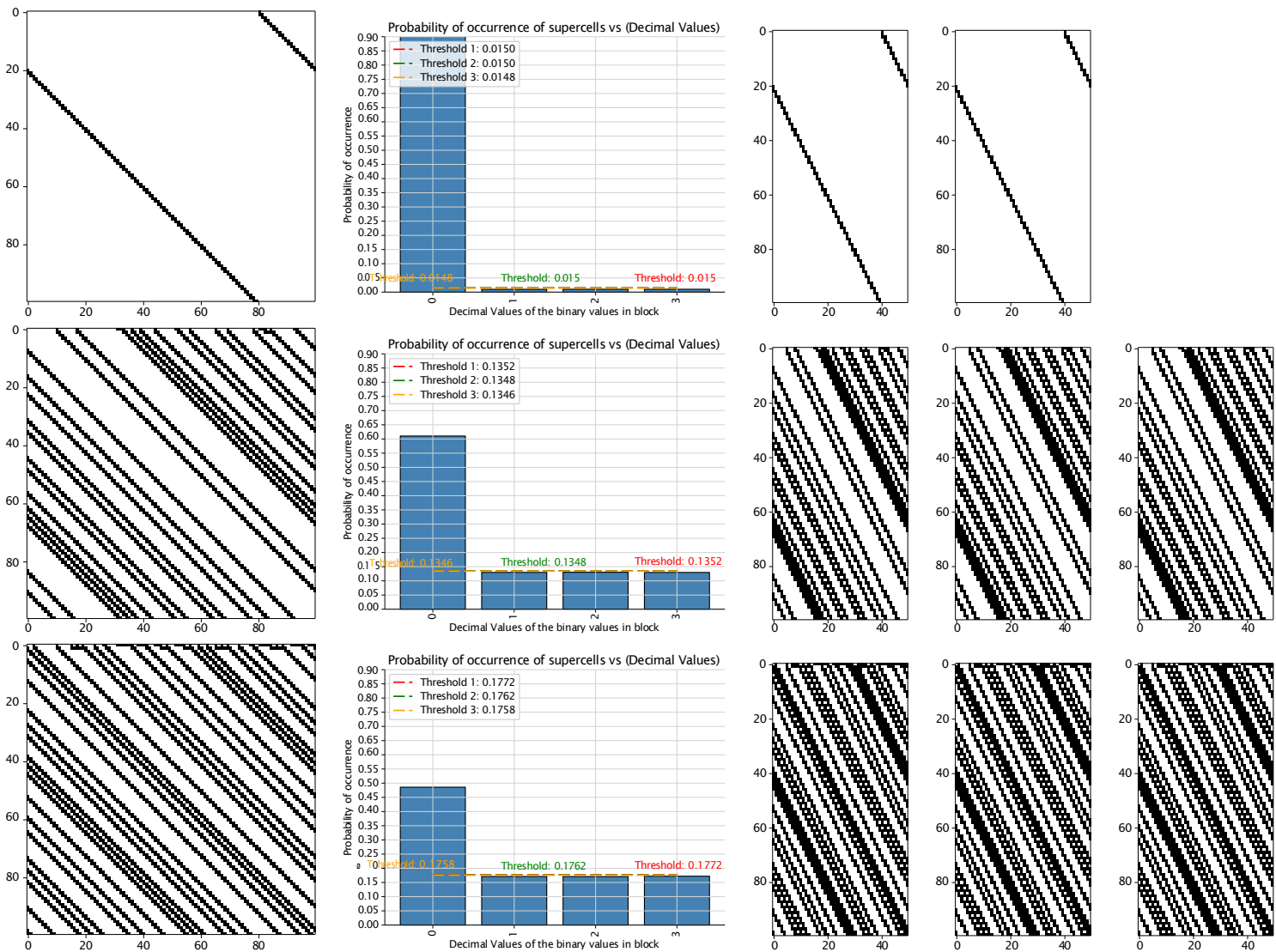

Table 42: FHCG plots for ECA Rule 50.

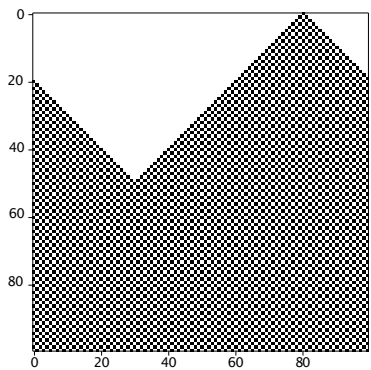 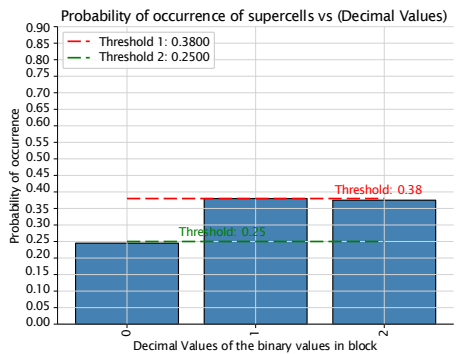 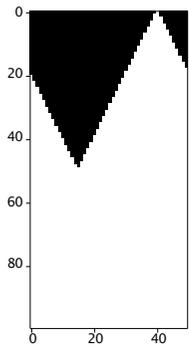

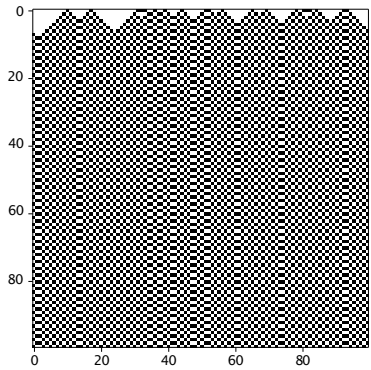 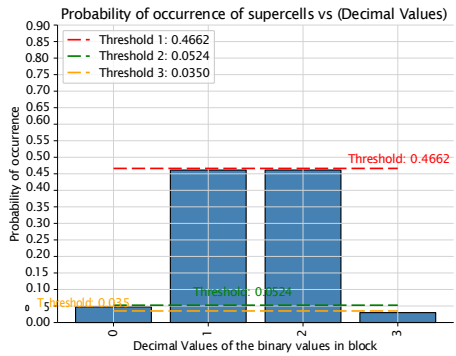 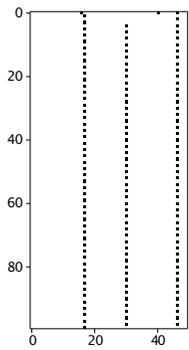 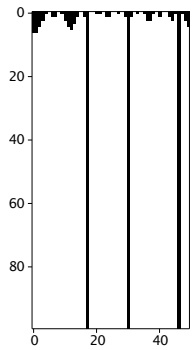

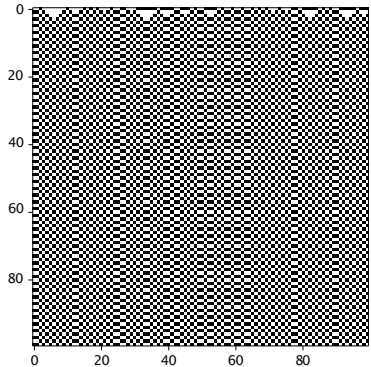 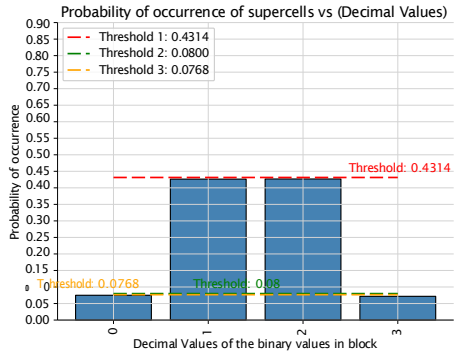 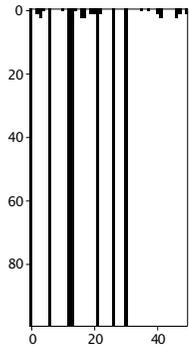 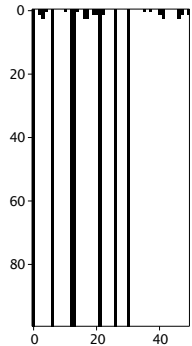

Table 43: FHCG plots for ECA Rule 50.

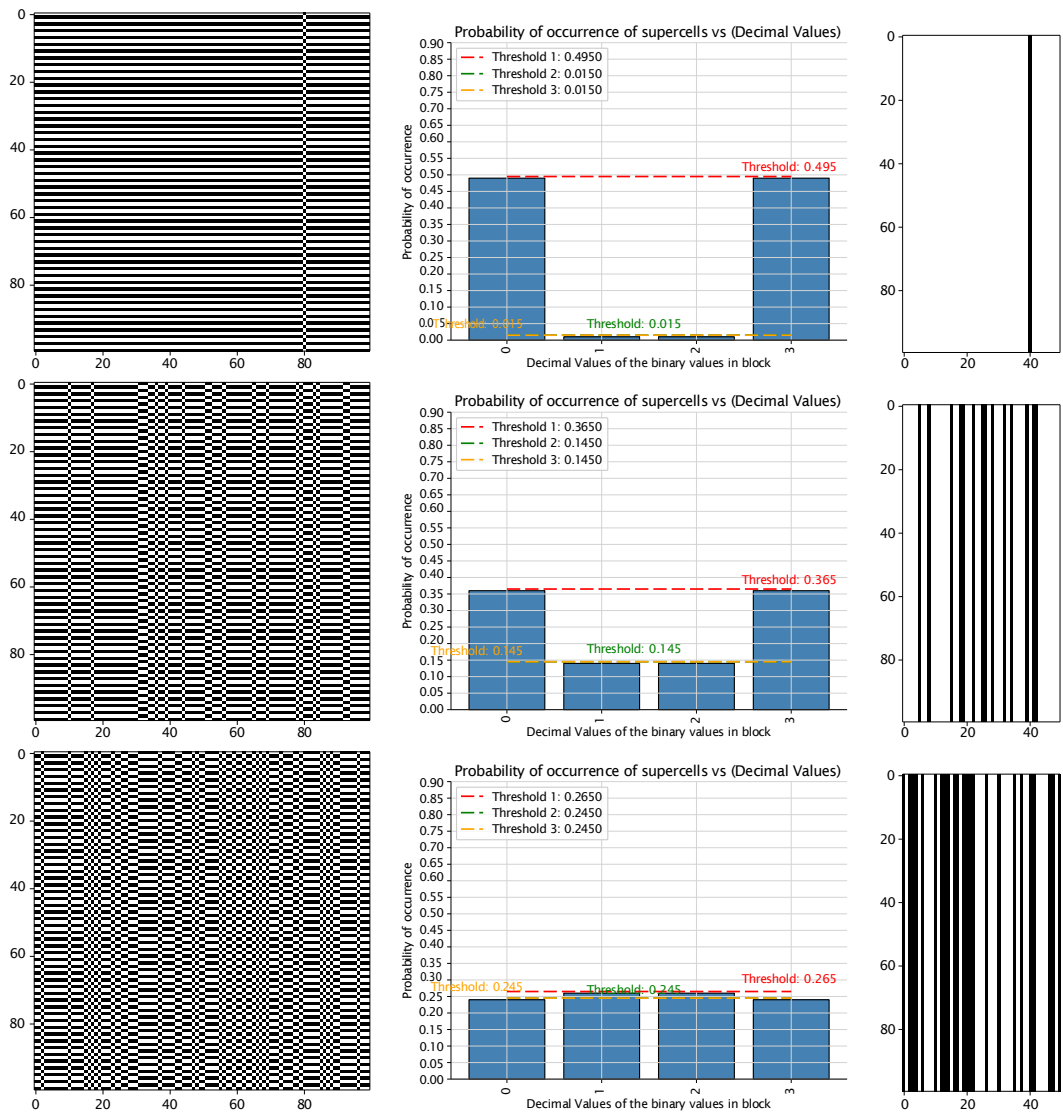

Table 44: FHCG plots for ECA Rule 50.

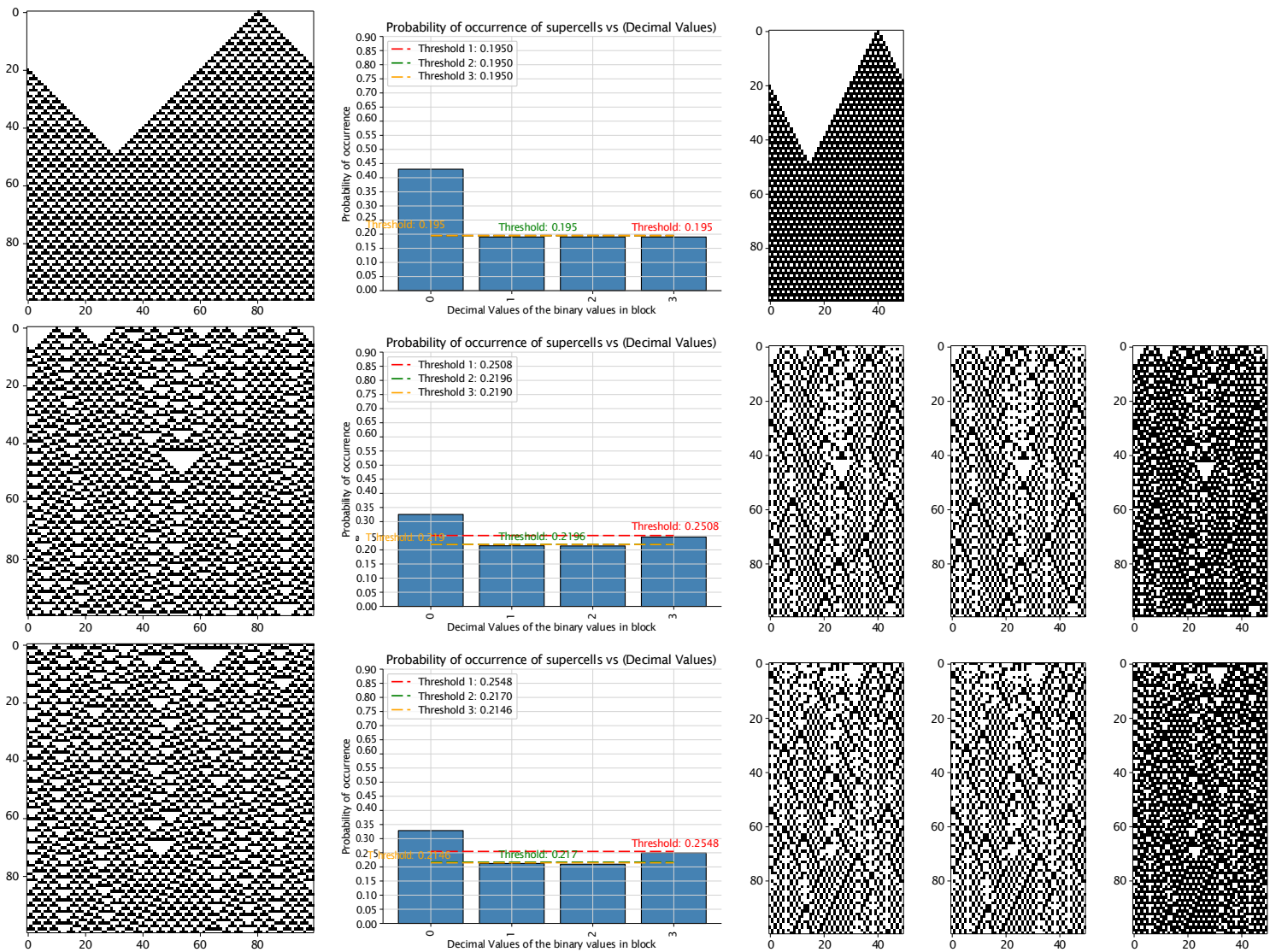

Table 45: FHCG plots for ECA Rule 50.

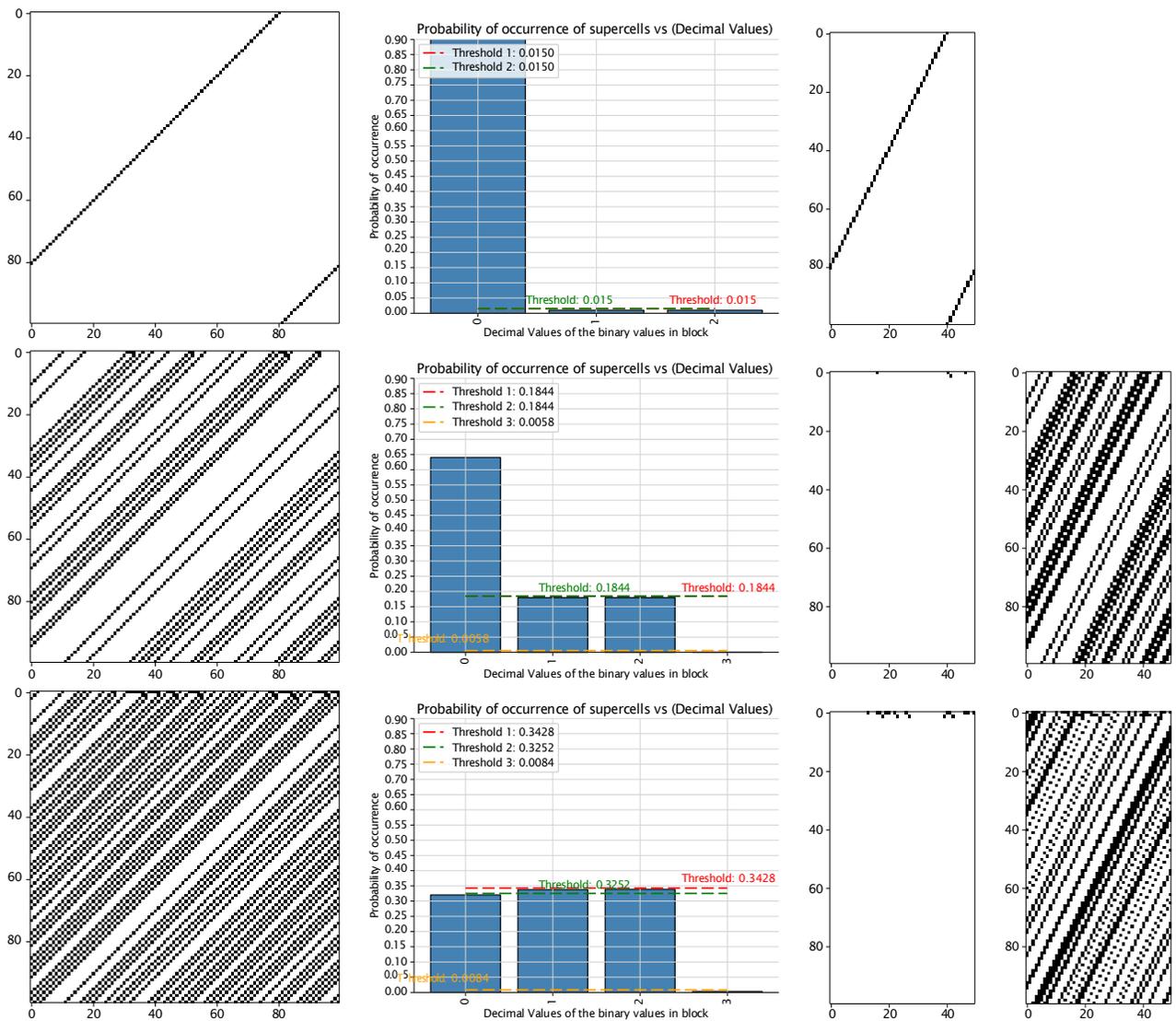

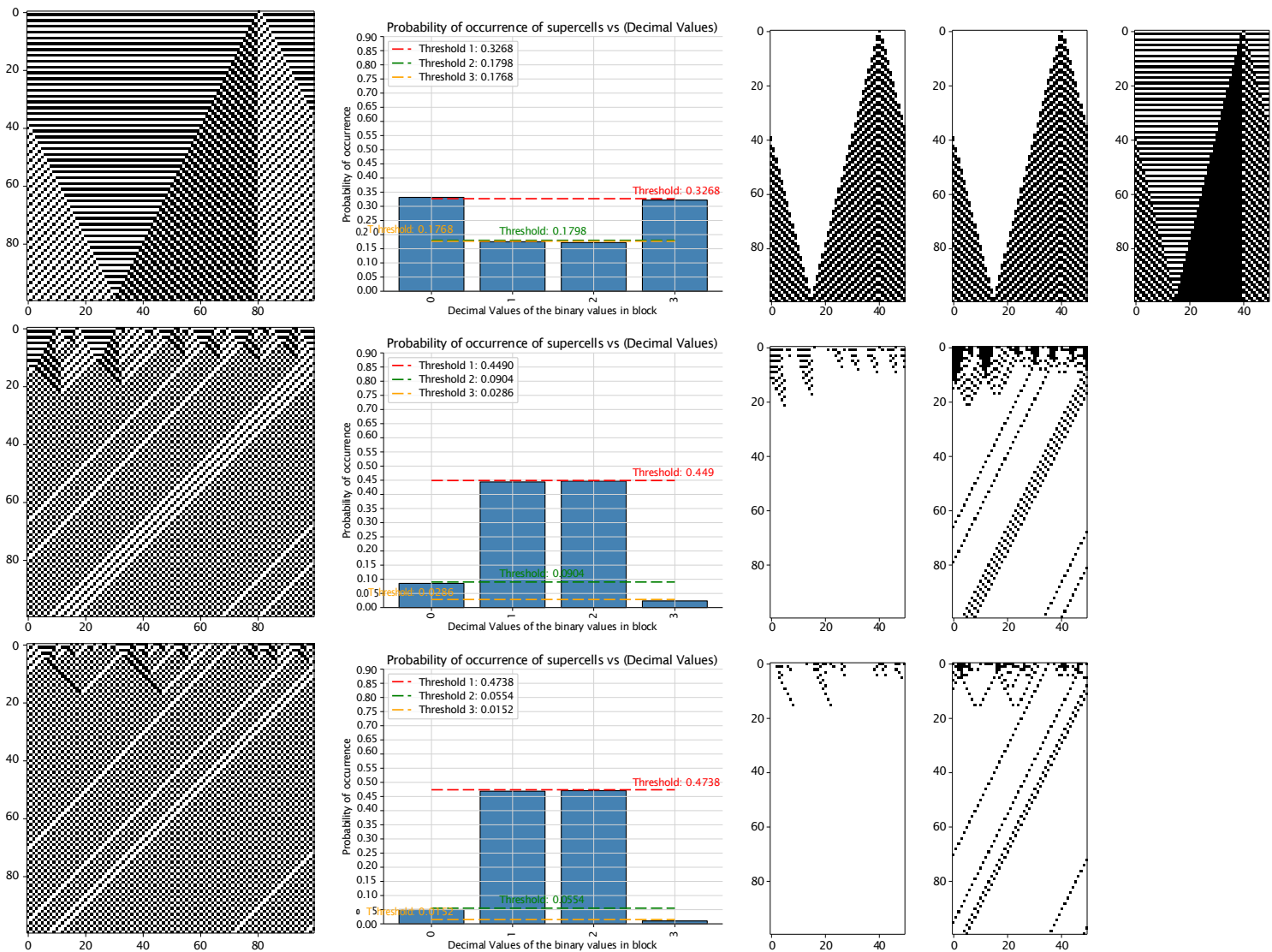

Table 47: FHCG plots for ECA Rule 50.

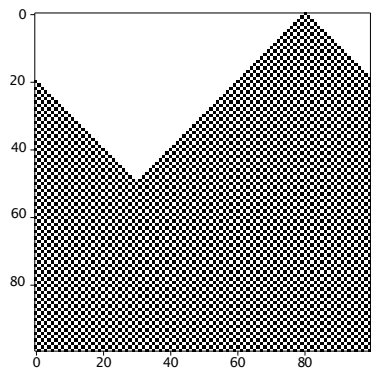
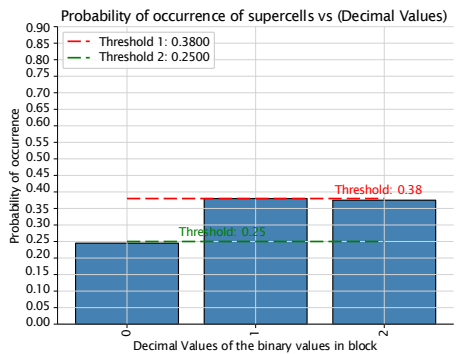
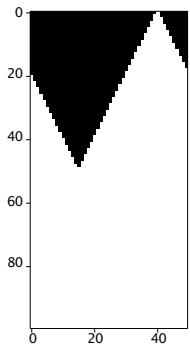
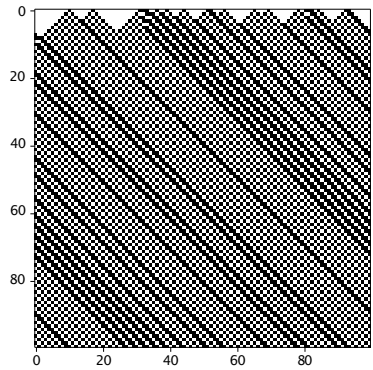
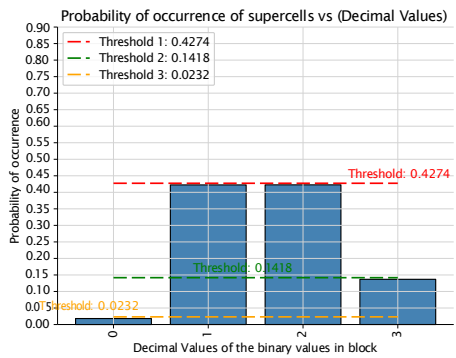
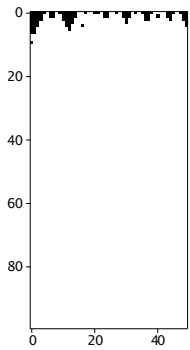
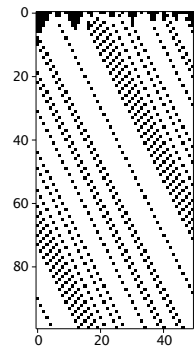
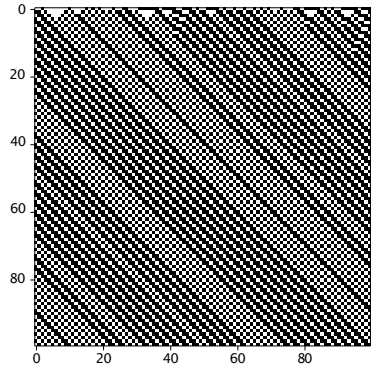
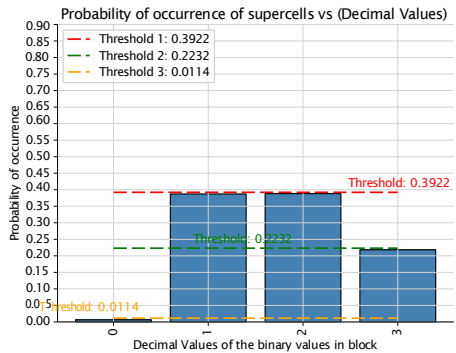
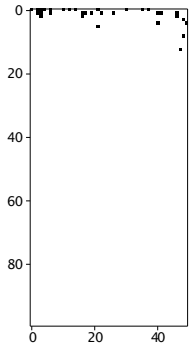
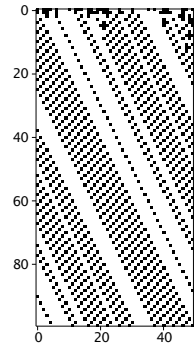

53Jain and Nichele: Coarse Graining in CA, NMI, 10, 1–96, 2023

Table 48: FHCG plots for ECA Rule 50.

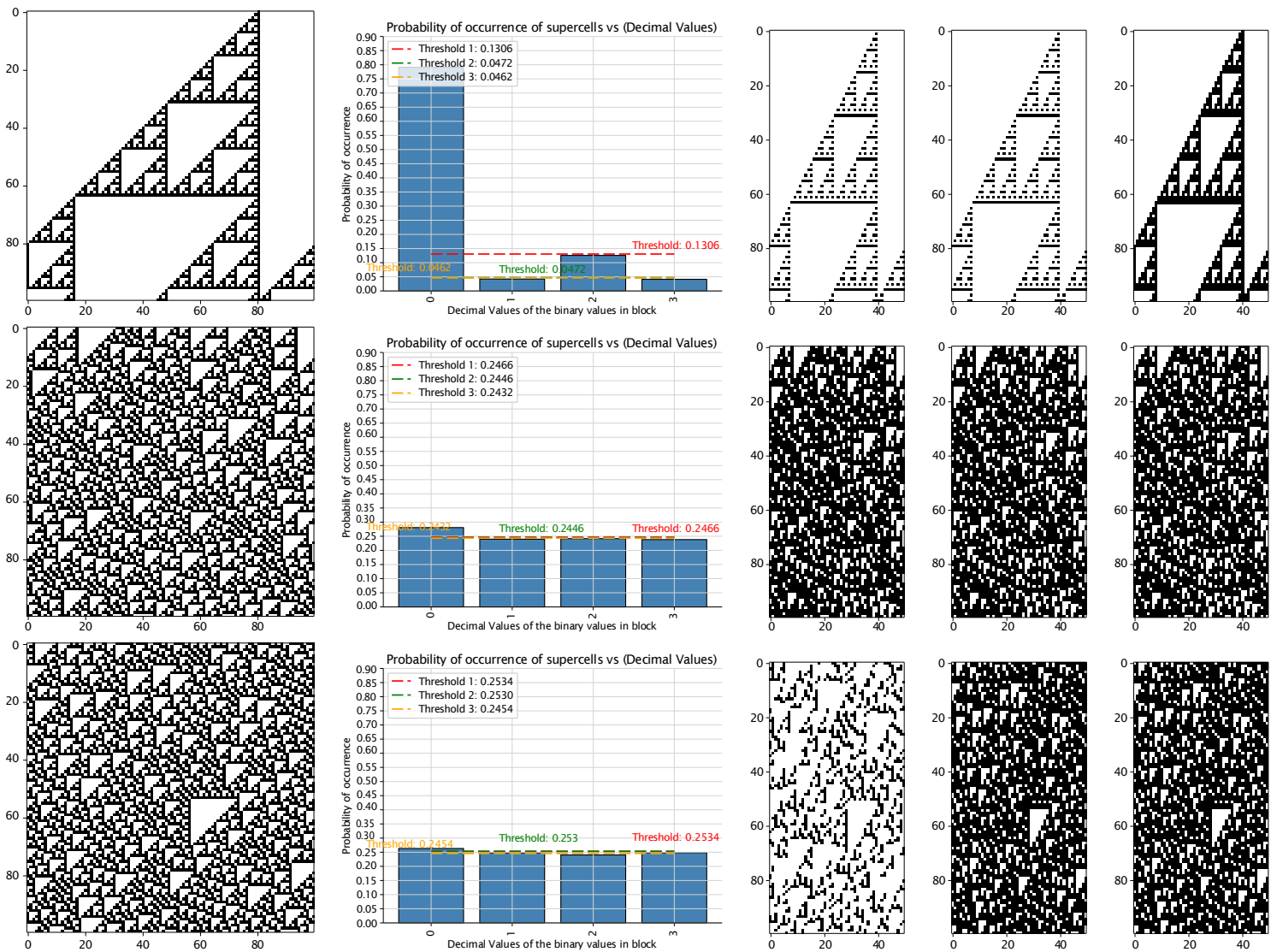

Table 49: FHCG plots for ECA Rule 50.

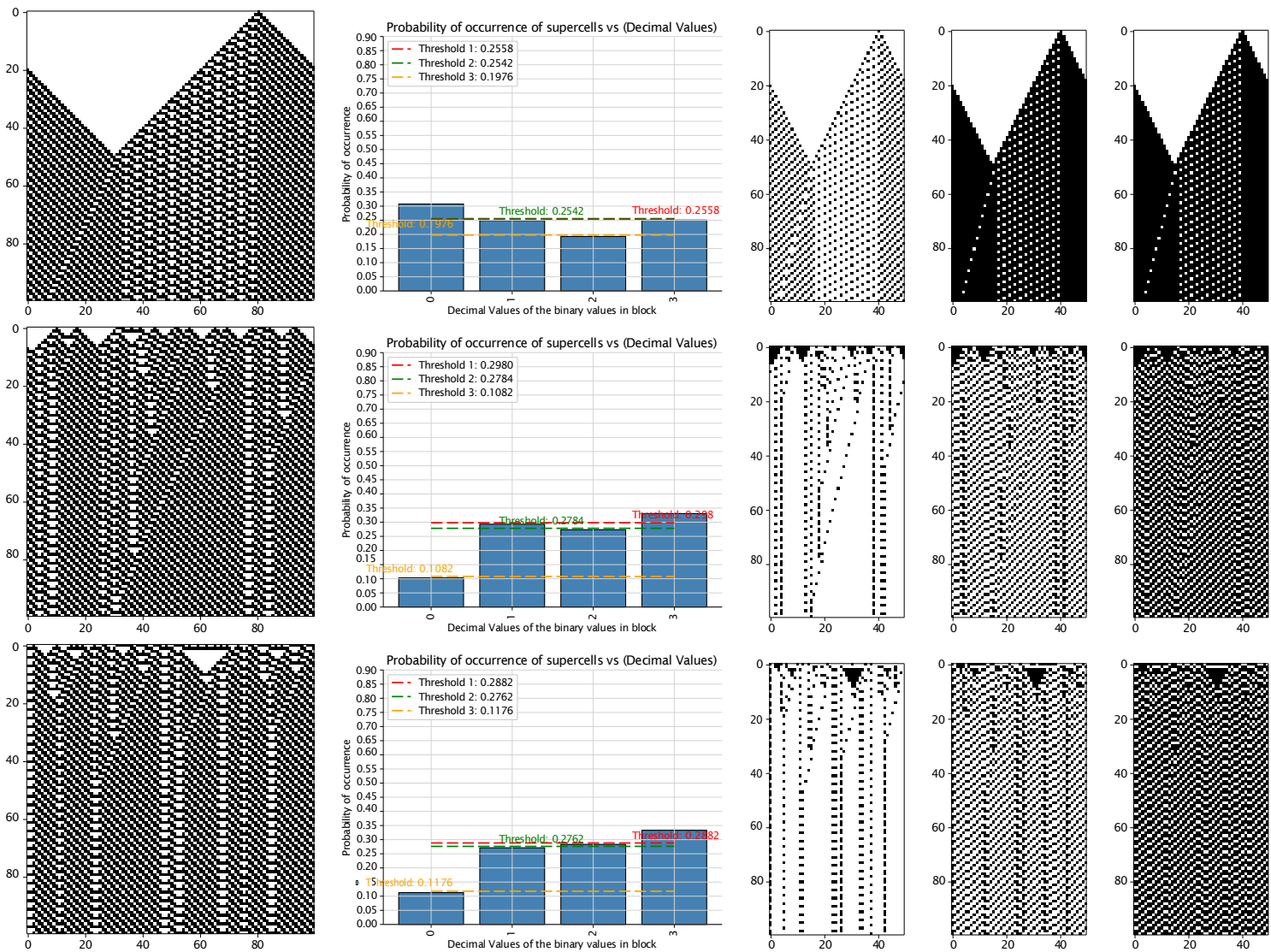

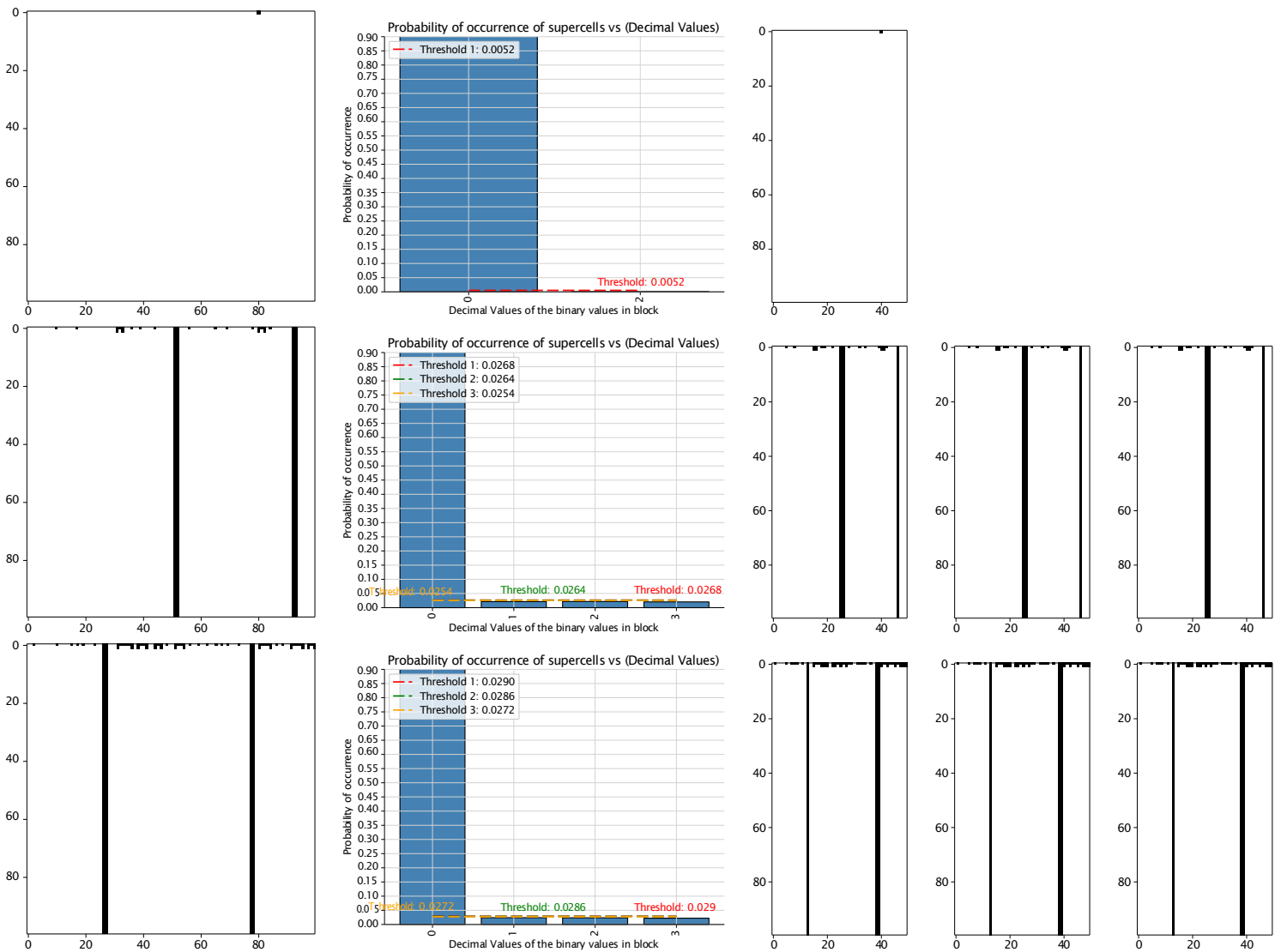

Table 50: FHCG plots for ECA Rule 50.

Table 51: FHCG plots for ECA Rule 50.

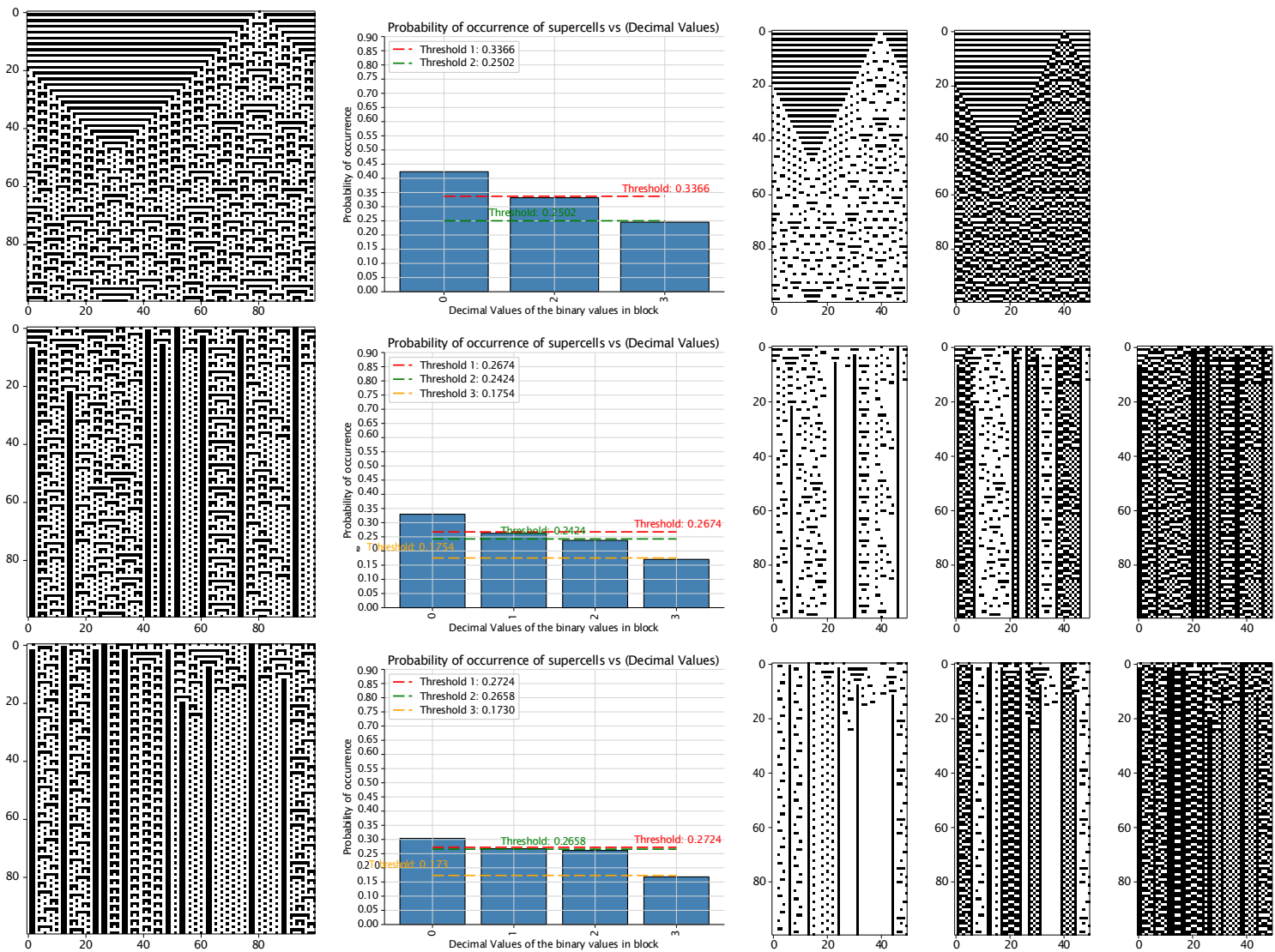

Table 52: FHCG plots for ECA Rule 50.

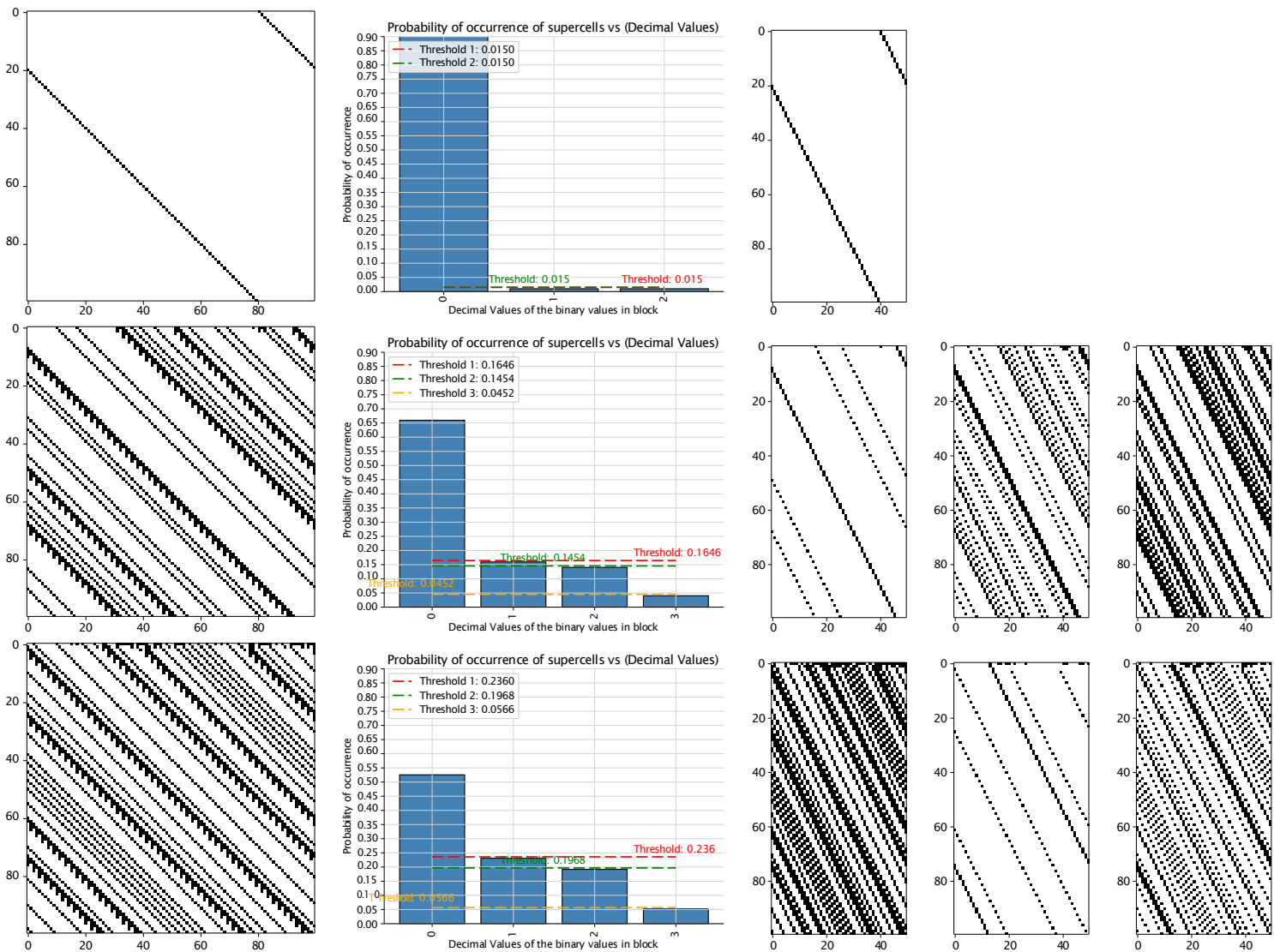

Table 53: FHCG plots for ECA Rule 50.

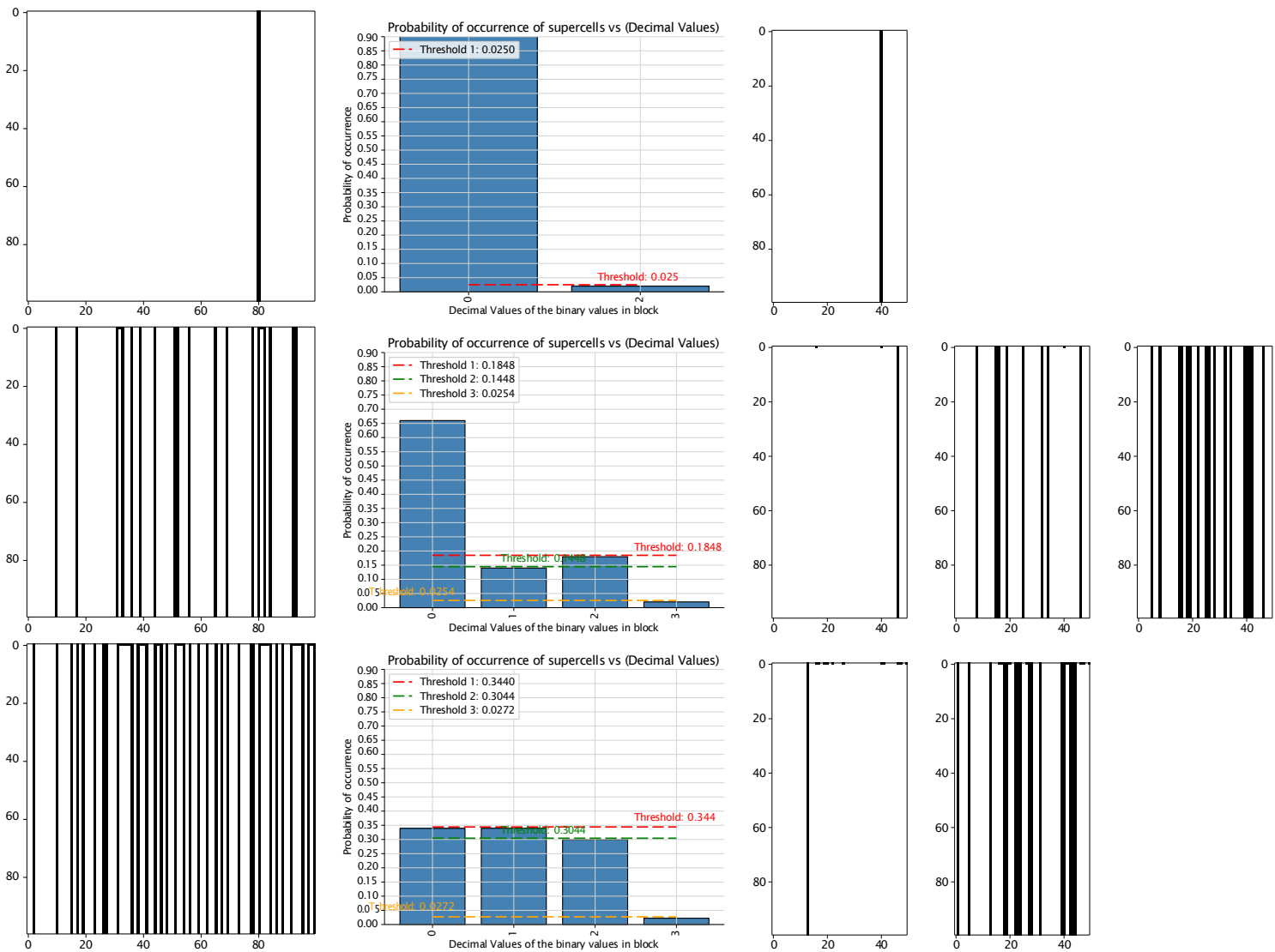

Table 54: FHCG plots for ECA Rule 50.

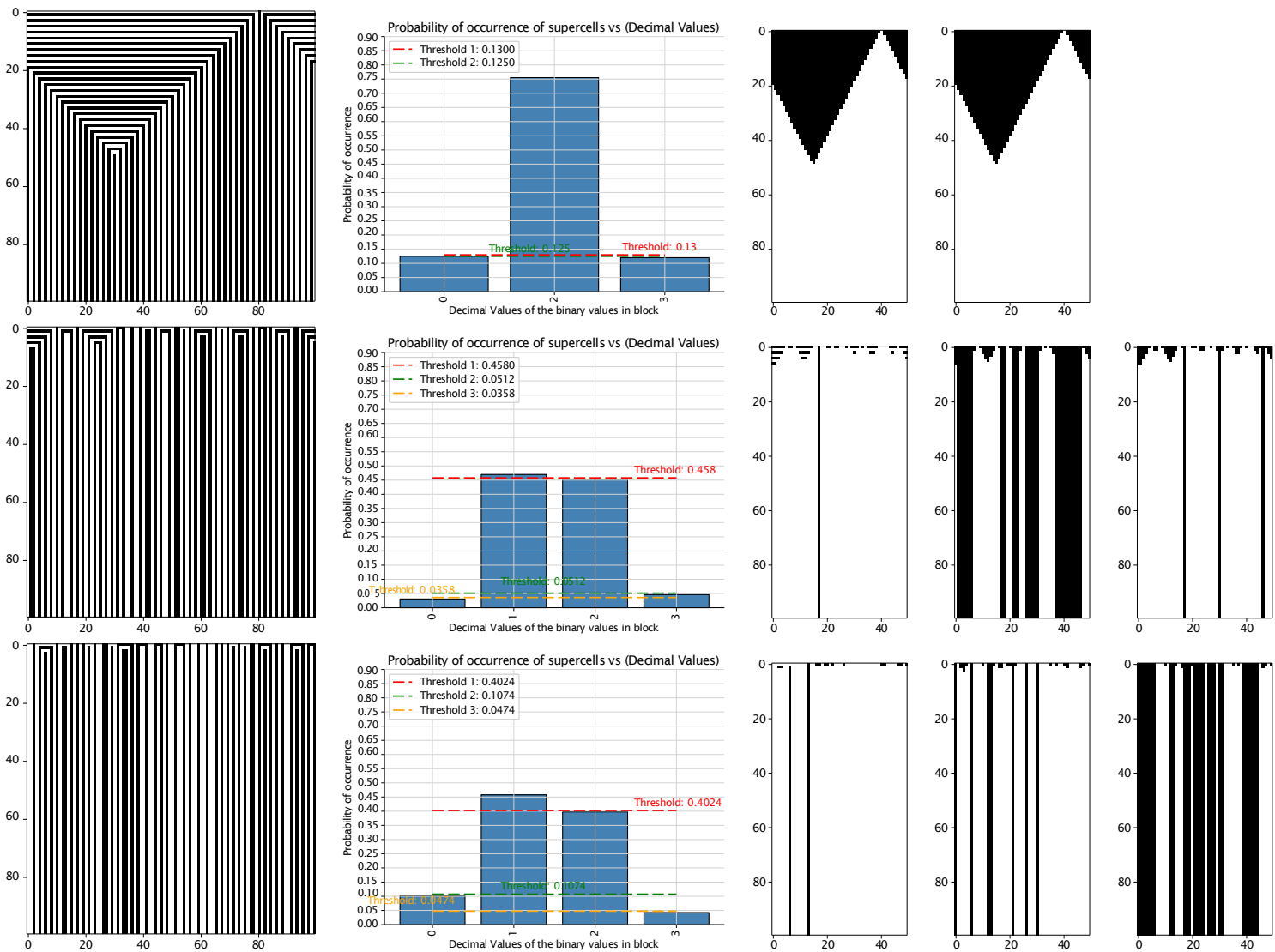

Table 55: FHCG plots for ECA Rule 50.

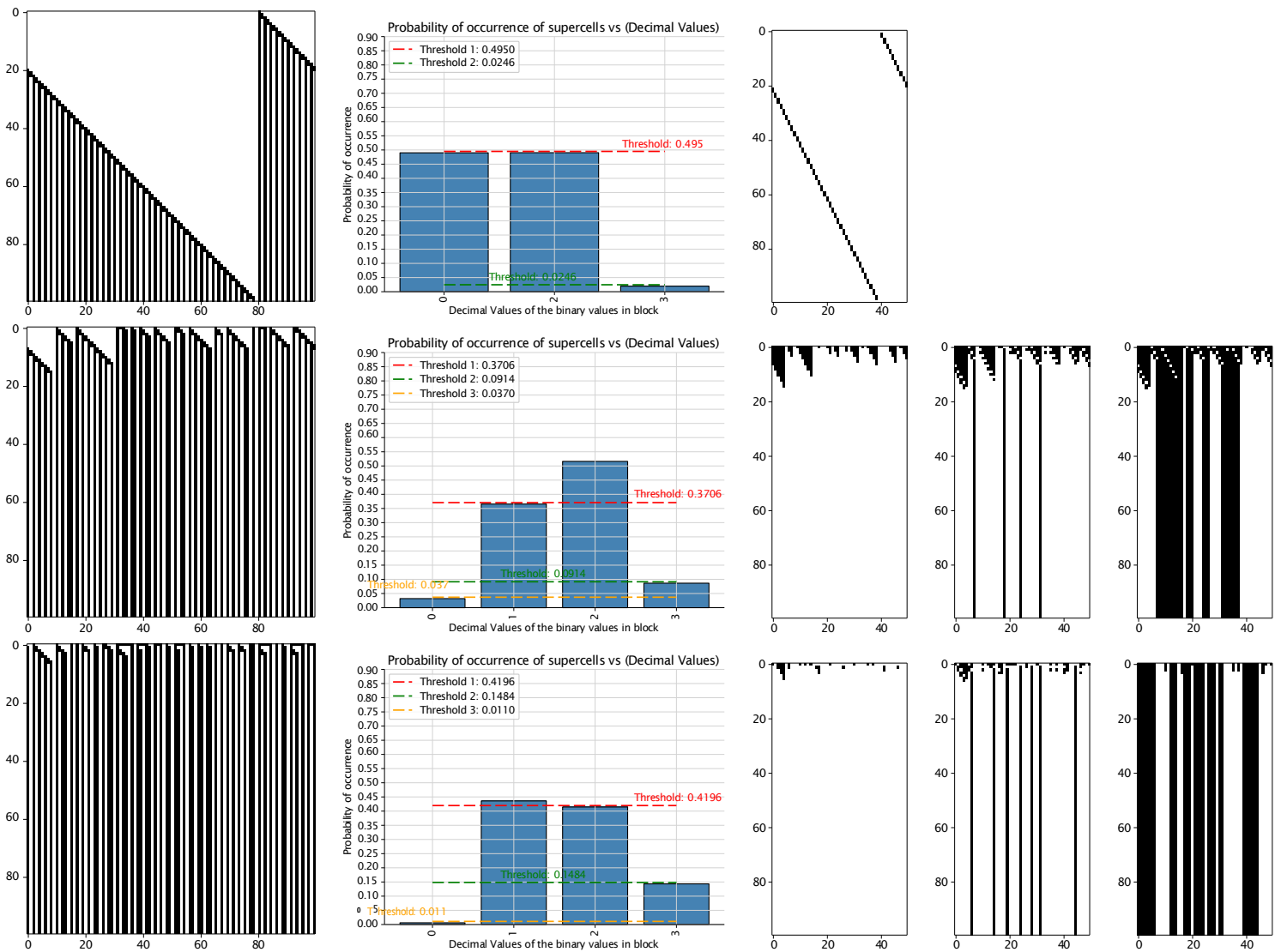

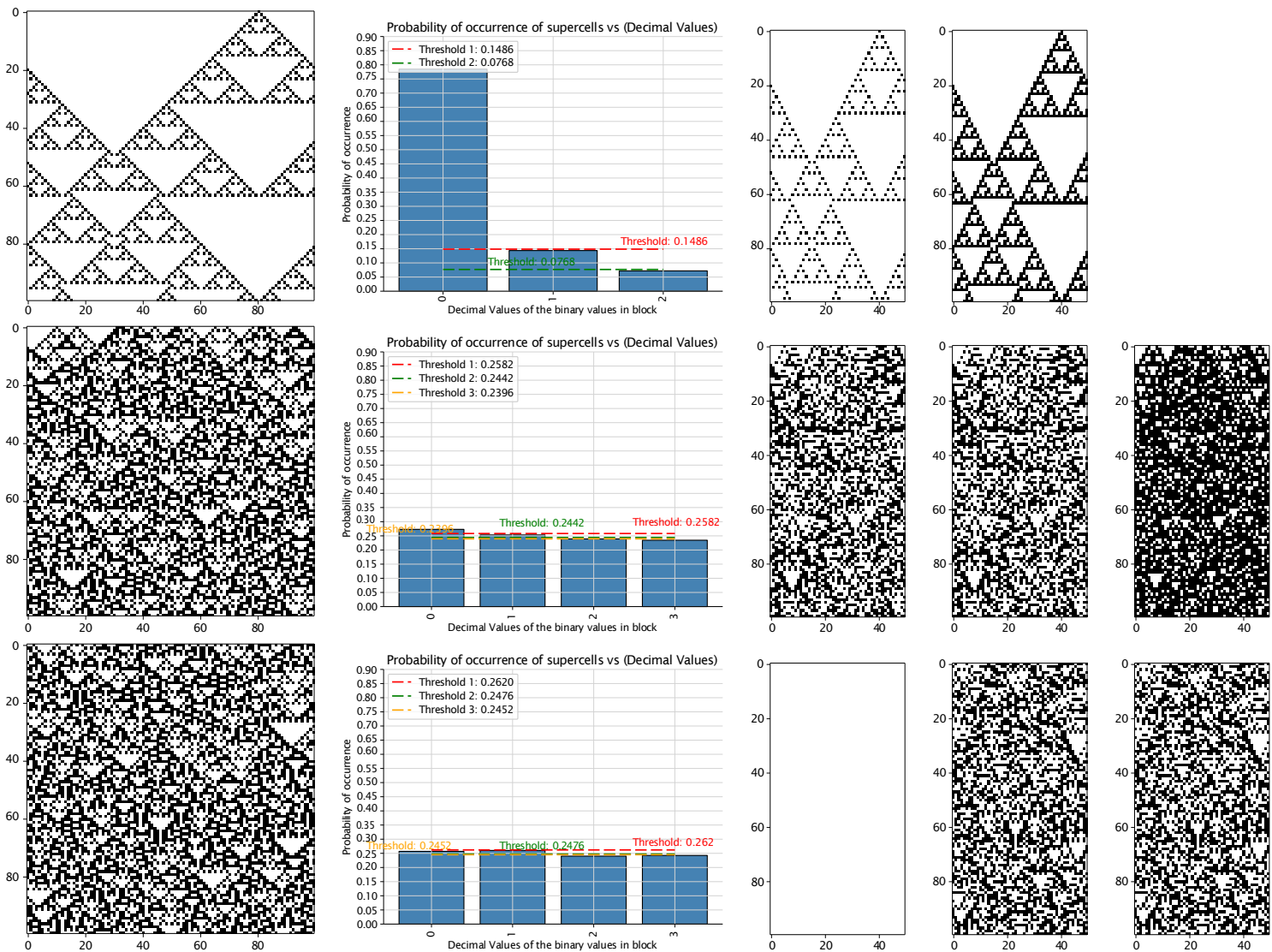

Table 57: FHCG plots for ECA Rule 50.

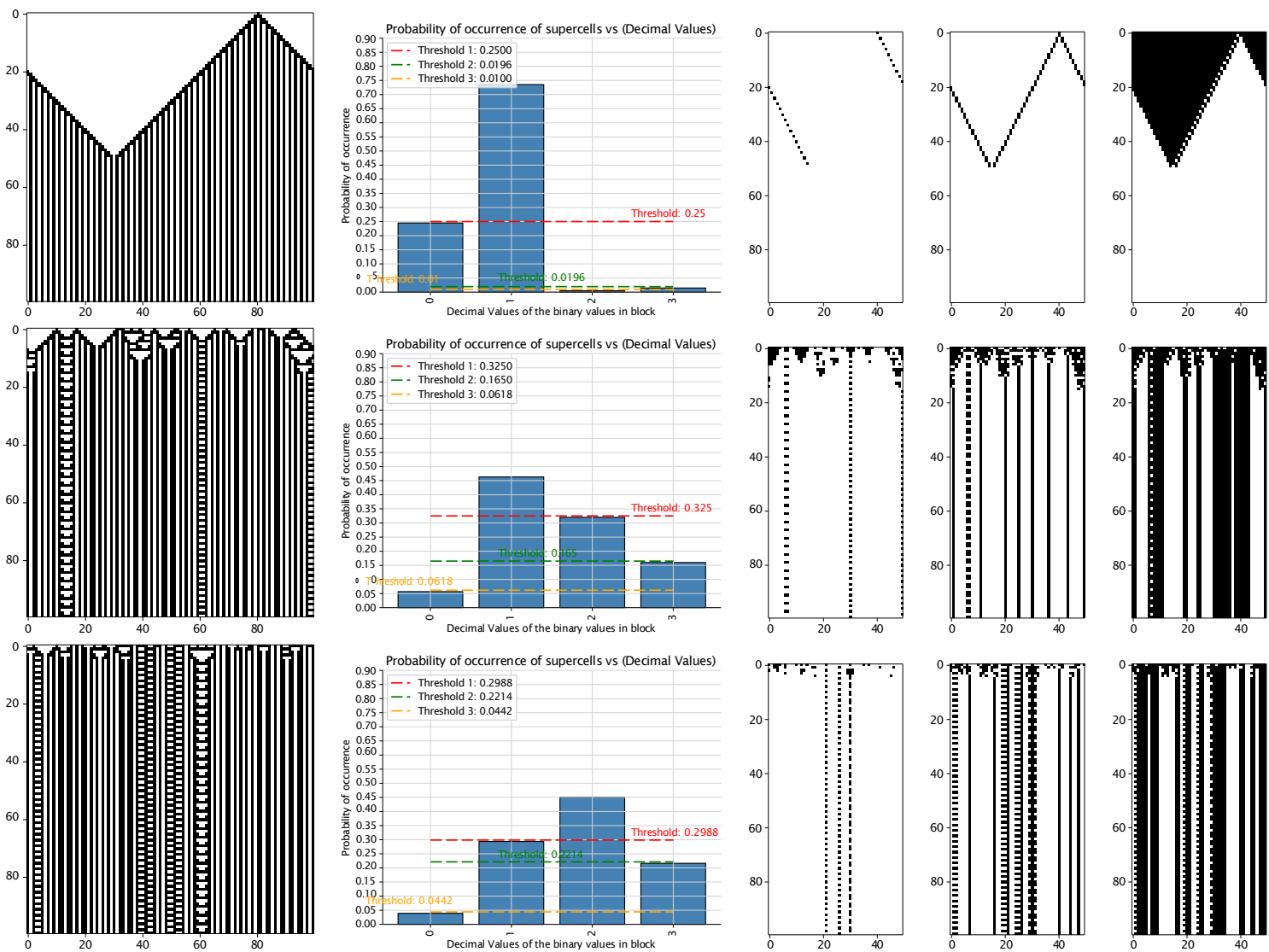

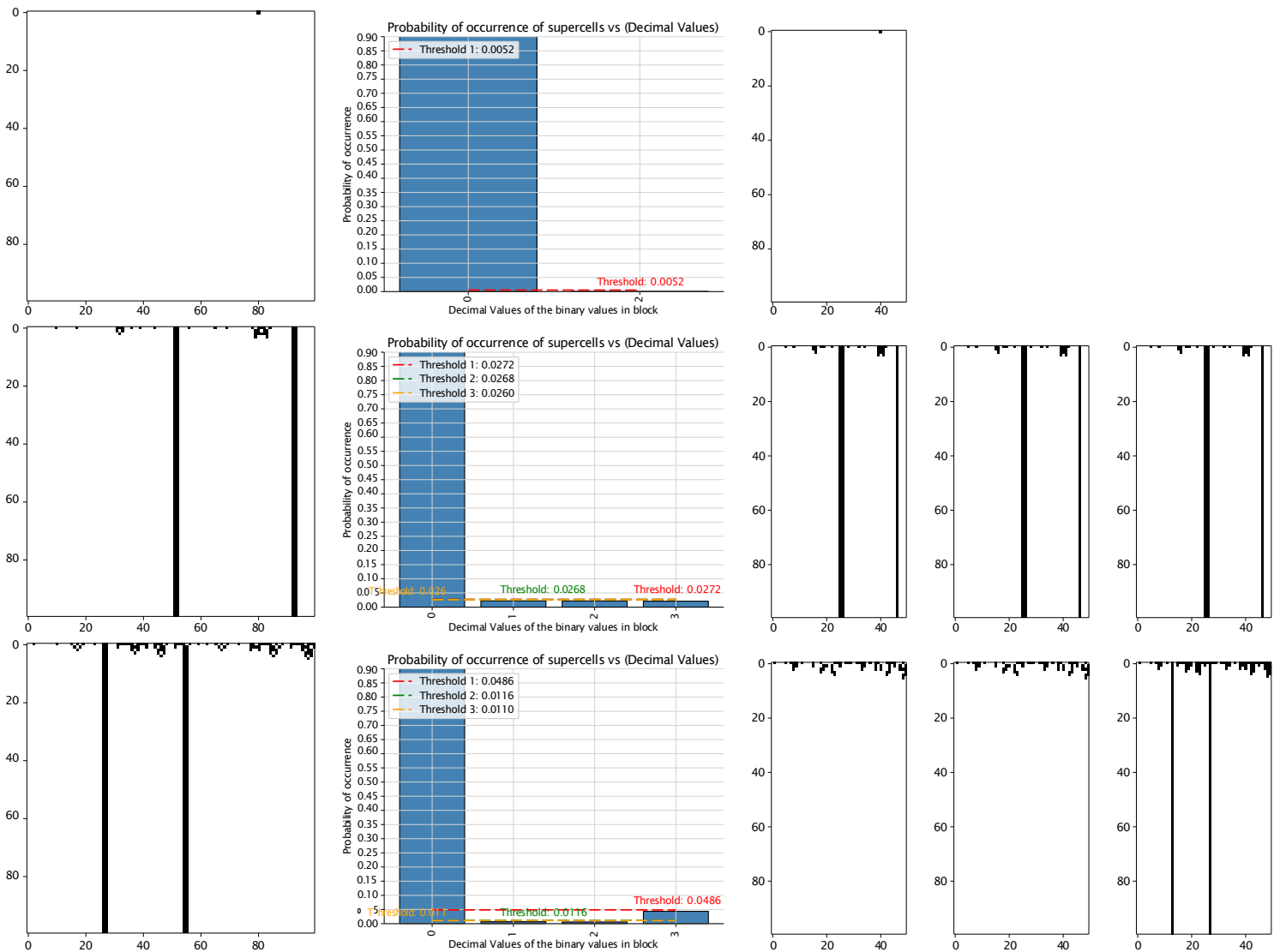

Table 58: FHCG plots for ECA Rule 50.

*Jain and Nichele: Coarse Graining in CA, NMI, 10, 1–96, 2023*

Table 59: FHCG plots for ECA Rule 50.

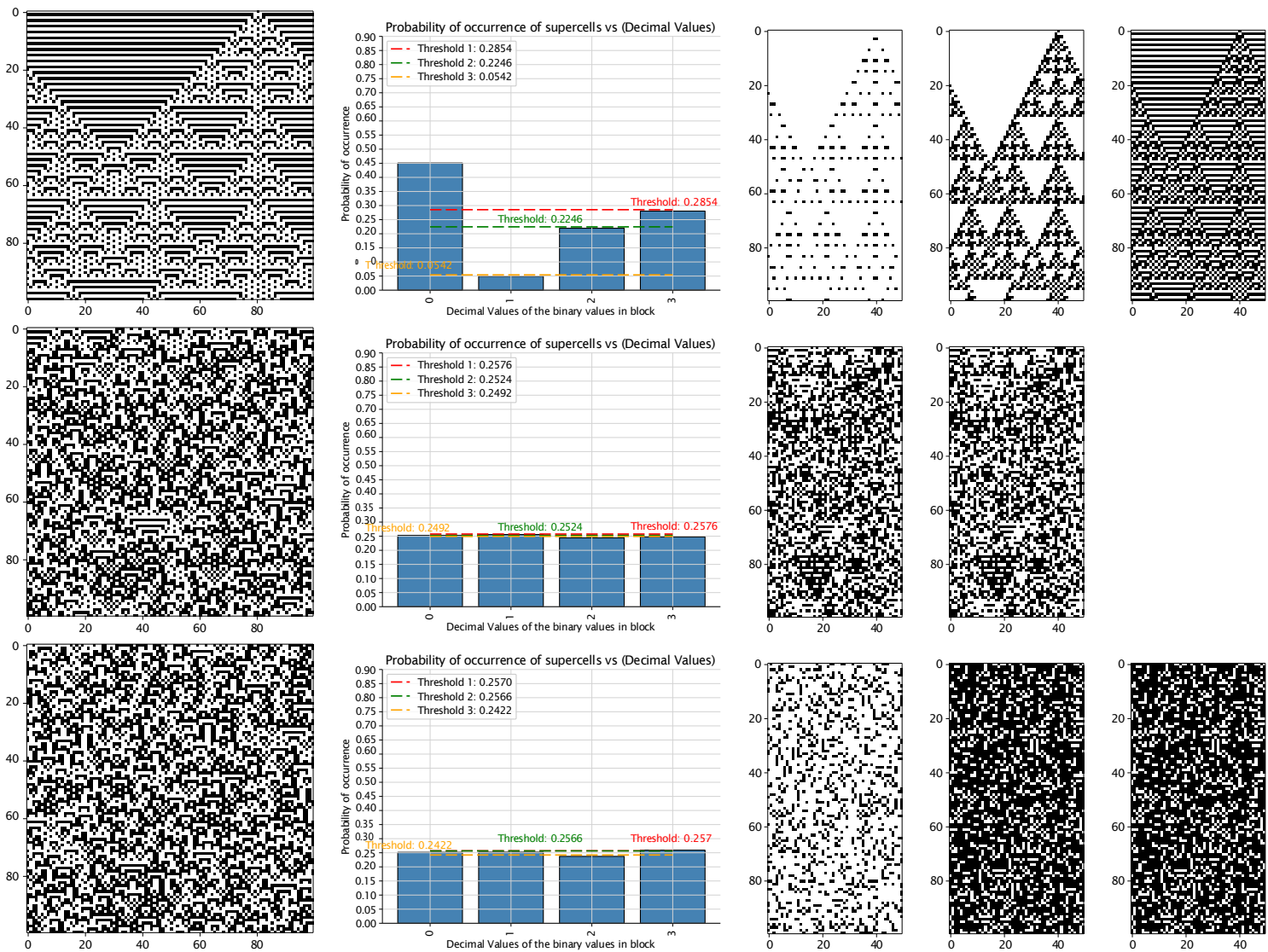

Table 60: FHCG plots for ECA Rule 50.

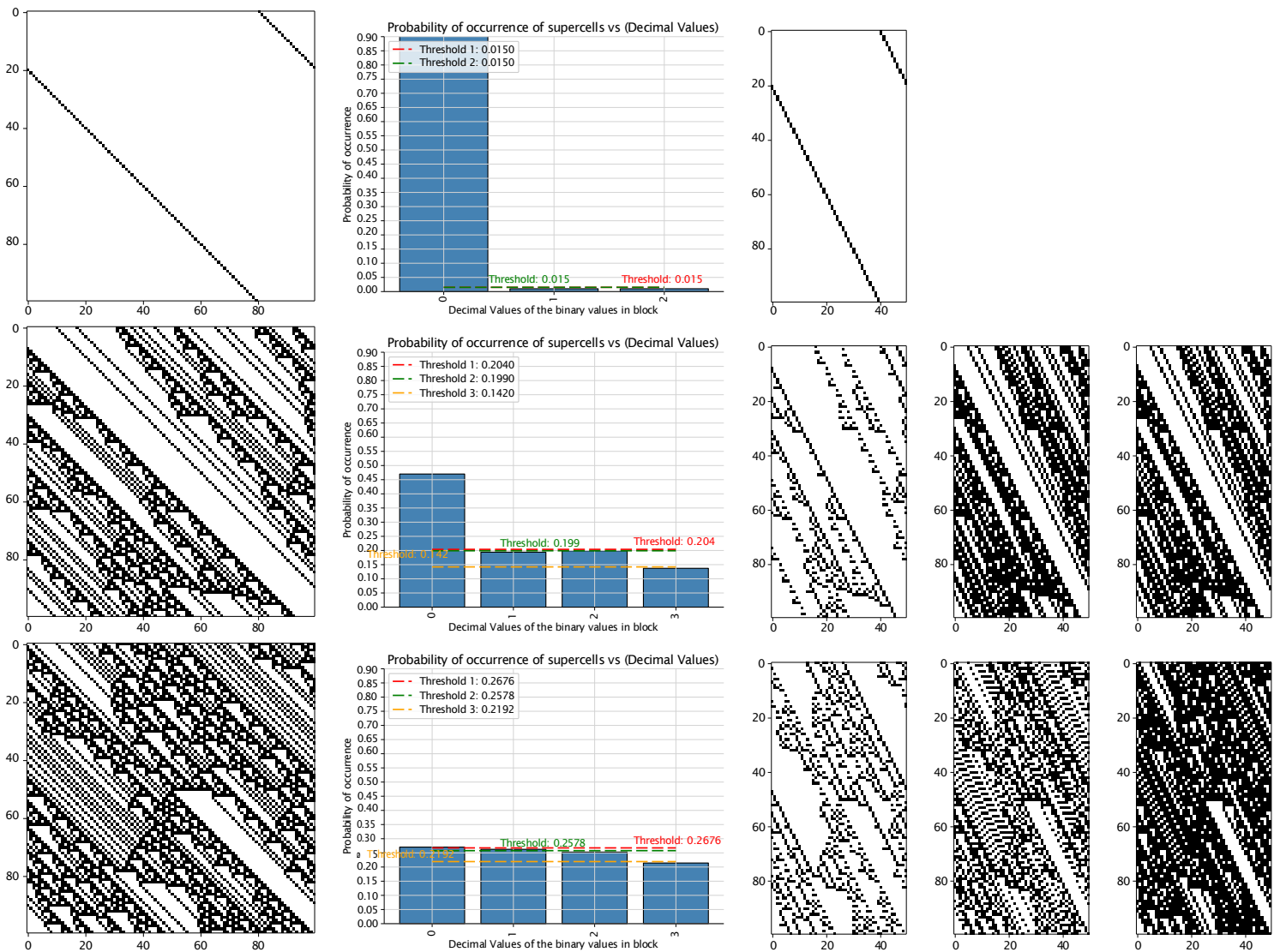

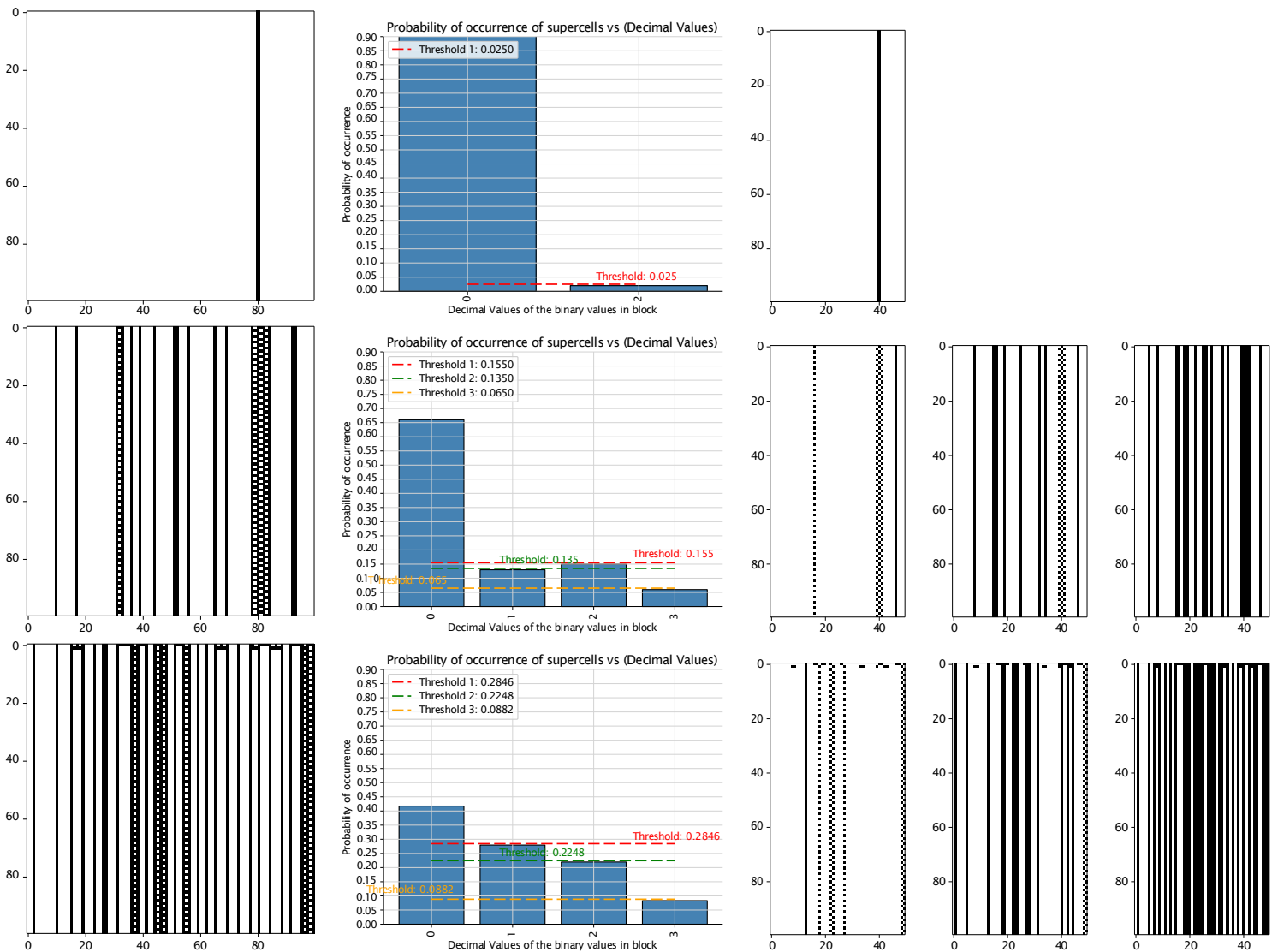

Table 61: FHCG plots for ECA Rule 50.

Table 62: FHCG plots for ECA Rule 110.

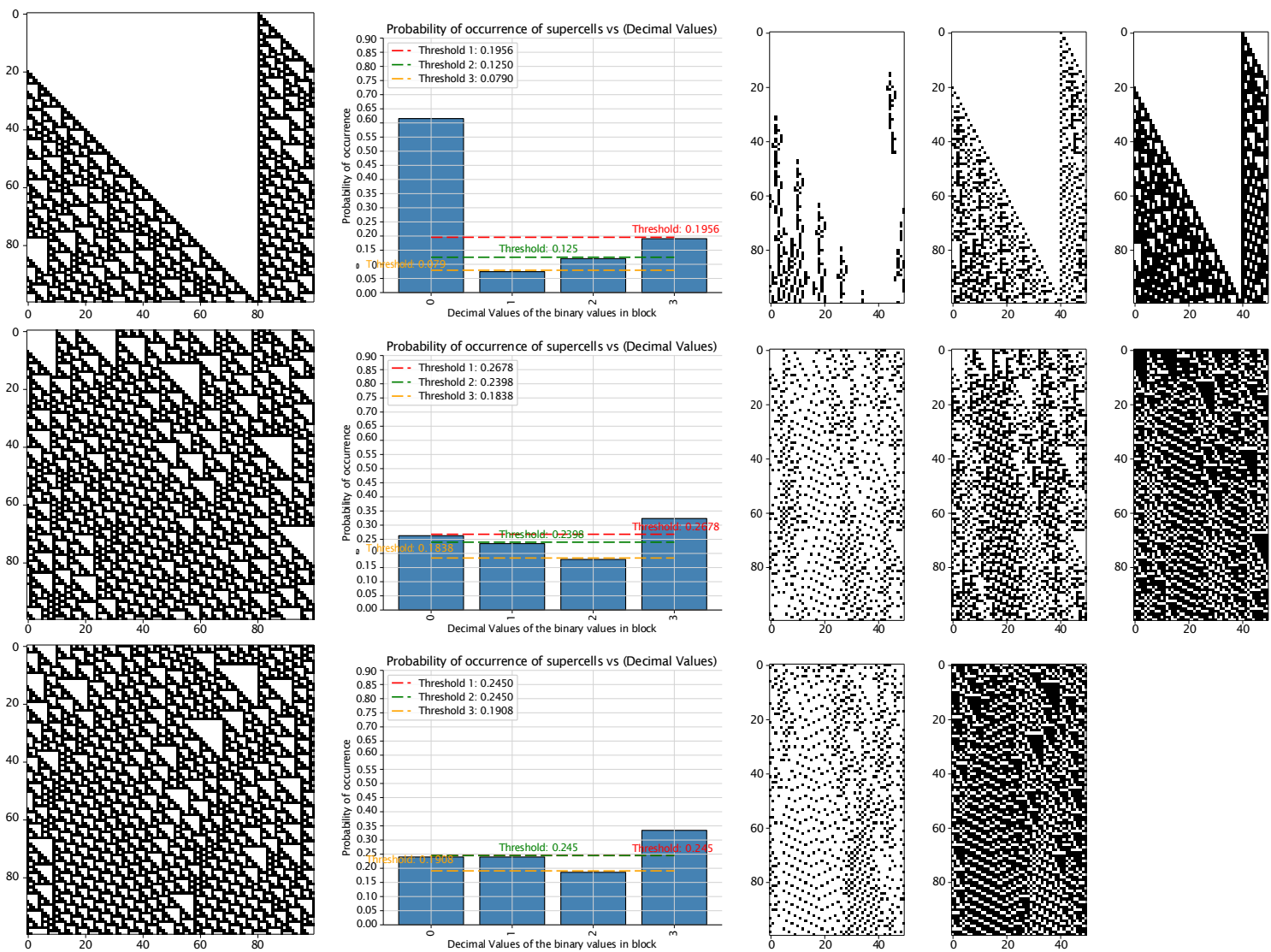

Table 63: FHCG plots for ECA Rule 122.

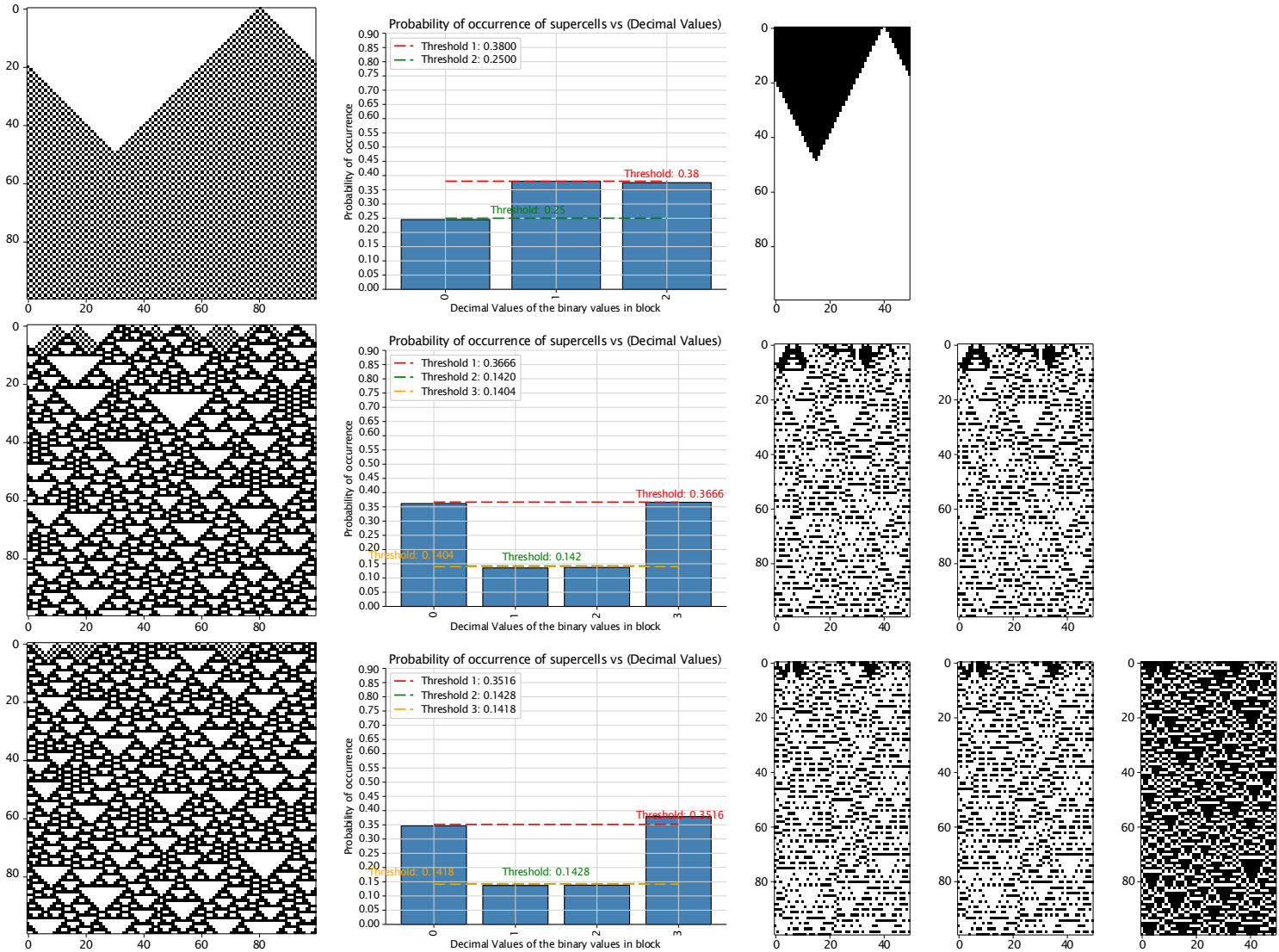

Table 64: FHCG plots for ECA Rule 126.

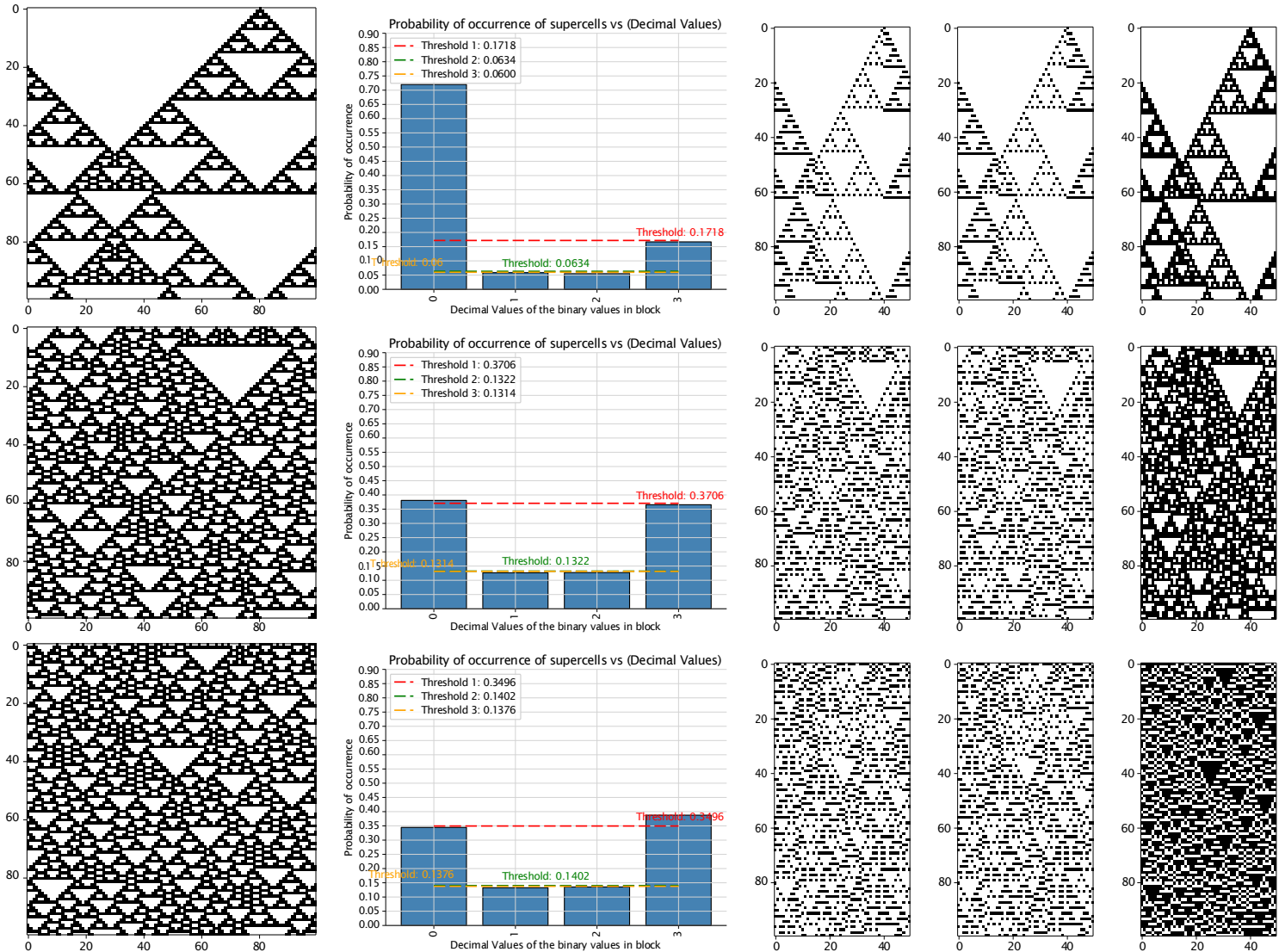

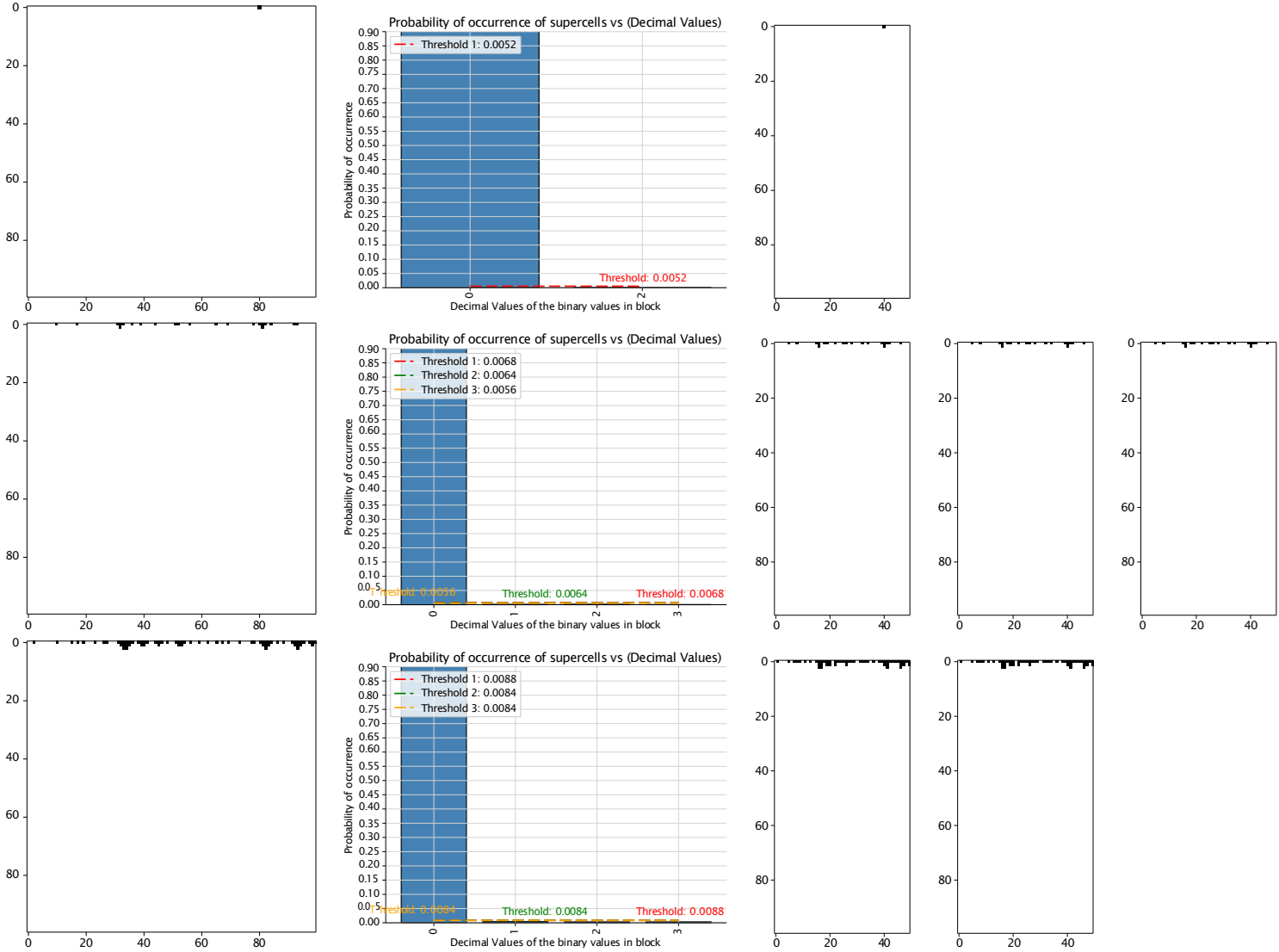

Table 65: FHCG plots for ECA Rule 126.

Table 66: FHCG plots for ECA Rule 126.

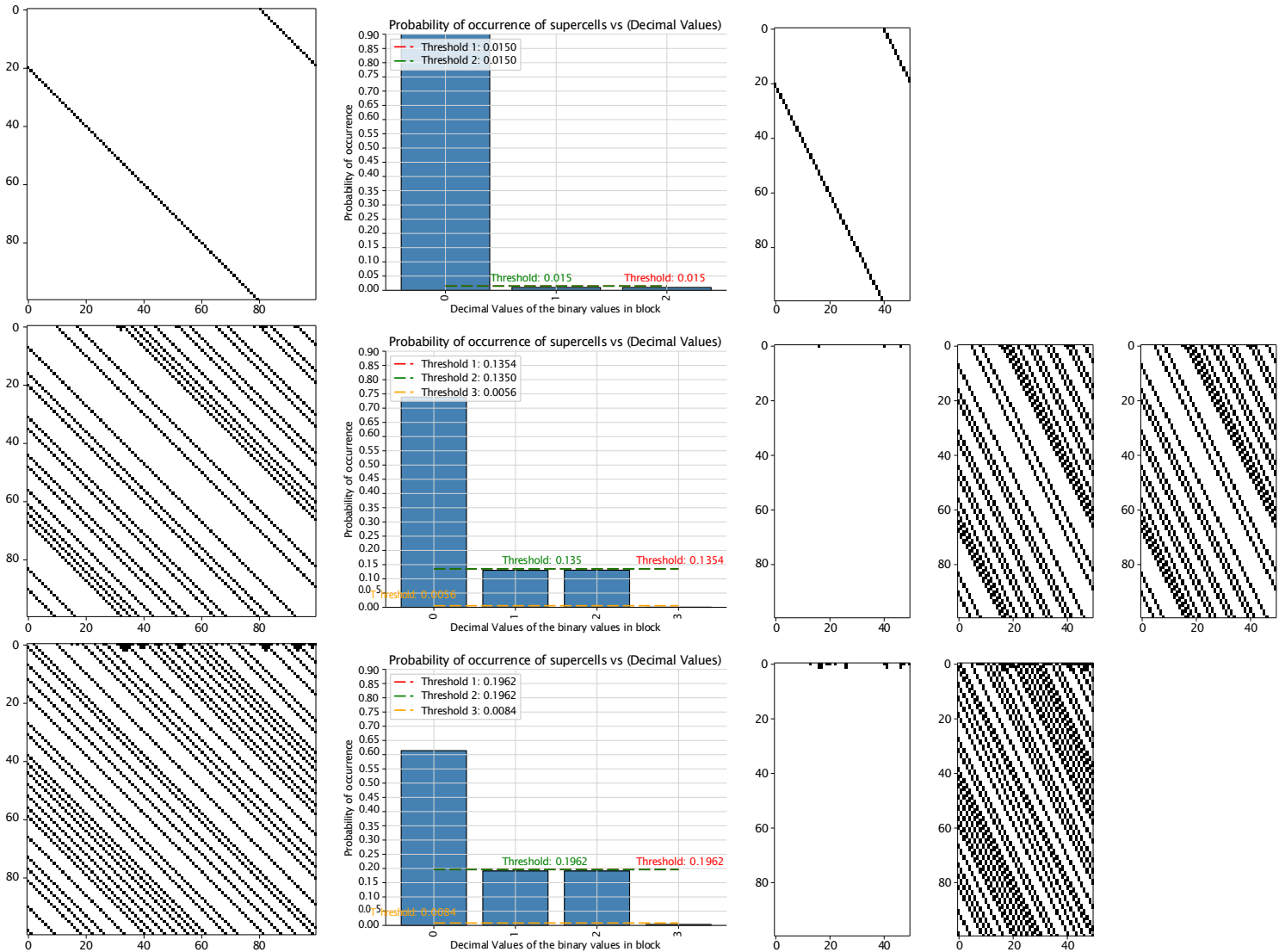

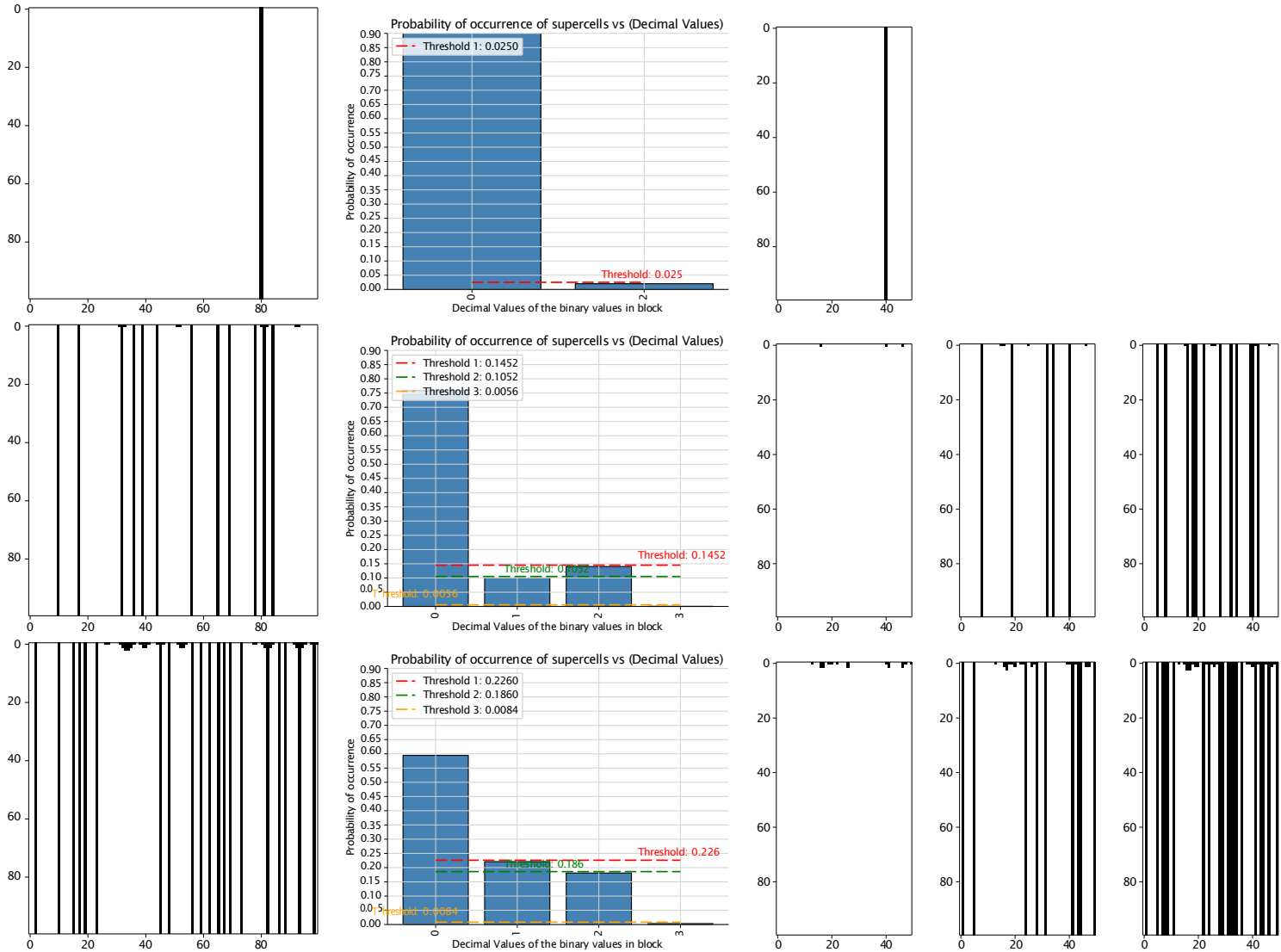

Table 67: FHCG plots for ECA Rule 126.

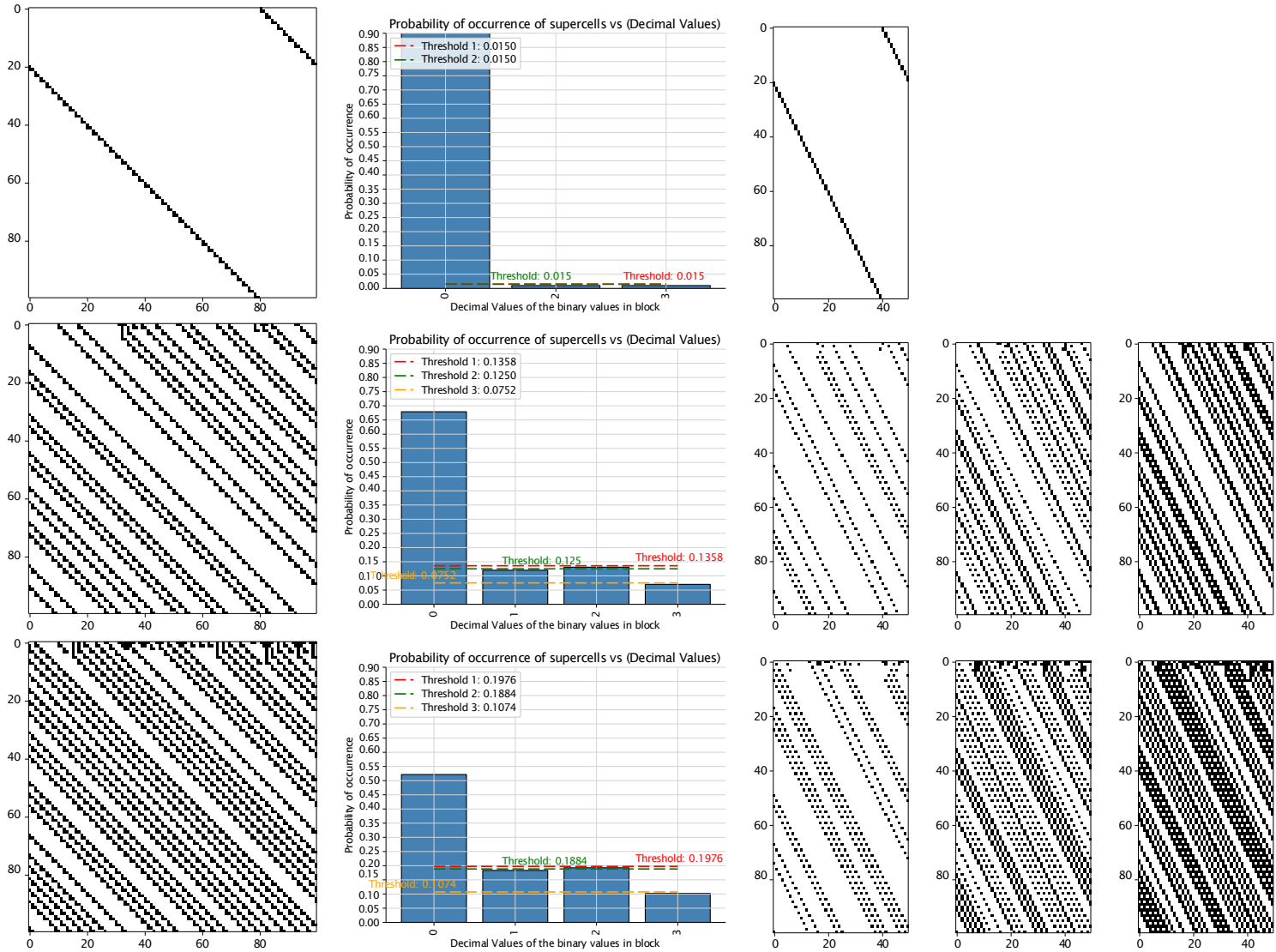

Table 68: FHCG plots for ECA Rule 126.

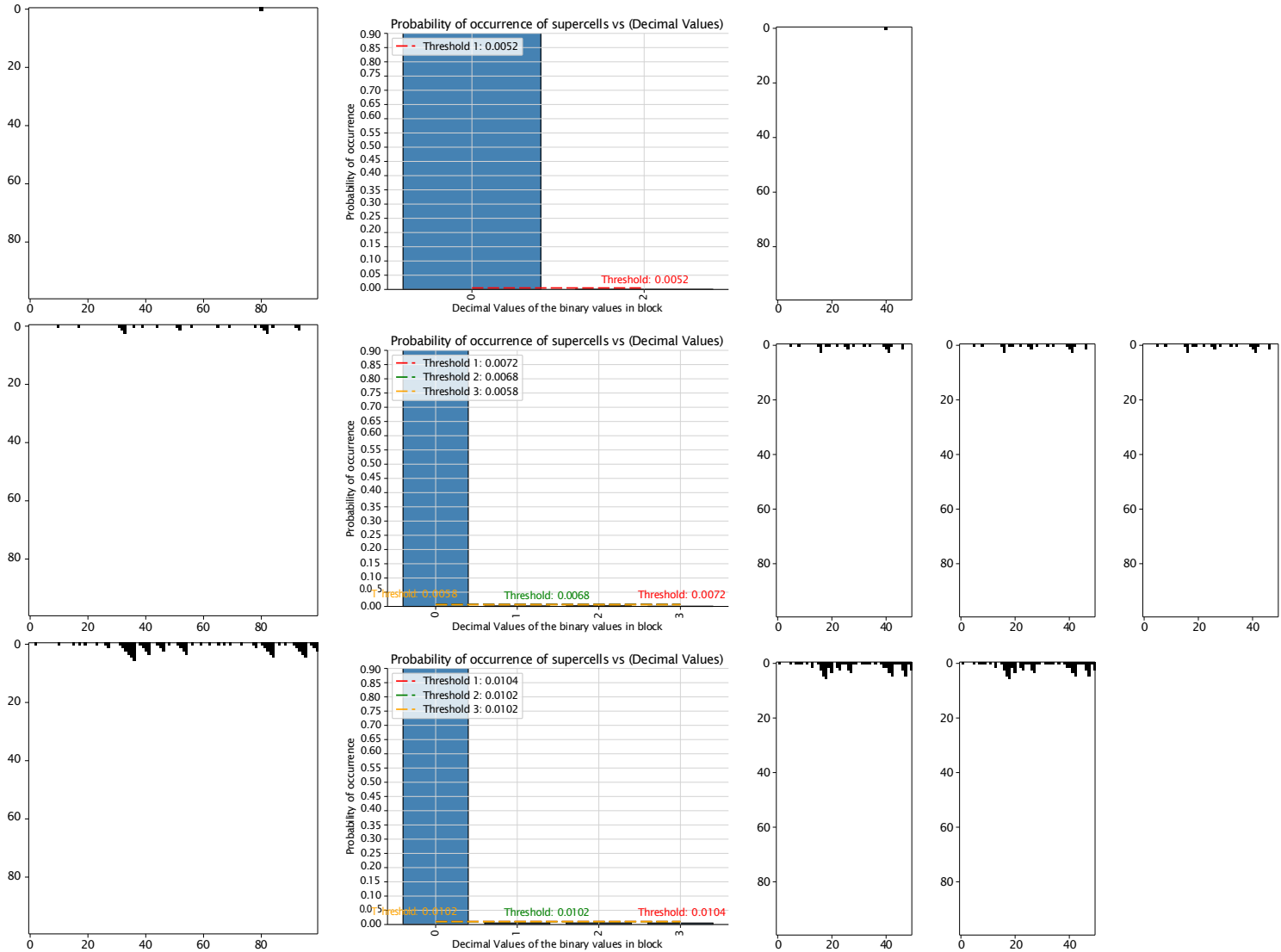

Table 69: FHCG plots for ECA Rule 126.

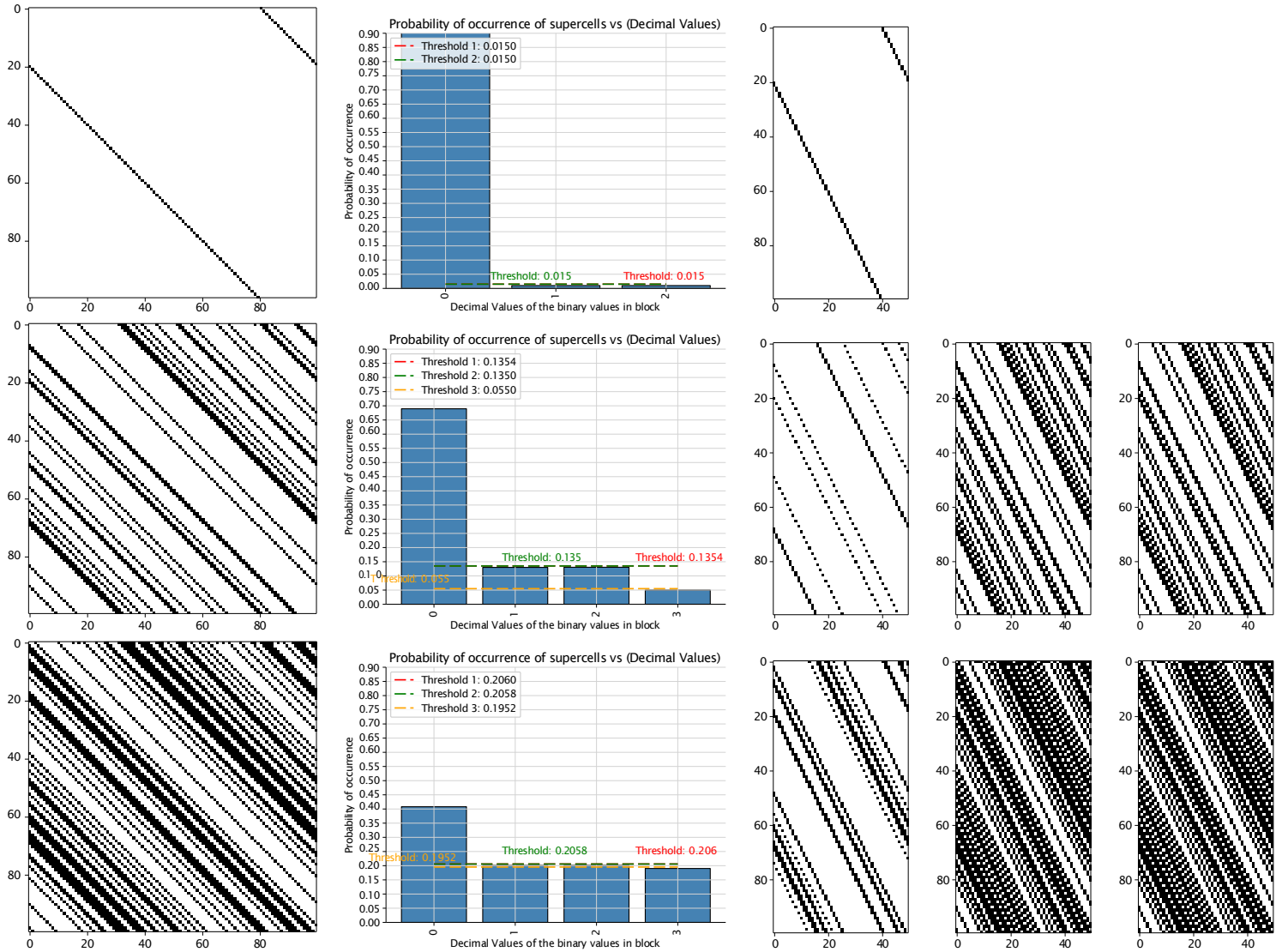

Table 70: FHCG plots for ECA Rule 126.

Table 71: FHCG plots for ECA Rule 126.

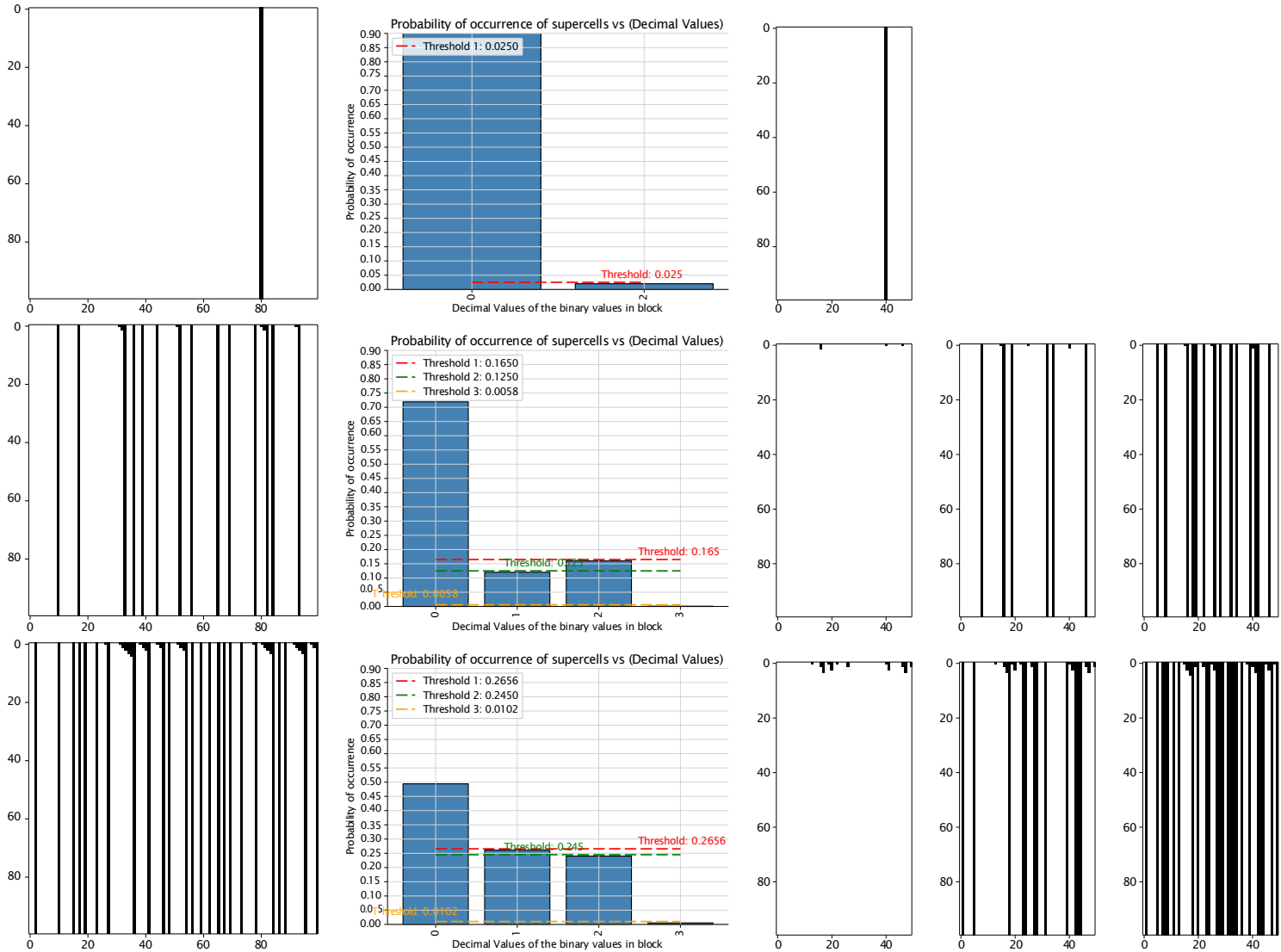

Table 72: FHCG plots for ECA Rule 126.

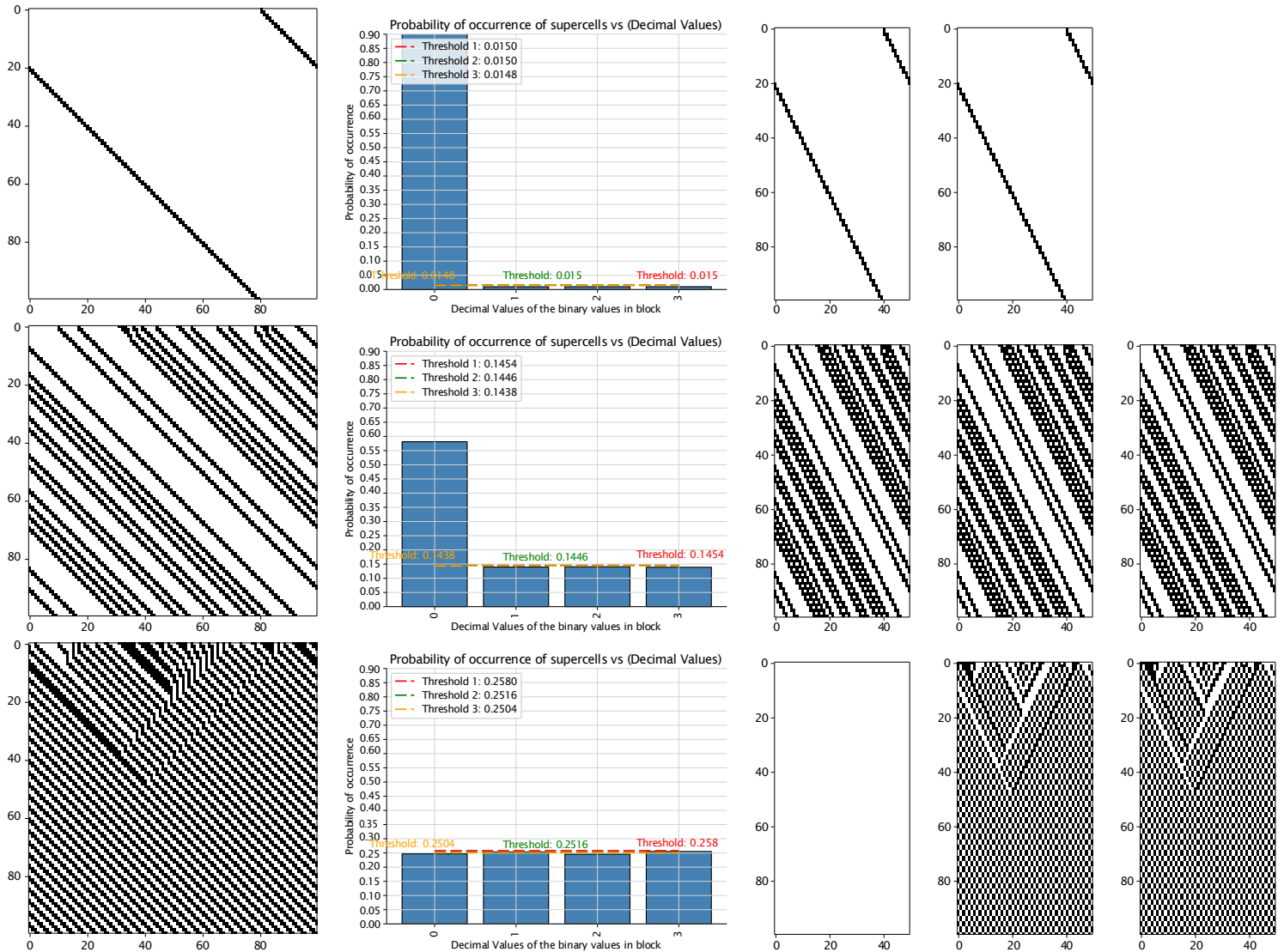

Table 73: FHCG plots for ECA Rule 126.

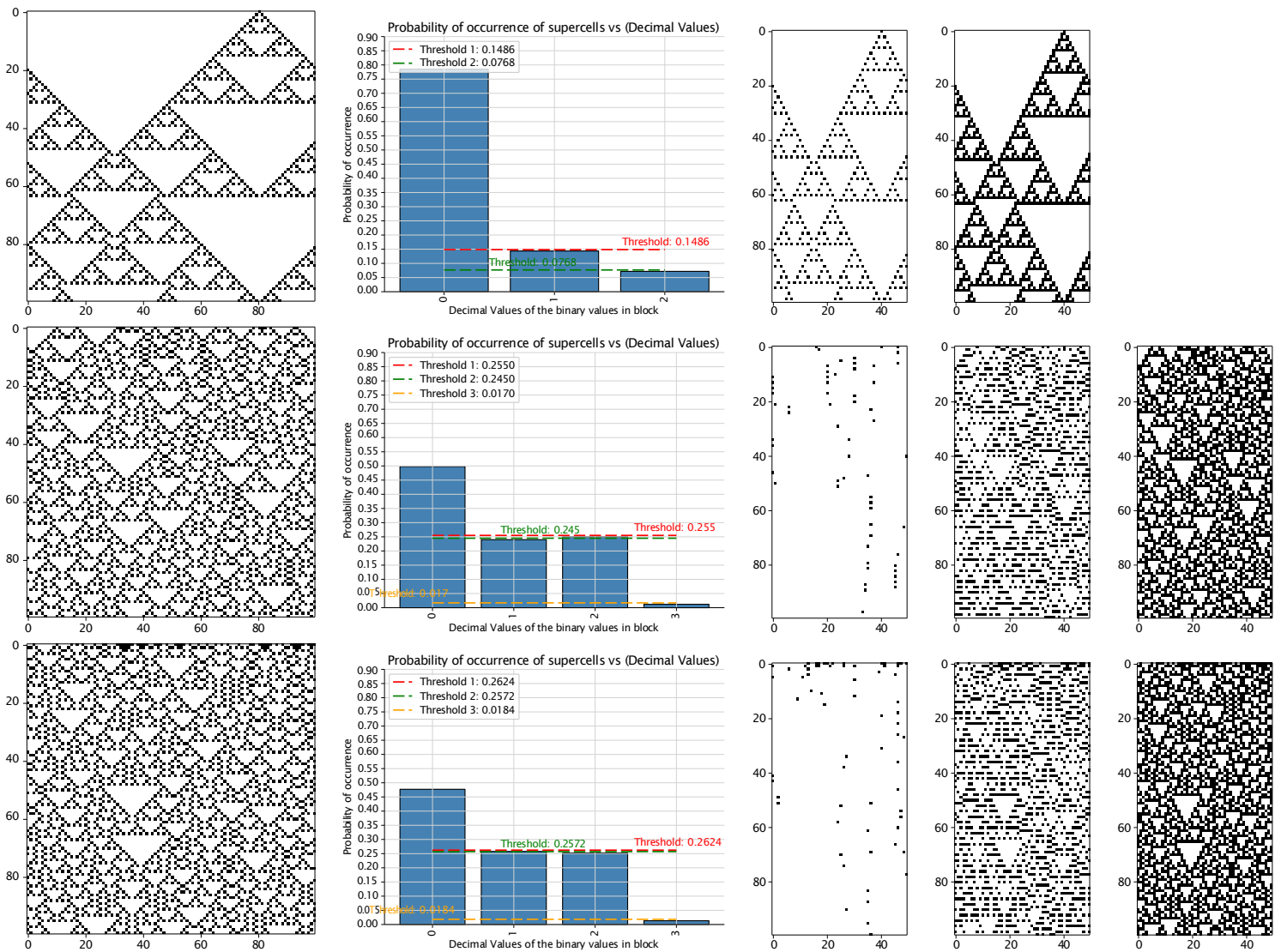

Table 74: FHCG plots for ECA Rule 126.

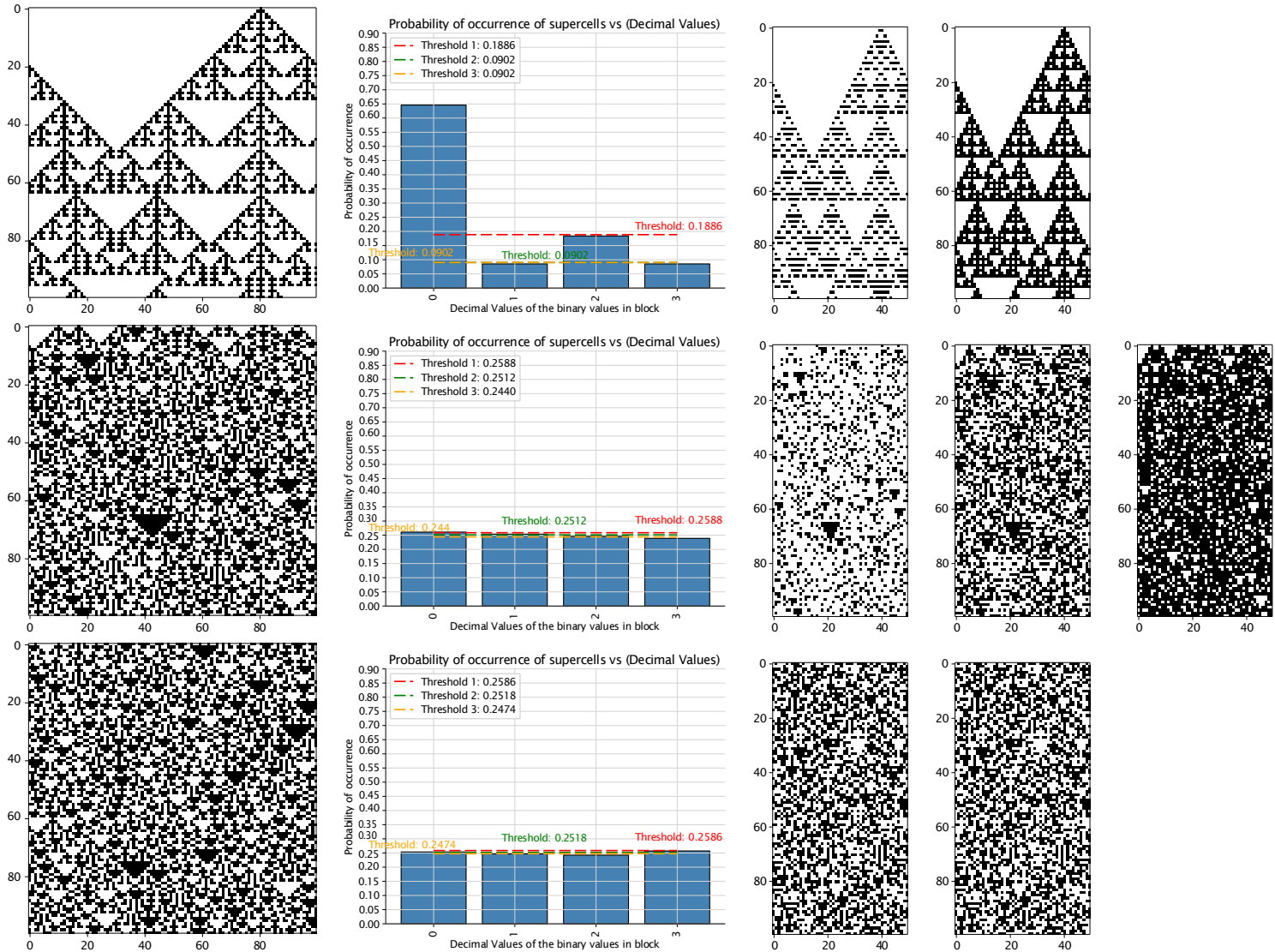

Table 75: FHCG plots for ECA Rule 126.

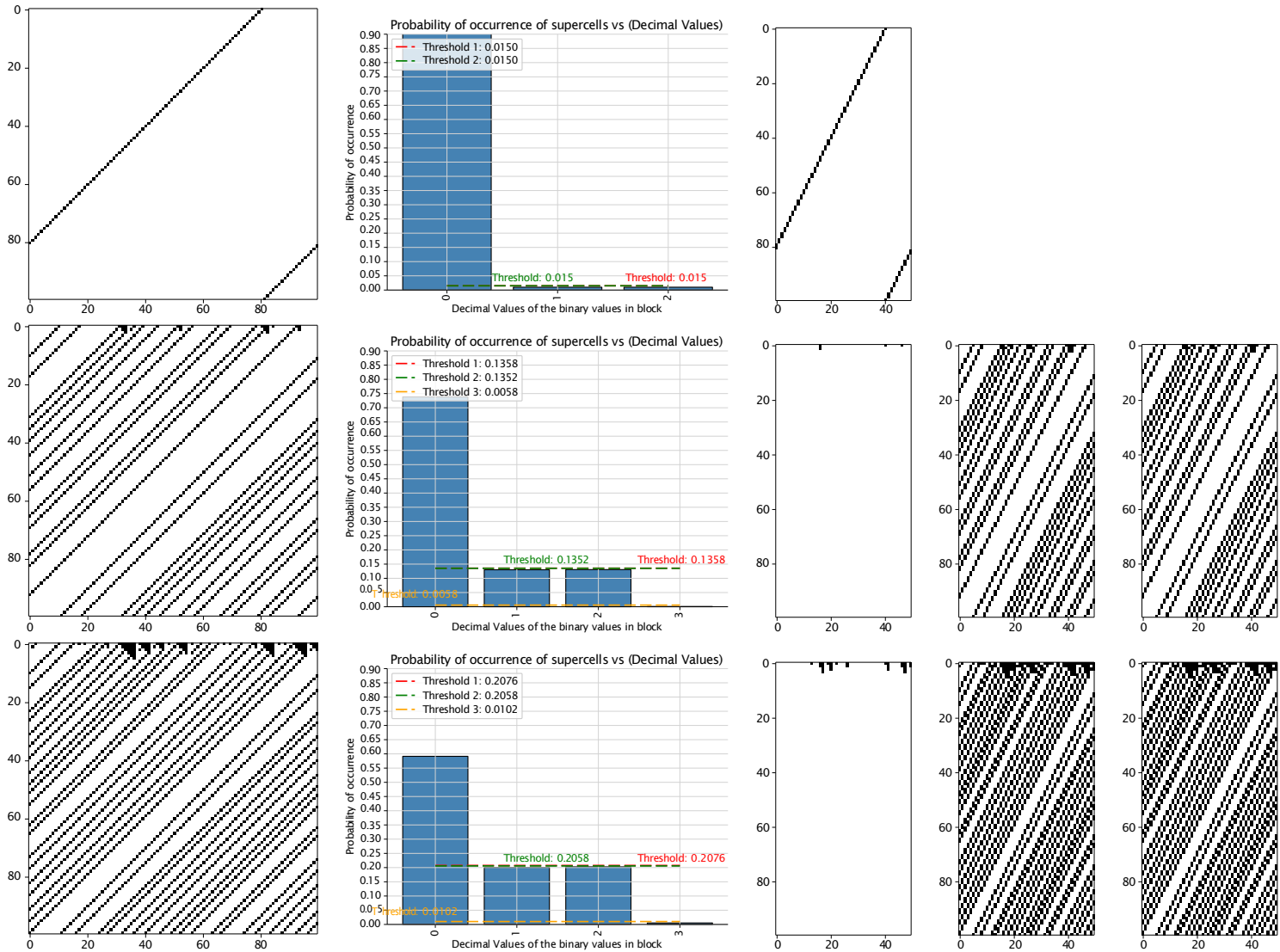

Table 76: FHCG plots for ECA Rule 126.

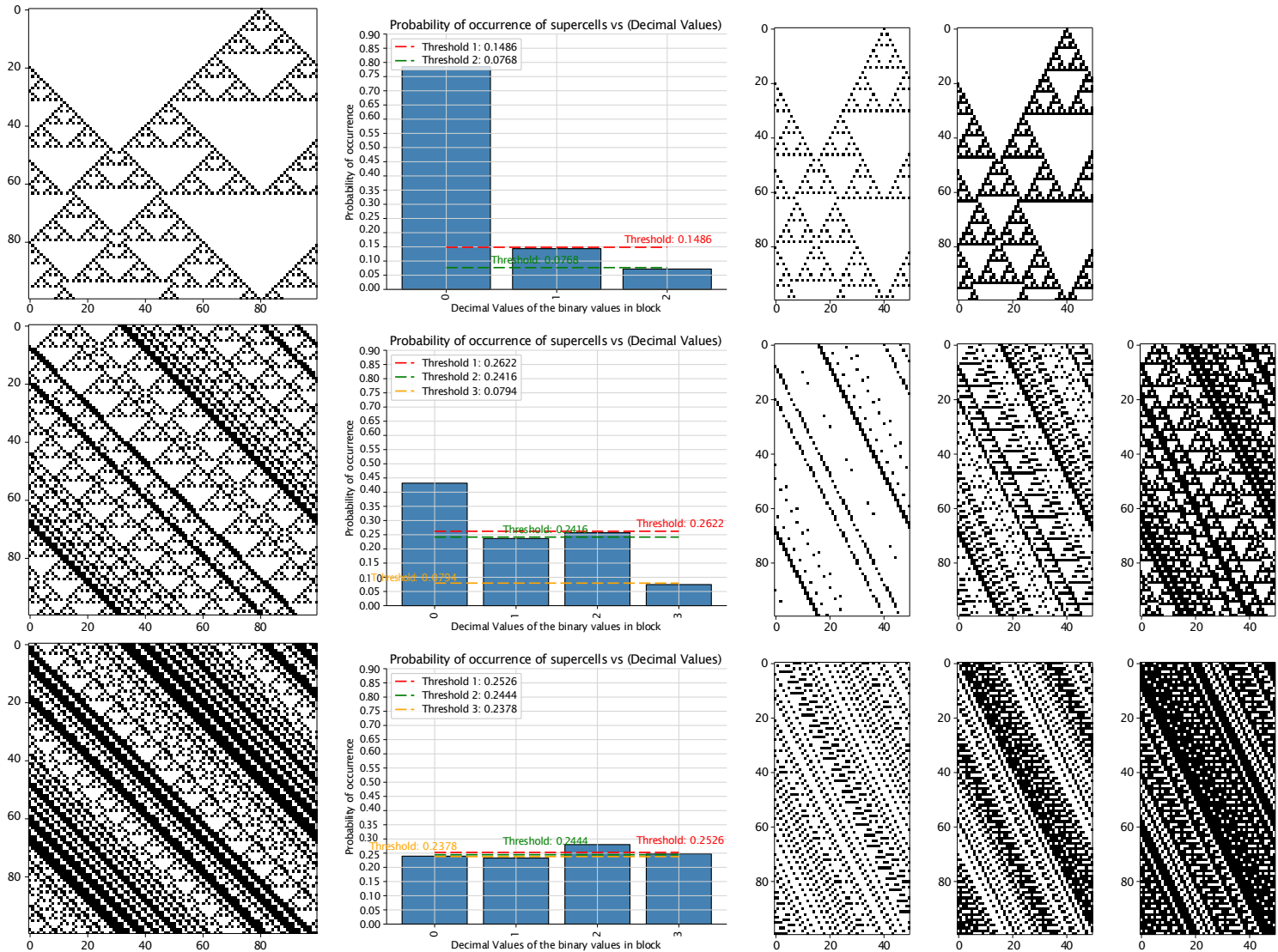

Table 77: FHCG plots for ECA Rule 126.

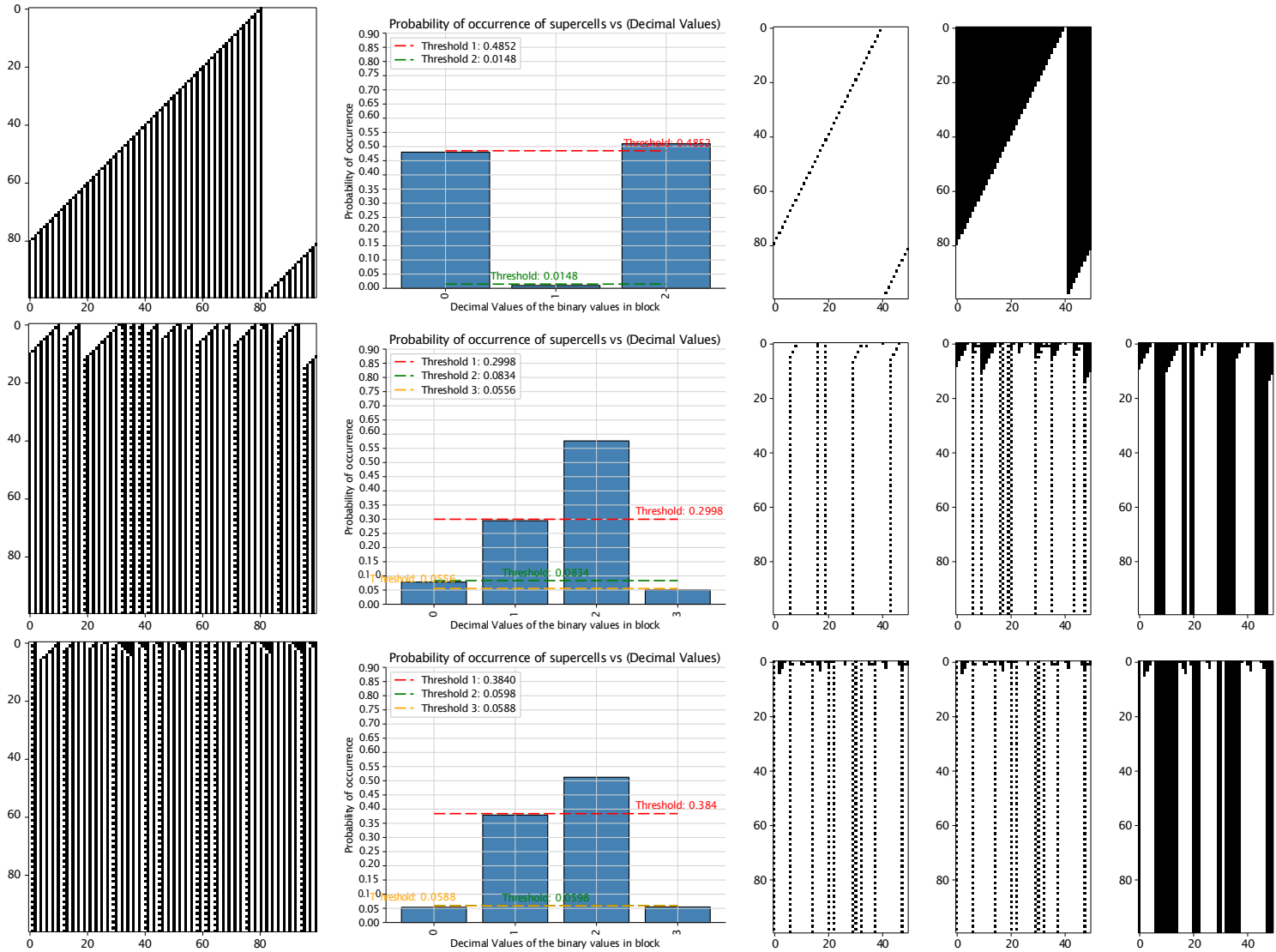

Table 78: FHCG plots for ECA Rule 126.

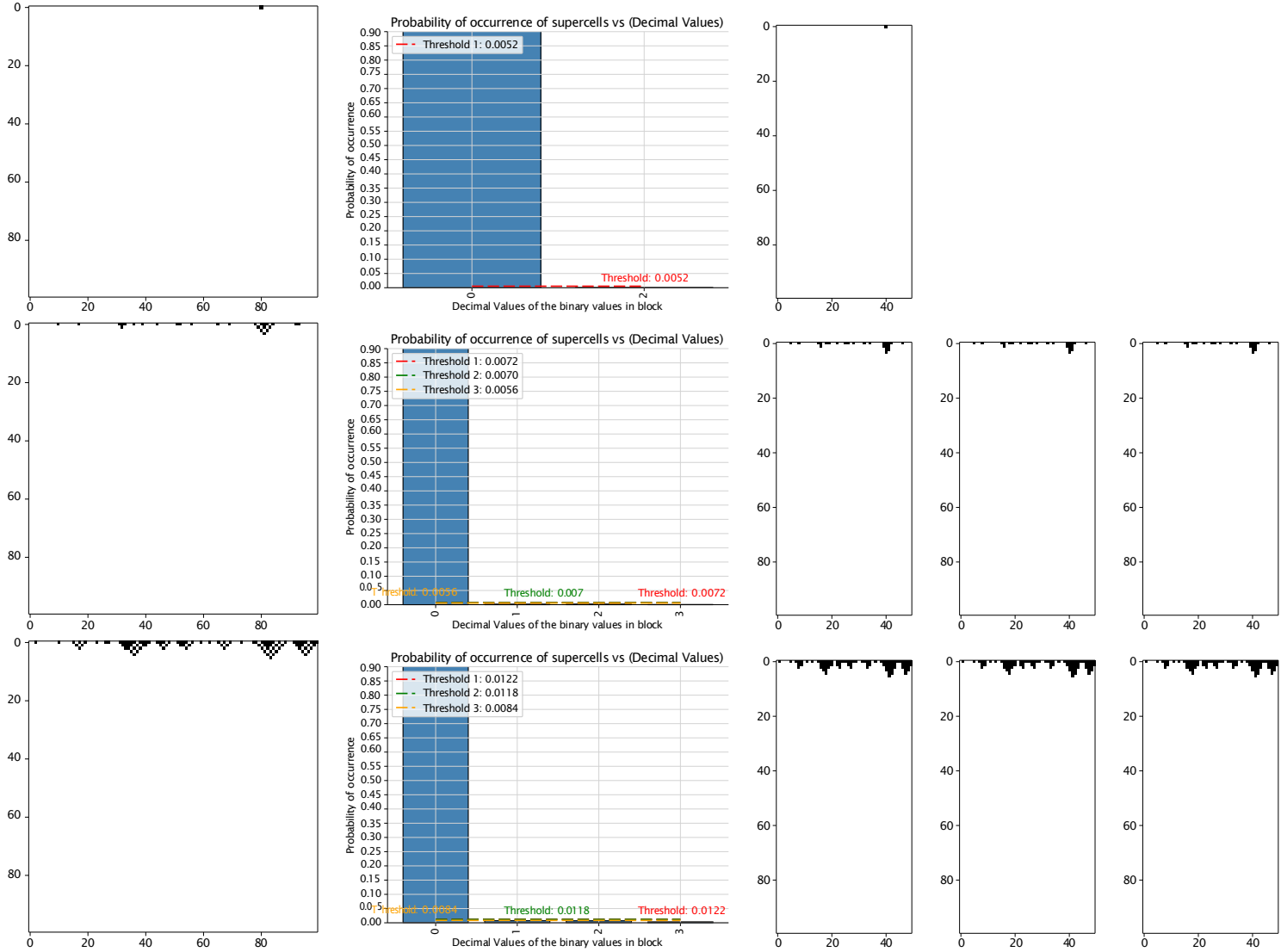

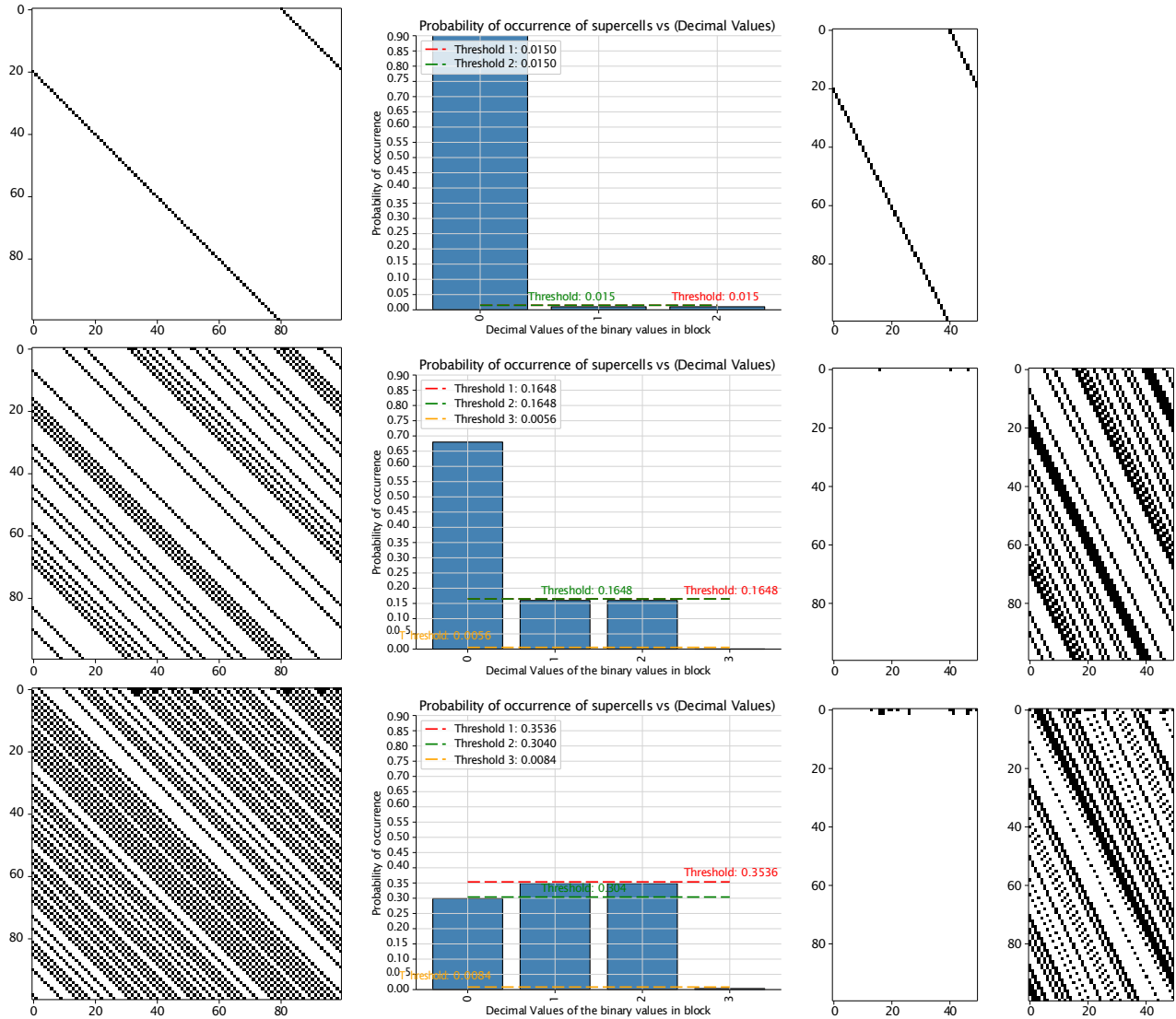

Table 79: FHCG plots for ECA Rule 126.

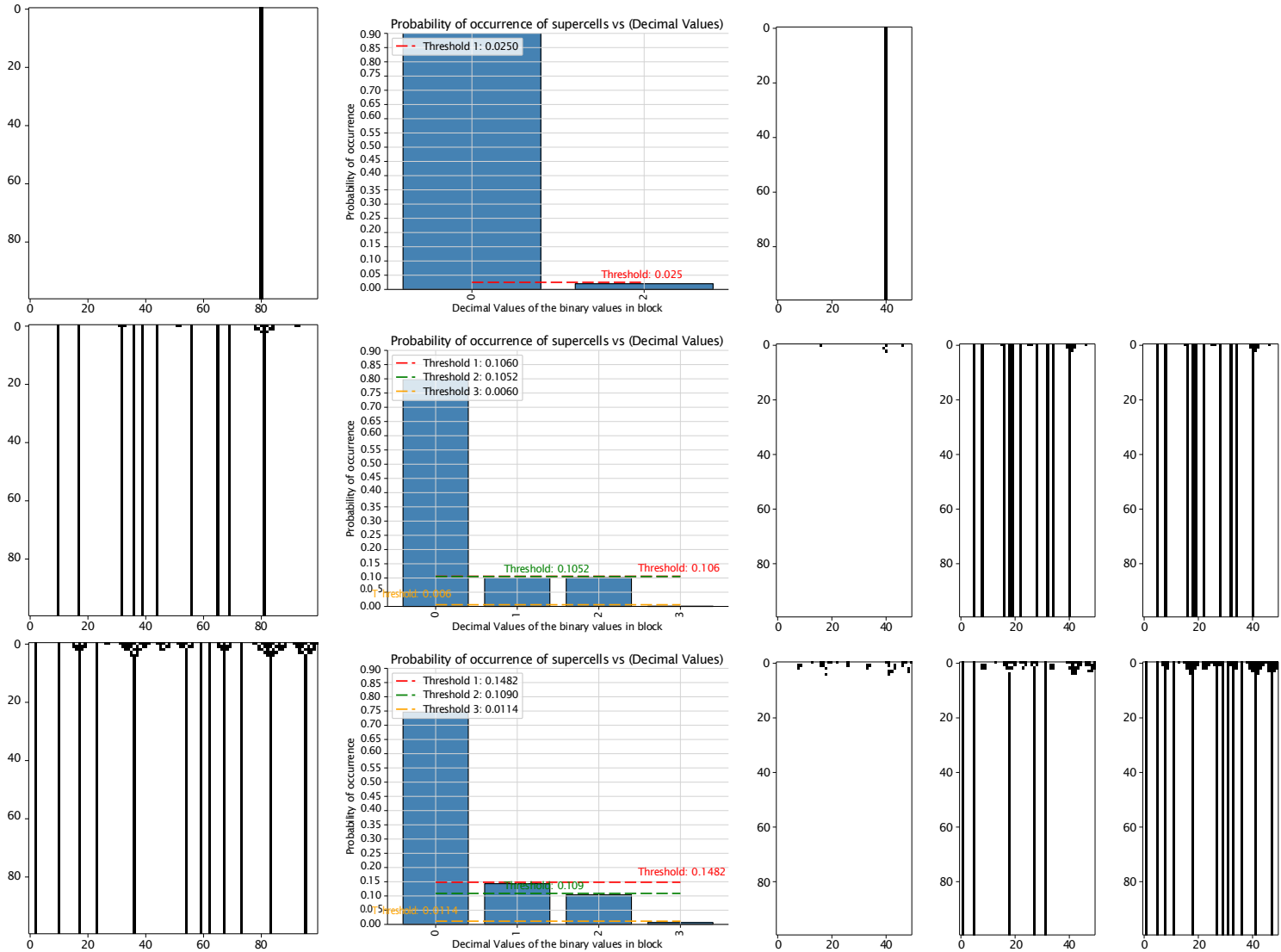

Table 80: FHCG plots for ECA Rule 126.

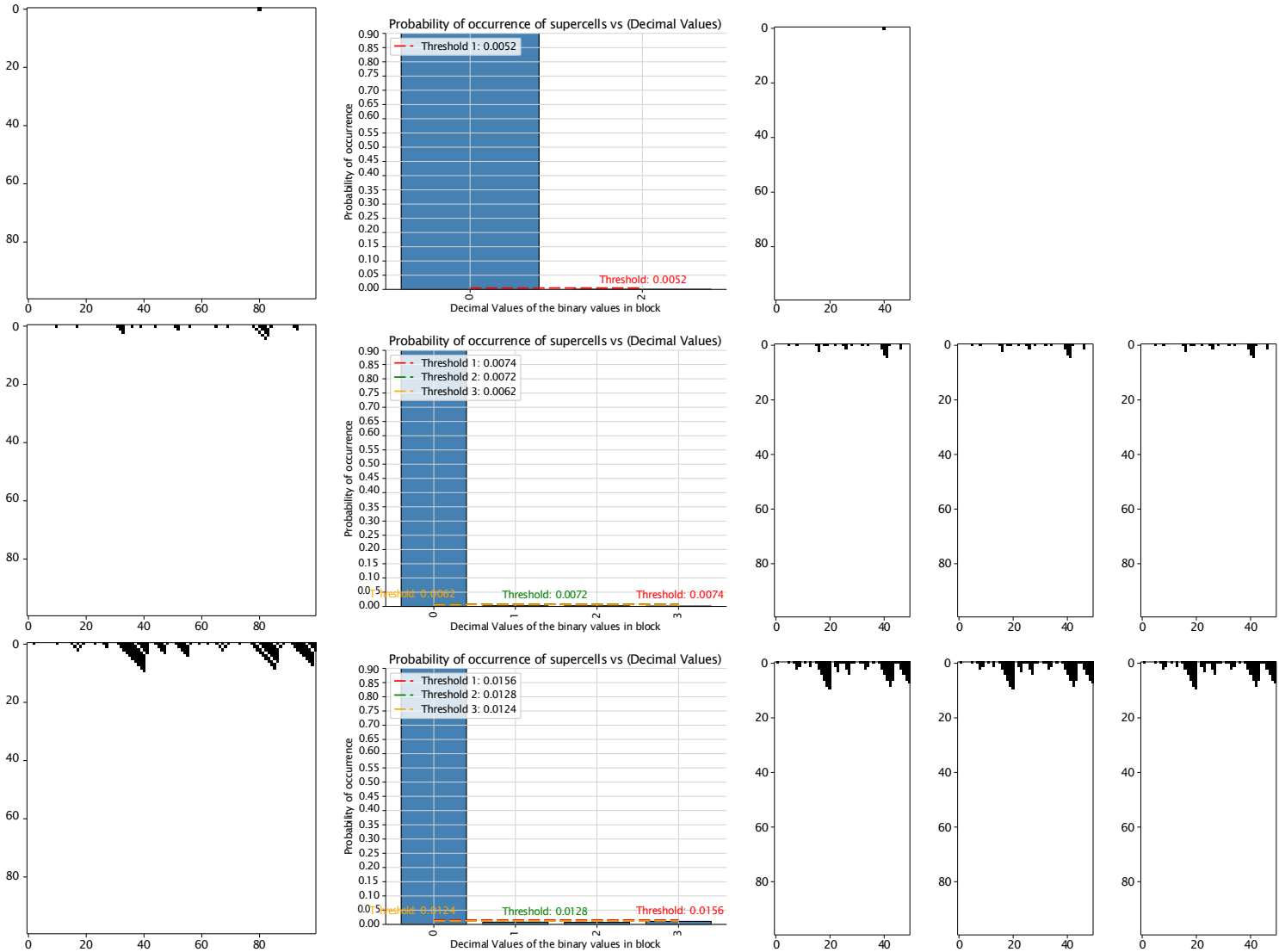

Table 81: FHCG plots for ECA Rule 126.

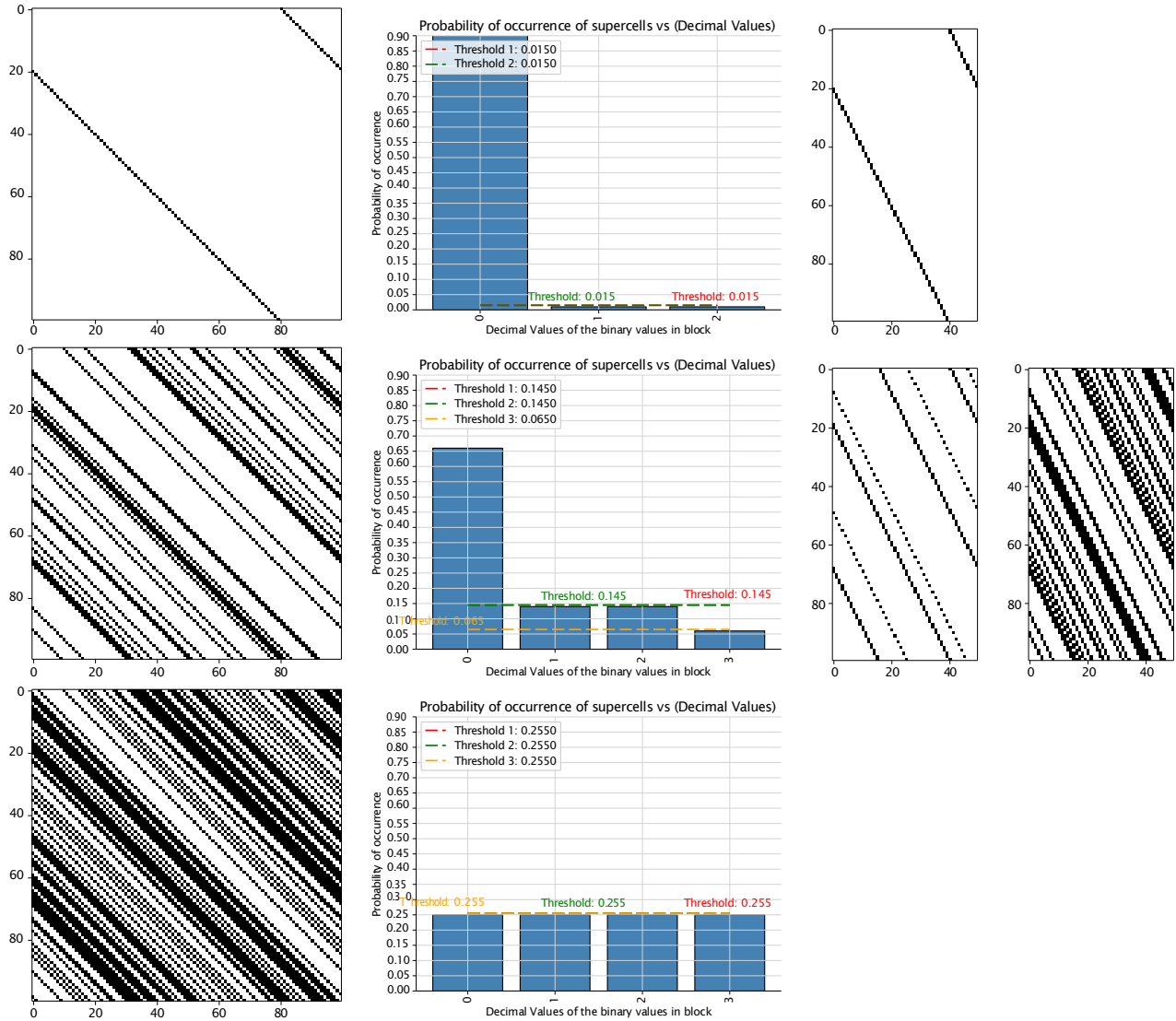

Table 82: FHCG plots for ECA Rule 126.

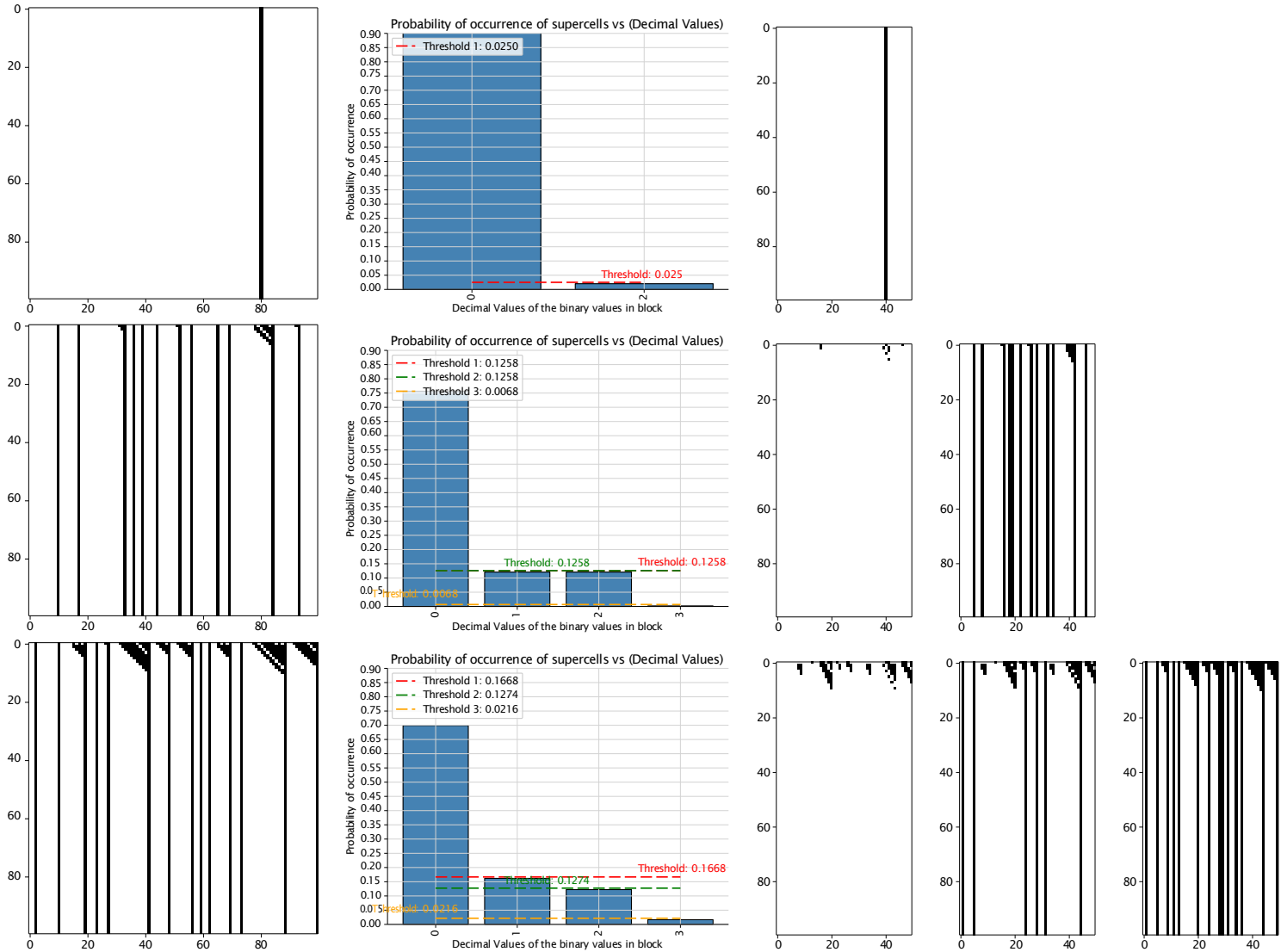

Table 83: FHCG plots for ECA Rule 126.

Table 84: FHCG plots for ECA Rule 126.

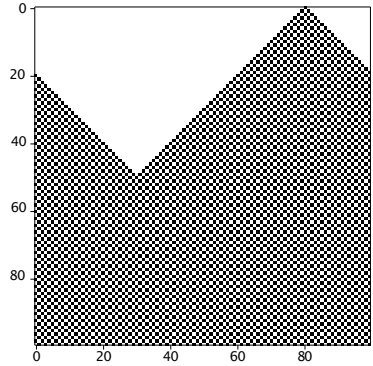 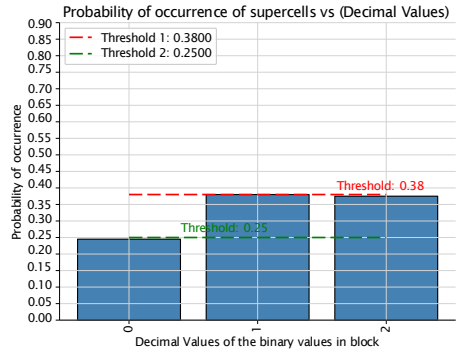 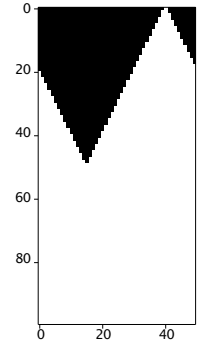

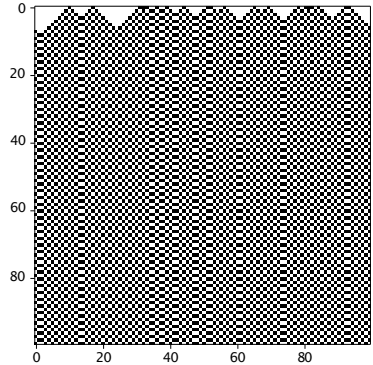 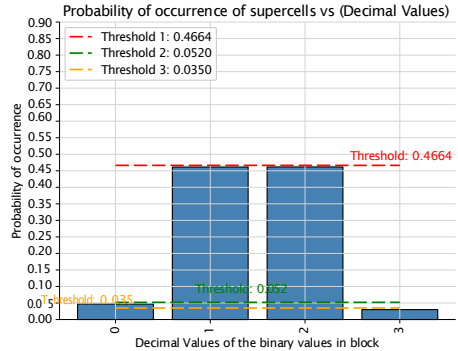 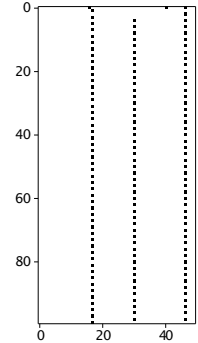 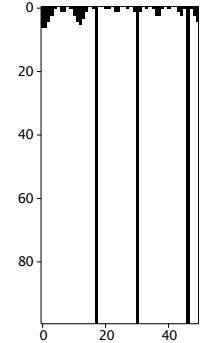

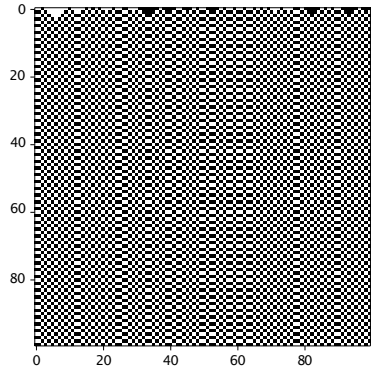 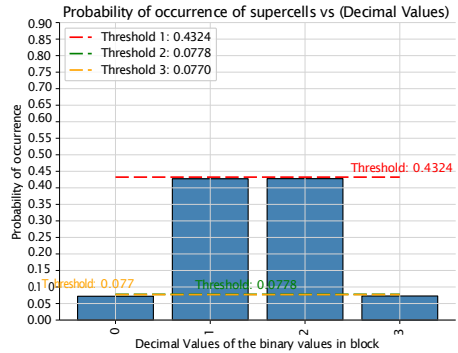 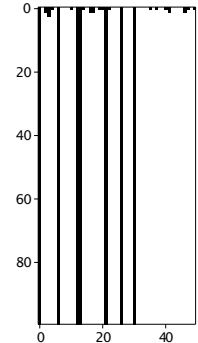 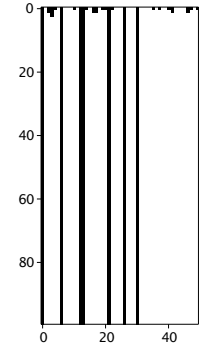

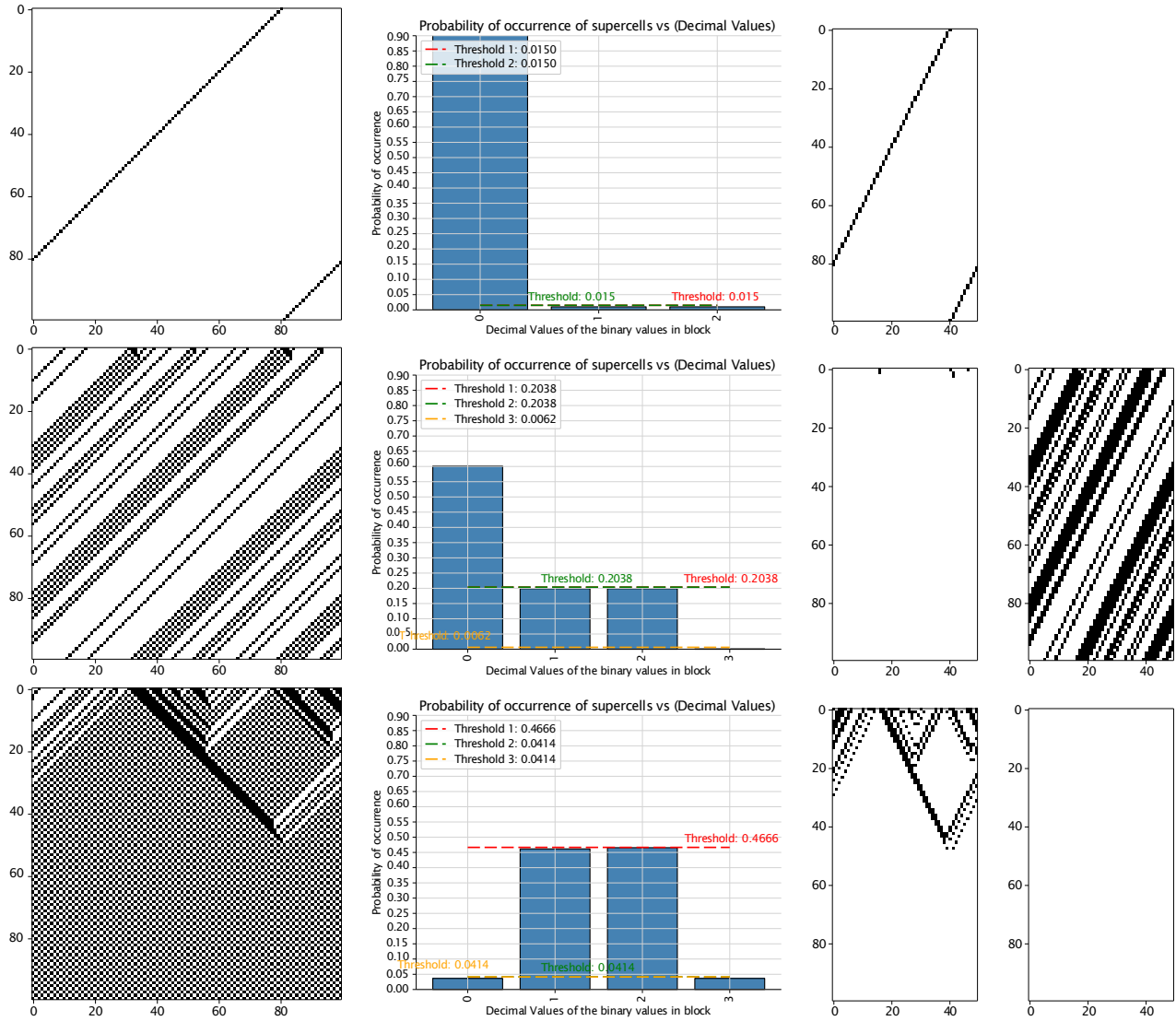

Table 85: FHCG plots for ECA Rule 126.

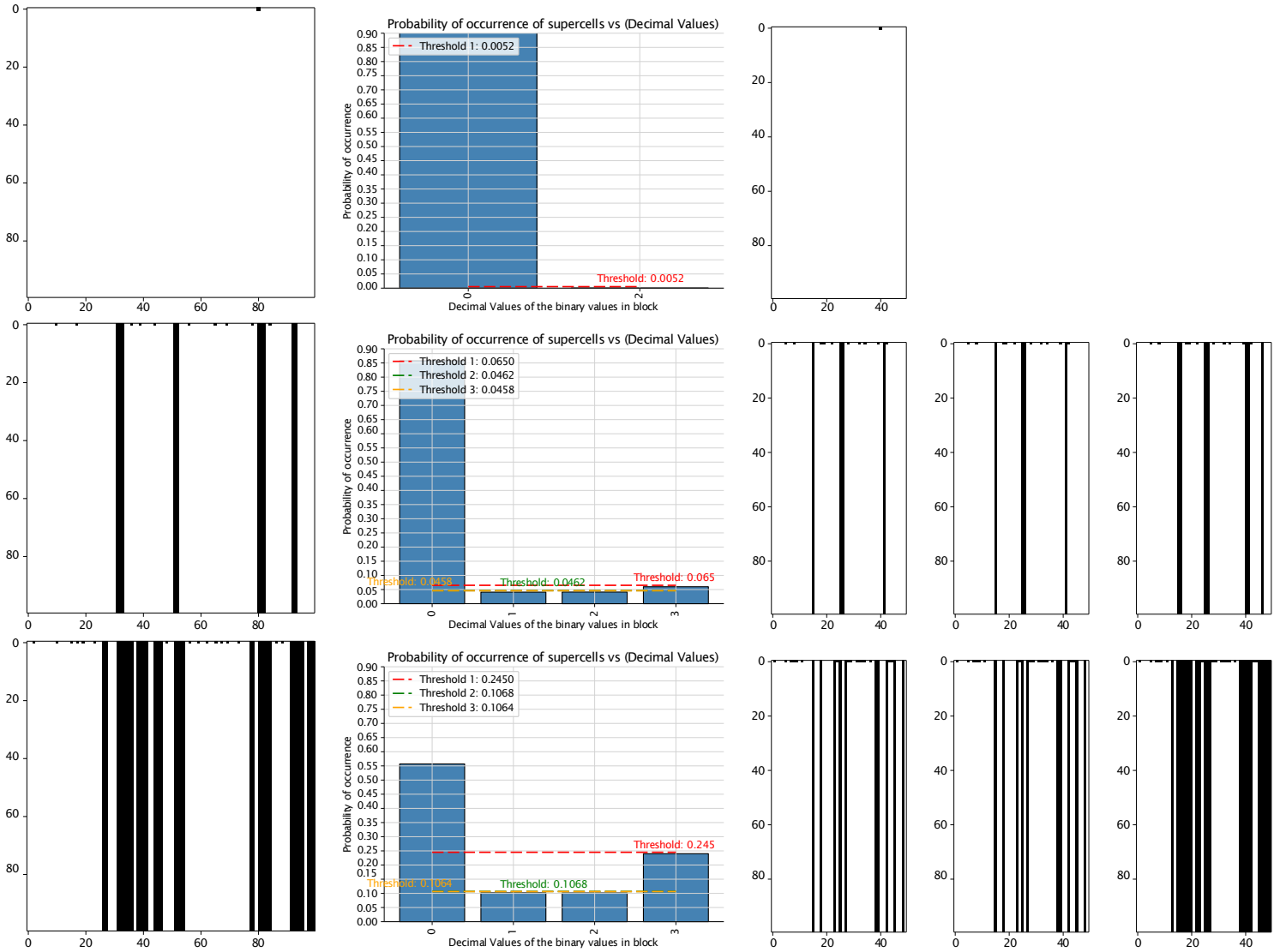

Table 86: FHCG plots for ECA Rule 126.

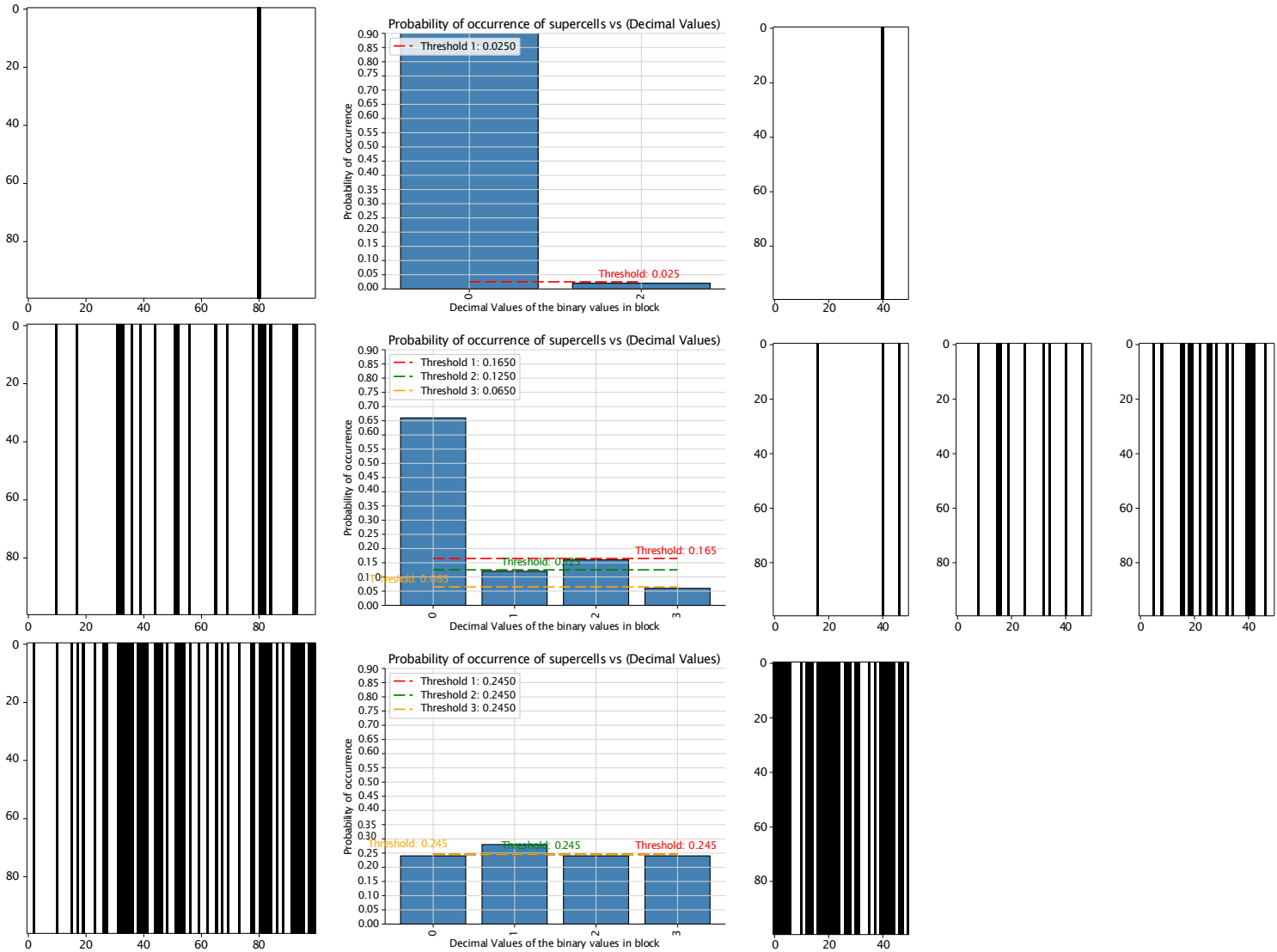

Table 87: FHCG plots for ECA Rule 126.

Table 88: FHCG plots for ECA Rule 126.

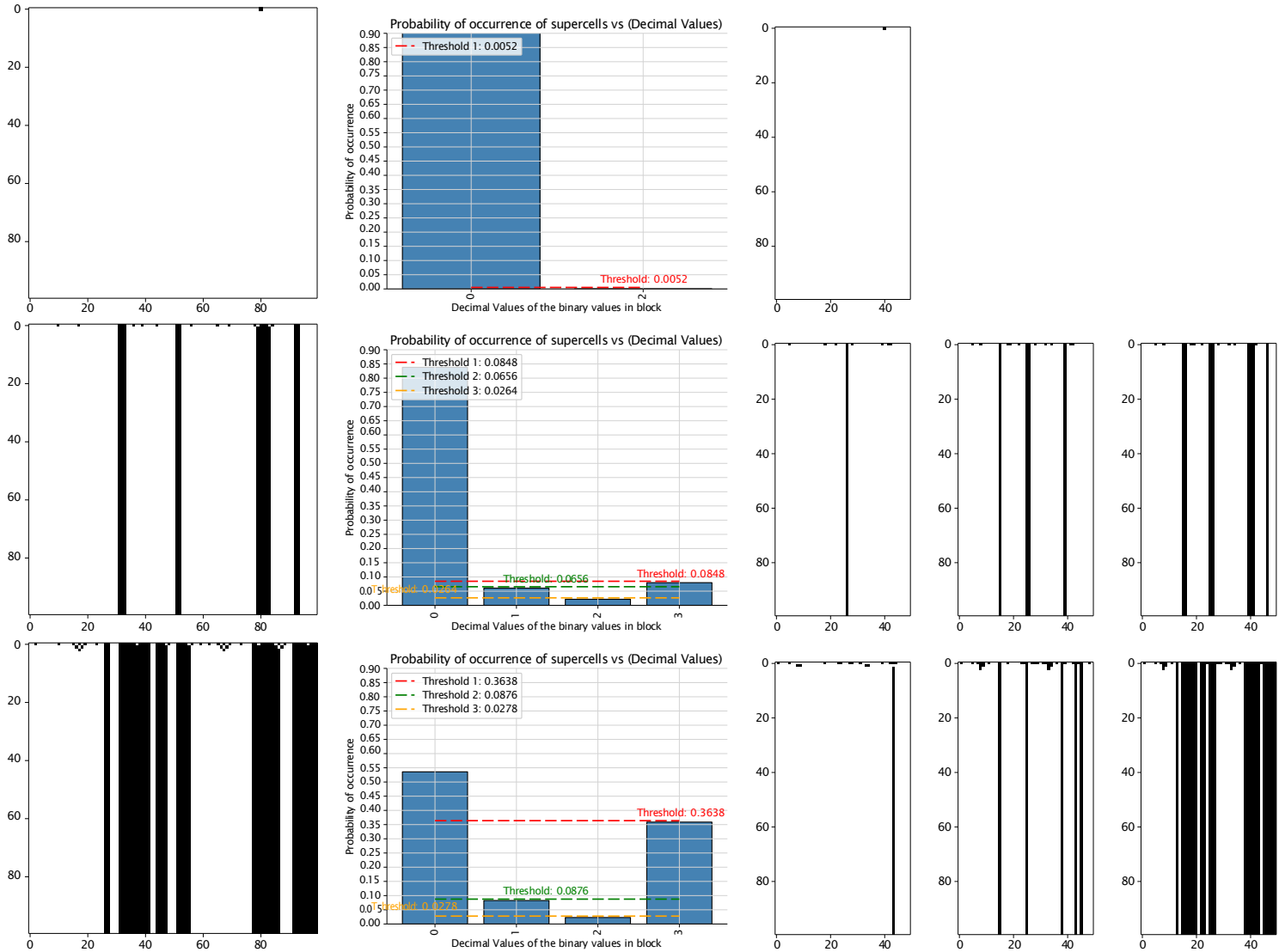

Table 89: Game of Life from MNCA with standard B3/S23 rule. The figure shows high definition (above) and FHCG (below). The chosen handpicked threshold for FHCG-Evo-MNCA is 0.011

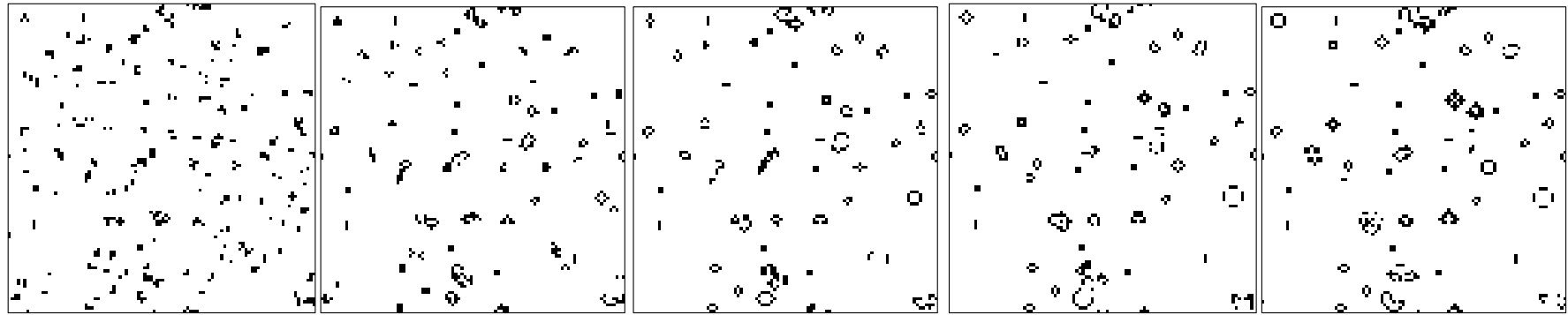

Game of Life (HD)

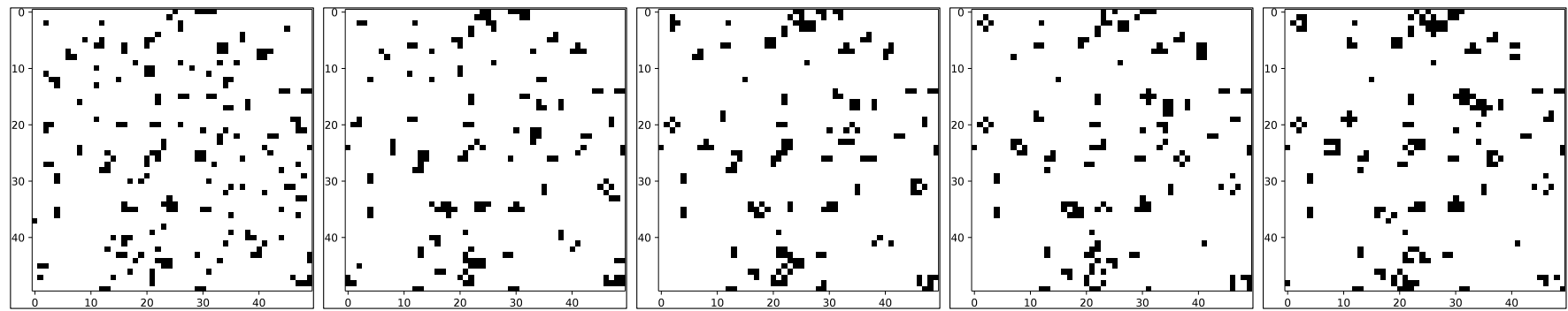

Game of Life (FHCG)

Table 90: Example of evolved rule from Evo-MNCA The values of the evolved genotype are shown on top of the image. The figure shows high definition (above) and FHCG (below). The chosen handpicked threshold for FHCG-Evo-MNCA is 0.025

[[(0.451, 0.713, 0), (0.449, 0.663, 1), (0.191, 0.52, 1), (0.089, 0.296, 0)], [(0.332, 0.51, 0), (0.084, 0.307, 0), (0.182, 0.465, 1)], [(0.799, 0.902, 0), (0.303, 0.456, 0), (0.835, 0.984, 1)]]

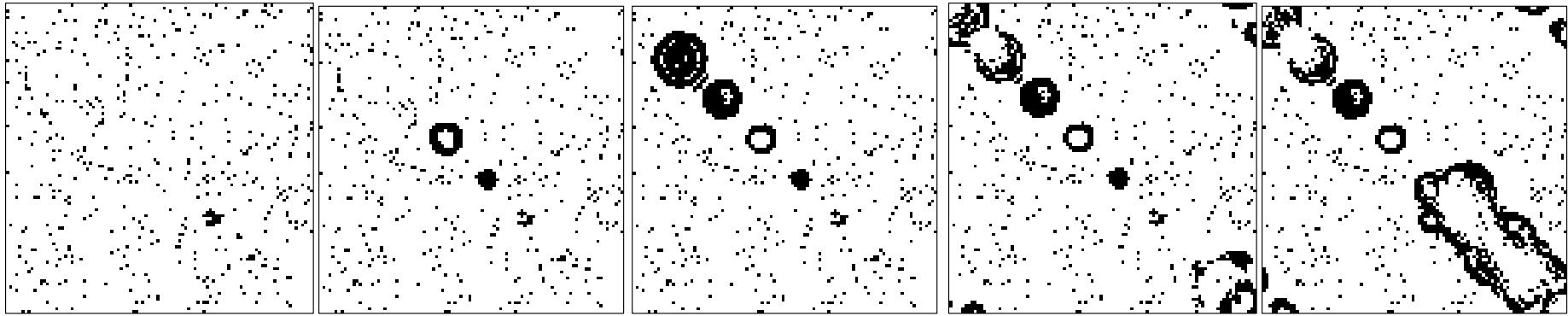

Evolved rule (HD)

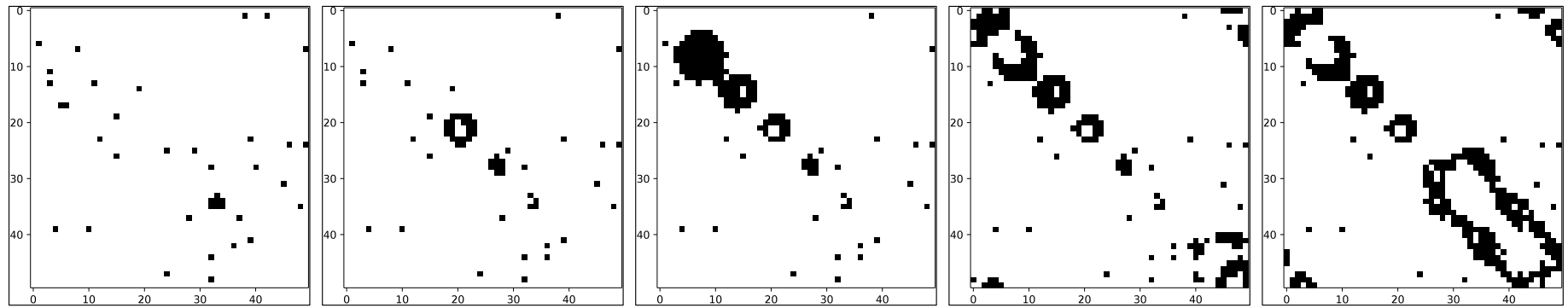

Evolved rule (FHCG)